\documentclass[11pt,a4paper]{article}\usepackage[]{graphicx}\usepackage[]{xcolor}
\makeatletter
\def\maxwidth{ %
  \ifdim\Gin@nat@width>\linewidth
    \linewidth
  \else
    \Gin@nat@width
  \fi
}
\makeatother

\definecolor{fgcolor}{rgb}{0.345, 0.345, 0.345}
\newcommand{\hlnum}[1]{\textcolor[rgb]{0.686,0.059,0.569}{#1}}%
\newcommand{\hlstr}[1]{\textcolor[rgb]{0.192,0.494,0.8}{#1}}%
\newcommand{\hlcom}[1]{\textcolor[rgb]{0.678,0.584,0.686}{\textit{#1}}}%
\newcommand{\hlopt}[1]{\textcolor[rgb]{0,0,0}{#1}}%
\newcommand{\hlstd}[1]{\textcolor[rgb]{0.345,0.345,0.345}{#1}}%
\newcommand{\hlkwa}[1]{\textcolor[rgb]{0.161,0.373,0.58}{\textbf{#1}}}%
\newcommand{\hlkwb}[1]{\textcolor[rgb]{0.69,0.353,0.396}{#1}}%
\newcommand{\hlkwc}[1]{\textcolor[rgb]{0.333,0.667,0.333}{#1}}%
\newcommand{\hlkwd}[1]{\textcolor[rgb]{0.737,0.353,0.396}{\textbf{#1}}}%

\usepackage{framed}
\makeatletter
\newenvironment{kframe}{%
 \def\at@end@of@kframe{}%
 \ifinner\ifhmode%
  \def\at@end@of@kframe{\end{minipage}}%
  \begin{minipage}{\columnwidth}%
 \fi\fi%
 \def\FrameCommand##1{\hskip\@totalleftmargin \hskip-\fboxsep
 \colorbox{shadecolor}{##1}\hskip-\fboxsep
     \hskip-\linewidth \hskip-\@totalleftmargin \hskip\columnwidth}%
 \MakeFramed {\advance\hsize-\width
   \@totalleftmargin\z@ \linewidth\hsize
   \@setminipage}}%
 {\par\unskip\endMakeFramed%
 \at@end@of@kframe}
\makeatother

\definecolor{shadecolor}{rgb}{.97, .97, .97}
\definecolor{messagecolor}{rgb}{0, 0, 0}
\definecolor{warningcolor}{rgb}{1, 0, 1}
\definecolor{errorcolor}{rgb}{1, 0, 0}
\newenvironment{knitrout}{}{} % an empty environment to be redefined in TeX

\usepackage{alltt}
\usepackage{fullpage}
\usepackage[utf8]{inputenc}
\usepackage[T1]{fontenc}
\usepackage{amsmath}
\usepackage{amsthm}
\usepackage{amsfonts}
\usepackage{amssymb}
\usepackage{xcolor}
\usepackage{array,multirow,makecell}
\usepackage{caption}
\usepackage{subcaption}
\usepackage{authblk}
\usepackage{float}
\usepackage{tikz}
\usepackage{graphicx}
\usepackage[ruled,lined,noend,linesnumbered]{algorithm2e}

\usepackage{hyperref}
\usepackage{longtable}
\usepackage{booktabs}

\newtheorem{definition}{Definition}[section]
\theoremstyle{plain}
\newtheorem{assumption}{Assumption}
\newtheorem{theo}[definition]{Theorem}
\newtheorem{lemma}[definition]{Lemma}
\newtheorem{prop}[definition]{Proposition}
\newtheorem{cor}[definition]{Corollary}
\theoremstyle{remark}
\newtheorem{remark}{Remark}[section]

\def\cX{{\mathcal{X}}}
\def\R{{\mathbb{R}}}
\def\N{{\mathbb{N}}}
\def\P{{\mathbb{P}}}
\def\E{{\mathbb{E}}}
\def\d{\mathrm{d}}
\newcommand{\J}{\mathcal{J}}
\newcommand{\cE}{\mathcal{E}}
\newcommand{\cI}{\mathcal{I}}

\renewcommand{\r}[1]{\texttt{#1}\xspace}
\newcommand{\RR}{\r{R}}
\newcommand{\IBMPopSim}{\r{IBMPopSim}}
\newcommand{\Rcpp}{\r{Rcpp}}
\newcommand{\cpp}{\r{C++}}

\title{Efficient simulation of individual-based population models: the \RR Package \IBMPopSim}
\date{}

\author[1]{Daphné Giorgi}
\author[2]{Sarah Kaakai\footnote{The research of the author is funded by the European Union (ERC, SINGER, 101054787). Views and opinions
expressed are however those of the author(s) only and do not necessarily
reflect those of the European Union or the European Research Council. Neither
the European Union nor the granting authority can be held responsible
for them.}}
\author[1]{Vincent Lemaire}
\affil[1]{Laboratoire de Probabilités, Statistique et Modélisation, Sorbonne Université and Université Paris Cité, CNRS, F-75005 Paris, France.}
\affil[2]{Laboratoire Manceau de Mathématiques \& FR CNRS 2962,  Le Mans Université,  Le Mans,  France.}

\IfFileExists{upquote.sty}{\usepackage{upquote}}{}
\begin{document}
\maketitle

\begin{abstract}
    The \RR Package \IBMPopSim aims to simulate the random evolution of heterogeneous populations using stochastic Individual-Based Models (IBMs). 
    The package enables users to simulate population evolution, in which individuals are characterized by their age and some characteristics, and the population is modified by different types of events, including births/arrivals, death/exit events, or changes of characteristics. The frequency at which an event can occur to an individual can depend on their age and characteristics, but also on the characteristics of other individuals (interactions). 
    Such models have a wide range of applications in fields including  actuarial science, biology, ecology or epidemiology. 

\IBMPopSim overcomes the limitations of time-consuming IBMs simulations by implementing new efficient algorithms  based on thinning methods, which are compiled using the \Rcpp package while providing a user-friendly interface.

\vspace*{1em}
\noindent \textbf{Keywords:} Individual-based models, stochastic simulation, population dynamics, Poisson measures, thinning method, actuarial science, insurance portfolio simulation.
\end{abstract}

\section*{Introduction}

In various fields, advances in probability have contributed to the development of a new mathematical framework for so-called individual-based stochastic population dynamics, also called stochastic Individual-Based Models (IBMs).

%These models are initially developed in a Markovian setup in view of applications in mathematical biology and ecology.  After the pioneer works~\cite{FouMel04, Tra06,champagnat2006unifying} there is a large community that has used this formalism for the study of the evolution of structured populations (see e.g. \cite{FerTra09, collet2013rigorous, BanMel15, costa2016stochastic, billiard2016effect,  lavallee2019stochastic, meleard2019birth, calvez2020horizontal}).
Stochastic IBMs allow the modeling in continuous time of populations dynamics structured by age and/or characteristics. In the field of mathematical biology and ecology, a large community has used this formalism for the study of the evolution of structured populations (see e.g. \cite{FerTra09, collet2013rigorous, BanMel15, costa2016stochastic, billiard2016effect,  lavallee2019stochastic, meleard2019birth, calvez2020horizontal}), after the pioneer works~\cite{FouMel04,  champagnat2006unifying, tran_2008}.

IBMs are also useful in demography and actuarial sciences, for the modeling of  human populations dynamics (see e.g. \cite{Ben10, Bou16, karoui2021simulating}). They allow the modeling of heterogeneous and complex population dynamics, which can be used to compute demographic indicators or simulate the evolution of insurance portfolios in order to study the basis risk, compute cash flows for annuity products or pension schemes, or for a fine assessment of mortality models (\cite{barrieu2012understanding}).
% In a quite different field, IBMs have been used to study human populations and to calculate demographic indicators for actuarial purposes or to take into account longevity risk in financial products. We refer to \cite{Ben10, Bou16, karoui2021simulating} for some applications in this field.
There are other domains in which stochastic IBMs can be used, for example in epidemiology with stochastic compartmental models, neurosciences, cyber risk, or Agent-Based Models (ABMs) in economy and social sciences, which can be seen as IBMs. \\[2mm]
\indent Many mathematical results have been obtained in the literature cited above, for quantifying  the limit behaviors of IBMs in long time or in large population. 
In particular, pathwise representations of IBMs have been introduced in \cite{FouMel04} (and extended to age-structured populations in \cite{tran_2008}), as measure-valued pure jumps Markov processes, solutions of SDEs driven by  Poisson measures. These pathwise representations are based on the \emph{thinning}  and projection of  Poisson random measures defined on extended spaces.   However, the simulation of large and interacting populations is often referred as computationally expensive.

The aim of the \RR package \IBMPopSim is to meet the needs of the various communities for efficient  tools in order to simulate the evolution of stochastic IBMs.  \IBMPopSim   provides a general framework for the simulation of a wide class of IBMs, where individuals are characterized by their age and/or a set of characteristics. Different types of events can be included in the modeling by users, depending on their needs: births, deaths, entry or exit in/to the population and changes of characteristics (swap events). Furthermore, the various events that can happen to individuals in the population can occur at a non-stationary frequency, depending on the individuals' characteristics and time, and also including potential interactions between individuals.

We introduce a unified mathematical and simulation framework for this class of IBMs, generalizing the pathwise representation of IBMs by thinning of Poisson measures, as well as the associated population simulation algorithm, based on an acceptance/rejection procedure. In particular, we provide general sufficient conditions on the event intensities  under which the simulation of a particular model is possible.

We opted to implement the algorithms of the \IBMPopSim package using the \Rcpp package, a tool facilitating the seamless integration of high-performance \cpp code into easily callable \RR functions (\cite{JSSv040i08}). With just a few lines of \cpp code, \IBMPopSim offers user-friendly R functions for defining IBMs. Once events and their associated intensities are specified, an automated procedure creates the model. This involves integrating the user's source code into the primary \cpp code using a template mechanism. Subsequently, \Rcpp is invoked to compile the model and integrate it into the \RR session. Following this process, the model becomes callable with varying parameters, enabling the generation of diverse population evolution scenarios. \\
Combined with the  design of the simulation algorithms, the package structure yields very competitive simulation runtimes for IBMs, while staying user-friendly for \RR users. Several outputs function are also implemented in \IBMPopSim. For instance the package allows the construction and visualization of age pyramids, as well as  the construction of death and exposures table from the censored individual data, compatible  with \RR packages concerned with mortality modelling, such as \cite{Rdemography} or \cite{Rstmomo}.  Several examples are provided in the form of \RR vignettes on the website~\url{https://daphnegiorgi.github.io/IBMPopSim/}, and in recent works of \cite{karoui2021simulating} and \cite{roget2022positive}.

Designed for applications in social sciences, the \RR package \texttt{MicSim}~\cite{Zin14} can be used for continuous time microsimulation. In continuous-time microsimulation, individual life-courses are usually specified by sequences of state transitions (events) and the time spans between these transitions. The state space is usually discrete and finite, which is no necessarily the case in \IBMPopSim, where individuals can have continuous characteristics. But most importantly, microsimulation does not allow for interactions between individuals. Indeed, microsimulation produces separately the life courses of all individuals in the populations, based on the computation of the distribution functions of the waiting times in the distinct states of the state space, for each individual (\cite{Zin14}). This can be slow in comparison to the simulation by thinning of event times occurring in the population, which is based on selecting event times among some competing proposed event times. Finally, \texttt{MicSim}  simplifies the Mic-Core microsimulation tool implemented in Java (\cite{zinn2009mic}). However, the  implementation in \RR of simulation algorithms yields longer simulation run times than when using \Rcpp. To the best of our knowledge, there are no other \RR packages currently available addressing the issue of IBMs efficient simulation.

In Section~\ref{section::IBM}, we introduce the mathematical framework that characterizes the class of Stochastic Individual-Based Models (IBMs) that can be implemented in  the \IBMPopSim package. In particular, a general pathwise representation of IBMs is presented. The population dynamics is obtained as the solution of an SDE driven by Poisson measures, for which we obtain existence and uniqueness results in Theorem~\ref{ThEqZ}.  Additionally, a succinct overview of the package is provided. In Section~\ref{sec::simulation}  the two  main algorithms for simulating the population evolution of an IBM across the interval $[0, T]$ are detailed.
In Section~\ref{sec::package} we present the main functions of the \IBMPopSim package, which allow for the definition of events and their intensities, the creation of a model, and the simulation of scenarios. 
Two examples are detailed in Sections~\ref{SectionInsurancePortofio} and~\ref{section:ExempleInteraction}, featuring applications involving an heterogeneous insurance portfolio characterized by entry and exit events, and an age and size-structured population with intricate interactions.

\section{Stochastic Individual-Based Models (IBMs) in IBMPopSim}
\label{section::IBM}
Stochastic Individual-Based Models (IBMs) represent a broad class of random population dynamics models, allowing the description of populations evolution on a microscopic scale. 
Informally, an IBM can be summarized by the description of the individuals constituting the population, the various types of events that can occur to these individuals, along with their respective frequencies.
% \IBMPopSim has been developed to efficiently simulate these models, even when they  include interactions. 
In  \IBMPopSim,  individuals can be  characterized by  their age and/or a collection of discrete or continuous characteristics. Moreover, the package enables users to simulate efficiently populations in which one or more of the  following event types may occur:
\begin{itemize}
    \item \textbf{Birth event}: addition of an individual of age 0 to the population.
    \item \textbf{Death event}: removal of an individual from the population.
    \item \textbf{Entry event}: arrival of an individual in the population.
    \item \textbf{Exit (emigration) event}: exit from the population (other than death).
    \item \textbf{Swap event}: an individual changes characteristics.
\end{itemize}

Each event type  is linked to an associated event kernel, describing how the population is modified following the occurrence of the event. For some event types, the event kernel requires explicit specification. This is the case for entry events when a new individual joins the population. Then,the model should specify how the age and characteristics of this new individual are chosen.  For instance, the characteristics of a new individual in the population can be chosen uniformly in the space of all characteristics, or can depend on the distribution of his parents or those of the other individuals composing the population.

The last component of an IBM are the event intensities.  Informally, an event intensity  is a function $\lambda^e_t(I, Z)$  describing the frequency at which an event $e$ can occur to an individual $I$ in a population $Z$ at a time $t$. Given a history of the population $(\mathcal{F}_t)$, the probability of event $e$ occurring to individual $I$ during a small interval of time $(t,t+dt]$ is proportional to $\lambda^e(I,t)$:
\begin{equation}
  \mathbb{P}(\text{event } e \text{ occurring to $I$ during } (t,t+dt] | \mathcal{F}_t) \simeq \lambda^e_t(I, Z)dt.
\end{equation}
The intensity function $\lambda^e$ can include  dependency on the individual's $I$ age and characteristics, the time $t$, or the population composition $Z$ in the presence of interactions.\\

\subsection{Brief package overview}
Prior to providing a detailed description of an Individual-Based Model (IBM), we present a simple model of birth and death in an age-structured ``human'' population.
We assume no interactions between individuals, and individuals are characterized by their gender, in addition to their age.
In this simple model, all individuals, regardless of gender, can give birth when their age falls between 15 and 40 years, with a constant birth rate of 0.05. The death intensity is assumed to follow a Gompertz-type intensity depending on age. 
The birth and death intensities are then given by  
\begin{equation*}
\lambda^b(t, I) = 0.05 \times \mathbf{1}_{[15,40]}(a(I,t)), \quad \lambda^d(I,t) = \alpha\exp(\beta a(I,t)), 
\end{equation*} 
with $a(I,t)$ the age of individual $I$ at time $t$. 
Birth events are also characterized with a kernel determining the gender of the newborn, who is male with  probability $p_{male}$. 

\paragraph{Model creation}  
To implement this model in \IBMPopSim, it is necessary to individually define each event type. In this example, the \r{mk\_event\_individual} function is used. The creation of an event involves  a few lines of \cpp instructions defining the intensity and, if applicable, the kernel of the event. For a more in depth description of the event creation step and its parameters, we refer to Section~\ref{sec::package_events}.

The events of this simple model are for example defined through the following calls.
\begin{knitrout}
\definecolor{shadecolor}{rgb}{0.969, 0.969, 0.969}\color{fgcolor}\begin{kframe}
\begin{alltt}
\hlstd{birth_event} \hlkwb{<-} \hlkwd{mk_event_individual}\hlstd{(}
  \hlkwc{type} \hlstd{=} \hlstr{"birth"}\hlstd{,}
  \hlkwc{intensity_code} \hlstd{=} \hlstr{"result = birth_rate(I.age(t));"}\hlstd{,}
  \hlkwc{kernel_code} \hlstd{=} \hlstr{"newI.male = CUnif(0,1) < p_male;"}\hlstd{)}

\hlstd{death_event} \hlkwb{<-} \hlkwd{mk_event_individual}\hlstd{(}
  \hlkwc{type} \hlstd{=} \hlstr{"death"}\hlstd{,}
  \hlkwc{intensity_code} \hlstd{=} \hlstr{"result = alpha * exp(beta * I.age(t));"}\hlstd{)}
\end{alltt}
\end{kframe}
\end{knitrout}
In the \cpp codes, the names \r{birth\_rate}, \r{p\_male}, \r{alpha} and \r{beta} refer to the model parameters defined in the following list. 
\begin{knitrout}
\definecolor{shadecolor}{rgb}{0.969, 0.969, 0.969}\color{fgcolor}\begin{kframe}
\begin{alltt}
\hlstd{params} \hlkwb{<-} \hlkwd{list}\hlstd{(}
  \hlstr{"alpha"} \hlstd{=} \hlnum{0.008}\hlstd{,} \hlstr{"beta"} \hlstd{=} \hlnum{0.02}\hlstd{,}
  \hlstr{"p_male"} \hlstd{=} \hlnum{0.51}\hlstd{,}
  \hlstr{"birth_rate"} \hlstd{=} \hlkwd{stepfun}\hlstd{(}\hlkwd{c}\hlstd{(}\hlnum{15}\hlstd{,} \hlnum{40}\hlstd{),} \hlkwd{c}\hlstd{(}\hlnum{0}\hlstd{,} \hlnum{0.05}\hlstd{,} \hlnum{0}\hlstd{)))}
\end{alltt}
\end{kframe}
\end{knitrout}

In a second step, the model is created by calling the function \r{mk\_model}. A \cpp source code is automatically created through a template mechanism based on the events and parameters, subsequently compiled using the \r{sourceCpp} function from the \Rcpp package.
\begin{knitrout}
\definecolor{shadecolor}{rgb}{0.969, 0.969, 0.969}\color{fgcolor}\begin{kframe}
\begin{alltt}
\hlstd{birth_death_model} \hlkwb{<-} \hlkwd{mk_model}\hlstd{(}
  \hlkwc{characteristics} \hlstd{=} \hlkwd{c}\hlstd{(}\hlstr{"male"} \hlstd{=} \hlstr{"bool"}\hlstd{),}
  \hlkwc{events} \hlstd{=} \hlkwd{list}\hlstd{(death_event, birth_event),}
  \hlkwc{parameters} \hlstd{= params)}
\end{alltt}
\end{kframe}
\end{knitrout}

\paragraph{Simulation} 
Once the model is created and compiled, the \r{popsim} function is called to simulate the evolution of a population according to this model. To achieve this, an initial population must be defined. In this example, we extract a population from a dataset specified in the package (a sample of $100\,000$ individuals based on the population of England and Wales in 2014). It is also necessary to set bounds for the events intensities. In this example, they are obtained by assuming that the maximum age for an individual is 115 years. 
\begin{knitrout}
\definecolor{shadecolor}{rgb}{0.969, 0.969, 0.969}\color{fgcolor}\begin{kframe}
\begin{alltt}
\hlstd{a_max} \hlkwb{<-} \hlnum{115}
\hlstd{events_bounds} \hlkwb{=} \hlkwd{c}\hlstd{(}
  \hlstr{"death"} \hlstd{= params}\hlopt{$}\hlstd{alpha} \hlopt{*} \hlkwd{exp}\hlstd{(params}\hlopt{$}\hlstd{beta} \hlopt{*} \hlstd{a_max),}
  \hlstr{"birth"} \hlstd{=} \hlkwd{max}\hlstd{(params}\hlopt{$}\hlstd{birth_rate))}
\end{alltt}
\end{kframe}
\end{knitrout}
The function \r{popsim} can now be called to simulate the population starting from the initial population \r{population(EW\_pop\_14\$sample)} up to time $T = 30$. 
\begin{knitrout}
\definecolor{shadecolor}{rgb}{0.969, 0.969, 0.969}\color{fgcolor}\begin{kframe}
\begin{alltt}
\hlstd{sim_out} \hlkwb{<-} \hlkwd{popsim}\hlstd{(}
  \hlstd{birth_death_model,}
  \hlkwd{population}\hlstd{(EW_pop_14}\hlopt{$}\hlstd{sample),}
  \hlstd{events_bounds,}
  \hlkwc{parameters} \hlstd{= params,} \hlkwc{age_max} \hlstd{= a_max,}
  \hlkwc{time} \hlstd{=} \hlnum{30}\hlstd{)}
\end{alltt}
\end{kframe}
\end{knitrout}
The data frame \r{sim\_out\$population} contains the information (birth, death, gender) on individuals who lived in the population over the period $[0,30]$. Functions of the package allows to provide aggregated information on the population.

In the remainder of this section,  we define rigorously the class of IBMs that can be simulated in \IBMPopSim, along with the assumptions that are required in order for the population to be simulatable. The  representation of age-structured IBMs based  on measure-valued processes, as introduced in \cite{tran_2008}, is generalized to a wider class of abstract population dynamics. The modeling differs slightly here, since individuals are ``kept in the population'' after their death (or exit), by including the death/exit date as an individual trait. 

\subsection{Population}
\label{sec::population}

\paragraph{Notations} In the remainder of the paper, the filtered probability space is denoted by $(\Omega,\{\mathcal{F}_t \},\P)$, under the usual assumptions.  All processes are assumed to be càdlàg and adapted to the filtration $\{\mathcal{F}_t \}$ (for instance the history of the population) on a time interval $[0,T]$.  For a càdlàg process $X$, we denote  $X_{t^-} := \lim_{\genfrac{}{}{0pt}{2}{s\to t}{s<t}} X_s$.
\paragraph{Individuals} An individual is represented by a triplet $I = (\tau^b, \tau^d, x) \in \cI= \R \times \bar \R \times \cX$ with:
 \begin{itemize}
 \item $\tau^b \in \R$ the date of birth, 
 \item  $\tau^d \in \bar \R$  the death date, with $\tau^d = \infty$ if the individual is still alive, 
 \item  a collection $x \in \cX$ of characteristics  where $\cX$ is the space of characteristics.
 \end{itemize}
Note that in IBMs, individuals are usually characterized by their age $a(t) =t-\tau^b$ instead of their date of birth $\tau^b$. However, using the latter is actually easier for the simulation, as it remains constant over time.
\paragraph{Population process} The population at a given time $t$ is a random set 
$$Z_t=\{ I_k \in \cI ; \; k= 1,\dots, N_t\},$$ composed of all individuals (alive or dead) who have lived in the population before time $t$. As a random set, $Z_t$ can be represented by a random counting measure on $\cI$ , that is 
an integer-valued measure $Z: \Omega \times \cI \to \bar \N$  where  for $A \in \cI$, $Z(A)$ is the (random) number of individuals $I$ in the subset $A$. With this representation: 
\begin{align}
\label{eq::popZ}
& Z_t (\d \tau^b, \d \tau^d , \d x)= \sum_{k=1}^{N_t} \delta_{I_k} (\tau^b, \tau^d,x), \nonumber\text{ with }  \int_{\cI} f(\tau^b, \tau^d, x)Z_t (\d \tau^b, \d \tau^d , \d x) = \sum_{k=1}^{N_t} f(I_k).
\end{align}
The number of individuals present in the population \textit{before time} $t$ is obtained by taking $f\equiv 1$:
\begin{equation}
 N_t =  \int_{\cI}  Z_t(\d\tau^b, \d \tau^d, \d x) = \sum_{k=1}^{N_t} \boldsymbol{1}_{\cI} (I_k).
\end{equation}
Note that $(N_t)_{t\geq 0}$ is an increasing process since dead/exited individuals are kept in the population $Z$. The number of alive individuals  in the population at time $t$ is:
\begin{equation}
\label{eq:Nta}
N_t^a =  \int_{\cI}  \mathsf{1}_{\{\tau^d > t \} }Z_t(\d\tau^b, \d \tau^d, \d x) = \sum_{k=1}^{N_t} \mathsf{1}_{\{\tau^d_k > t \} }.
\end{equation}
Another example is the number of alive individuals  of age over $a$ is
\begin{equation*}
N_t([a,+\infty)) :=  \int_{\cI}  \boldsymbol{1}_{[a,+\infty)}(t-\tau^b)\mathsf{1}_{]t,\infty]}(\tau^d) Z_t(\d\tau^b, \d \tau^d, \d x) = \sum_{k=1}^{N_t}  \boldsymbol{1}_{\{ t -\tau_k^b \geq a \}}\mathsf{1}_{\{\tau^d_k \geq t \} }.
\end{equation*}

\subsection{Events}
\label{sec::events}
The population composition changes at random dates  following different types of events.  \IBMPopSim allows the simulation of IBMs with the following events types:
 \begin{itemize}
\item {\em A birth} event at time $t$  is the addition of a new individual  $I'=(t,\infty, X)$ of age $0$ to the population. Their  date of birth  is $\tau^b =t$, and characteristics is $X$, a random variable of distribution defined by the birth kernel $k^b(t,I,\d x)$ on $ \cX$, depending on $t$ and its parent $I$. The population size becomes $N_t = N_{t^-} + 1$, and the population composition after the event is
\begin{equation*}
Z_t  = Z_{t^-} +  \delta_{(t,\infty, X)}.
\end{equation*}
\item An {\em entry} event at time $t$ is also the addition of an individual $I'$ in the population. However, this individual is not of age $0$. The date of birth and characteristics of the new individual $I'= (\tau^b, \infty, X)$ are random variables of probability distribution defined by the entry kernel $k^{en}(t, \d s, \d x) $ on $\R \times \cX$. The population size becomes $N_t = N_{t^-} + 1$, and the population composition after the event is:
\begin{equation*}
Z_t  = Z_{t^-} +  \delta_{(\tau^b, \infty, X)}.
\end{equation*}
 \item  {\em A death} or {\em exit}  event of an individual  $I= (\tau^b,\infty, x)\in Z_{t^-}$  at time $t$   is the modification of its death date $\tau^d$ from $+\infty$ to $t$. This event results in the simultaneous addition of  the individual $(\tau^b,t,x)$ and removal of the individual $I$ from the population. The population size is not modified, and the population composition after the event is
\begin{equation*}
Z_t  = Z_{t^-} +\delta_{(\tau^b,t,x)}- \delta_{I}.
\end{equation*}
\item {\em A swap} event (change of characteristics) results in the simultaneous addition and removal of an individual. If an individual $ I= (\tau^b,\infty, x) \in Z_{t^-}$ changes of characteristics at time $t$, then it is removed from the population and replaced by $I' = (\tau^b,\infty, X)$. The new characteristics $X$ is  a random variable of  distribution $k^s(t, I,\d x)$ on $\cX$, depending on time, the individual's age and previous characteristics $x$.  In this case, the population size is not modified and the population becomes:
\begin{equation*}
Z_t  = Z_{t^-}   +  \delta_{(\tau^b,  \infty, X)} -  \delta_{(\tau^b, \infty, x)}.
\end{equation*}
\end{itemize}

To summarize, the space of event types is $E = \{ b, en, d, s \}$, and  the jump $\Delta Z_t = Z_t - Z_{t^-}$ (change in the population composition) generated by  an event of type $e \in \{ b, en, d, s \}$ is denoted by $\phi^e(t, I)$, with:\\

\begin{table}[H]
\begin{center}
\begin{tabular}{|c|c|c|c|}
\hline Event & Type &   $\phi^e(t, I)$  & New individual \\
\hline Birth & $b$ & $\delta_{(t, \infty,  X)}$ & $ \tau^b =t, \; X \sim k^b(t,I,\d x)$\\
\hline Entry & $en$& $\delta_{(\tau^b, \infty,  X)}$ & $(\tau^b, X) \sim k^{en}(t,\d s, \d x)$  \\
\hline  Death/Exit & $d$ &  $\delta_{(\tau^b, t,x)} - \delta_{(\tau^b, \infty, x)}$  & {$\tau^d = t$} \\
\hline Swap & $s$ & $\delta_{(\tau^b, \infty , X)} - \delta_{(\tau^b, \infty, x)}$ &  $ X \sim k^s(t,I,\d x)$\\
\hline
\end{tabular}
\end{center}
\caption{Events action}
\label{TableEvAction}
\end{table}
\begin{remark}
\label{remark::popfinale}
\begin{itemize}
\item At time $T$, the population $Z_T$ contains all individuals who lived in the population before $T$, including dead/exited individuals. If there are no swap events, or entries,the population state $Z_t$ for any  time $t\leq T$ can be obtained from  $Z_T$. Indeed, if $Z_T = \sum_{k=1}^{N_T}  \delta_{I_k}$, then the population at time $t\leq T$ is simply composed of the individuals born before $t$:
\begin{equation*}
Z_t  = \sum_{k=1}^{N_T} \boldsymbol{1}_{\{\tau^b_k \leq t \}}   \delta_{I_k}.
\end{equation*}
\item In the presence of entries (open population), a characteristic $x$ can track  the  individuals'  entry dates. Then, the previous equation can be easily modified in order to obtain the population $Z_t$ at time $t\leq T$ from $Z_T$.
%\item If the IBM includes swap events, then for an individual $I = (\tau^b, \tau^d, x) \in Z_T$, the characteristics $x  \in \mathcal{X}$ are the characteristics of individual $I$ at time $T$ only. However, an approximation of the population path can be obtained by considering the vector $(Z_{t_0}, Z_{t_1}, \dots, Z_{t_n})$, with $0=t_0  <t_1 < \dots< t_n=T$. This is detailed in Section \ref{simulation} and \ref{SectionInsurancePortofio}.
\end{itemize}
\end{remark}
\subsection{Events intensity}
\label{sec::event_intensity}

Once the different event types have been defined in the population model, the frequency at which each event occur in the population $e$ have to be specified. \\
Informally, the intensity $\Lambda^e_t(Z_t)$ at which an event $e$ can occur is defined by
\begin{equation*}
\mathbb P\big( \text{event } e \text { occurs in the population }  Z_t  \in (t,t+\d t] | \mathcal{F}_t \big) \simeq  \Lambda^e_t (Z_t)\d t.
\end{equation*}
For a more formal definition of stochastic intensities, we refer to \cite{bremaud1981point} or \cite{KaaElK20}.\\
The  form of the intensity  function $(\Lambda^e_t (Z_t))$ determines the population simulation algorithm in \IBMPopSim:
\begin{itemize}
\item When the event intensity does not depend on the population state,
\begin{equation}
\label{PoissonIntensity}
(\Lambda^e_t (Z_t))_{t\in [0,T]} = (\mu^e(t))_{t \in [0,T]},
\end{equation}
with $\mu^e$ a deterministic function, the events of type $e$ occur at the jump times of an inhomogeneous Poisson process of intensity function $(\mu^e(t))_{t \in [0,T]}$. When such an event occurs, the individual to whom the event happens to is drawn uniformly among alive individuals in the population. \\
In a given model, the set of events $e\in E$ with Poisson intensities will be  denoted by $\mathcal{P}$.
\item Otherwise, we assume that the global intensity $\Lambda^e_t(Z_t)$ at which the events  of type $e$ occur in the population can be written as  the sum of individual intensities $\lambda^e_t(I,Z_t)$:
\begin{align}
\label{eq:GlobalIntensity}
&\Lambda^e_t (Z_t) = \sum_{k=1}^{N_t} \lambda^e_t ( I_k,Z_t),  \\
& \nonumber \text{with } \mathbb P\big( \text{event } e \text { occurs to an individual } I \in (t,t+\d t] | \mathcal{F}_t \big) \simeq  \lambda^e_t (I,Z_t)\d t.
\end{align}
\end{itemize}
Obviously, nothing can happen to dead or exited individuals, i.e. individuals $I= (\tau^b, \tau^d, x)$  with $\tau^d \leq t$. Thus, individual event intensities are assumed to be null  for dead/exited individuals:
\begin{equation*}
\lambda^e_t ( I,Z_t) = 0, \text{ if }\tau^d \leq t, \text{ so that } \Lambda^e_t (Z_t) = \sum_{k=1}^{N_t^a} \lambda^e_t ( I_k,Z_t),
\end{equation*}
with $N^a_t$ the number of alive individuals at time $t$.

The event's individual  intensity  $\lambda^e_t (I,Z_t)$ can depend on time (for instance when there is a mortality reduction over time),  on the individual's age $t-\tau^b$ and characteristics, but also on the population composition $Z_t$.
The dependence of $\lambda^e$ on the population $Z$ models interactions between individuals in the populations.
Hence,  two types of individual intensity functions can be implemented in \IBMPopSim:
\begin{enumerate}
\item \textit{No interactions:} The intensity function $\lambda^e$ does not depend on the population composition. The intensity at which the event of type $e$ occur to an individual $I$ only depends on its date of birth and characteristics:
\begin{equation}
\label{eq:intensityNointeraction}
\lambda^e_t (I,Z_t) = \lambda^e(t, I),
\end{equation}
where $\lambda^e: \mathbb{R}_+ \times \cI \to \R^+$ is a deterministic function.
In a given model, we denote by $\mathcal{E}$ the set of event types with individual intensity \eqref{eq:intensityNointeraction}.
\item \textit{``Quadratic'' interactions:}  The intensity at which an event of type $e$  occurs to an individual $I$ depends on $I$ and on the population composition, through an interaction function $W^e$. The quantity $W^e(t, I,J)$ describes the intensity of interactions between two alive individuals $I$ and $J$ at time $t$, for instance in the presence of competition or cooperation.  In this case, we have
\begin{equation}
\label{eq:intensityInteraction}
\lambda^e_t(I,Z_t)=\sum_{j=1}^{N_t} W^e(t, I, I_j) = \int_{\cI} W^e(t, I, (\tau^b,\tau^d,x)) Z_t (\d\tau^b,\d\tau^d, \d x),
\end{equation}
where $W^e(t, I, (\tau^b,\tau^d,x))  = 0$ if the individual $J =(\tau^b,\tau^d,x)$ is dead, i.e. $\tau^d \leq t$. \\
In a given model, we  denote by $\mathcal{E}_W$ the set of event types with individual intensity \eqref{eq:intensityInteraction}.
\end{enumerate}

To summarize, an individual intensity in IBMPopSim can be written as:
\begin{equation}
\label{IndividualIntensity}
\lambda^e_t(I,Z_t) = \lambda^e(t, I) \mathbf{1}_{\{e \in \mathcal{E}\}} + \biggl( \sum_{j=1}^{N_t} W^e(t, I, I_j) \biggr) \mathbf{1}_{\{e \in \mathcal{E}_W\}}.
\end{equation}

\noindent \textit{Examples}\\
\textit{(i)} An example of death intensity without interaction for an individual $I=(\tau^b, \tau^d, x)$ alive at time $t$ ($t <\tau^d$) is:
\begin{equation}
\lambda^d (t,I) =  \alpha_x \exp(\beta_x a(I,t)), \text{ where }  a(I,t) = t-\tau^b
\end{equation}
is the age of the individual $I$ at time $t$.
In this classical case, the death rate of an individual $I$  is an exponential (Gompertz) function of  the individual's age, with coefficients depending on the individual's characteristics $x$. \\
\textit{(ii)} In the presence of competition between individuals, the death intensity of an individual $I$ also depend on other individuals $J$ in the population. For example, if $I=(\tau^b,\tau^d, x)$, with $x$ its size, then we can have:
\begin{equation}
\label{ex:interaction}
W^d(t,I,J) = (x_J - x)^+ \mathbf{1}_{\{\tau^d_J > t\}}, \quad \forall \; J=(\tau^b_J,\tau^d_J , x_J).
\end{equation}
This can be interpreted as follows: if the individual $I$ meets randomly an individual $J$ alive at time $t$, and  of  bigger size $x_J > x$, then he can die  at the intensity $x_J-x$. If $J$ is smaller than $I$, then he cannot kill $I$. The bigger is the size $x$ of $I$, the lower is his
death intensity $\lambda^d_t(I,Z_t) $ defined by
\begin{equation*}
\lambda^d_t(I,Z_t) = \sum_{\genfrac{}{}{0pt}{2}{J\in Z_t,}{x_J > x}} (x_J -x)\mathbf{1}_{\{\tau^d_J > t\}}.
\end{equation*}
\textit{(iii)} \IBMPopSim can simulate IBMs that include intensities expressed as a sum of Poisson intensities and individual intensities of the form  $\Lambda^e(Z_t) =\mu^e_t + \sum_{k=1}^{N_t} \lambda^e(I_k, Z_t)$.  \\
Other examples are provided in Section \ref{SectionInsurancePortofio} and Section \ref{section:ExempleInteraction}.\\

Finally, the global intensity at which an event can occur in the population is defined by:
\begin{equation}
\label{eq:globalEvIntensity}
\Lambda_t(Z_t) = \sum_{e\in \mathcal{P}} \mu^e(t) + \sum_{e \in \mathcal E} \Big(\sum_{k=1}^{N_t} \lambda^e(t, I_k)\Big) + \sum_{e \in \mathcal E_W} \Big(\sum_{k=1}^{N_t}\sum_{j=1}^{N_t} W^e(t, I_k, I_j)\Big).
\end{equation}
 An important point is that for events $e \in \mathcal E$ without interactions, the global event intensity $\Lambda^e_t(Z_t) = \sum_{k=1}^{N_t} \lambda^e(t, I_k)$ is ``of order'' $N_t^a$ defined in \eqref{eq:Nta} (number of alive individuals at time $t$). On the other hand, for events $e \in \mathcal{E}_W$ with interactions, $\Lambda^e_t(Z_t) = \sum_{k=1}^{N_t}\sum_{j=1}^{N_t} W^e(t, I_k, I_j)$ is of order $(N_t^a)^2$. Informally, this means that when the population size increases, events with interaction are  more costly to simulate. Furthermore, the numerous computations of the interaction kernel $W^e$ can also be quite costly. The randomized Algorithm \ref{algo::rzndomized}, detailed in Section \ref{sec::simulation_algo_randomized}, allows us to overcome these limitations.

\paragraph{Events intensity bounds} The simulation algorithms implemented in \texttt{IBMPopSim} are based on an acceptance/rejection procedure, which requires to specify bounds for the various events intensities $\Lambda^e_t(Z_t)$. These bounds are defined differently depending on the expression of the intensity. 
\begin{assumption}
\label{AssumptionIntensityPoisson}
For all events $e \in \mathcal{P}$ with Poisson  intensity \eqref{PoissonIntensity},  the  intensity  is assumed to be bounded on $[0,T]$:
\begin{equation}
    \forall t \in [0,T], \quad \Lambda^e_t(Z_t) = \mu^e(t) \leq \bar \mu^e.
\end{equation}
\end{assumption}
When $e \in \mathcal{E} \cup \mathcal{E}_W$ ($\Lambda^e_t(Z_t) =\sum_{k=1}^{N_t} \lambda^e_t(I_k,Z_t)$), assuming that  $\Lambda^e_t(Z_t)$ is uniformly bounded is too restrictive since the event intensity depends on the  population size. In this case, the assumption is made on the individual intensity or interaction function $W^e$, depending on the situation:
\begin{assumption}
\label{AssumptionIntensity1}
For all event types $e \in \mathcal{E}$,  the  associated individual event intensity  $\lambda^e$ with no interactions ($\lambda^e$ verifies \eqref{eq:intensityNointeraction}) is assumed to be  uniformly bounded:
\begin{equation}
\lambda^e(t, I) \leq \bar \lambda^e, \quad \forall \;  t\in [0, T],  \;   I \in \cI.
\end{equation}
In particular,
\begin{equation}
\label{EqdefbarLambda}
\forall t \in [0,T], \quad \Lambda^e_t (Z_t) = \sum_{k=1}^{N_t} \lambda^e(t, I) \leq \bar \lambda^e  N_t .
\end{equation}
\end{assumption}
\begin{assumption}
\label{AssumptionIntensity2}
For all event types $e \in \mathcal{E}_W$,  the  associated interaction function $W^e$ is assumed to be  uniformly bounded:
\begin{equation}
W^e(t, I, J) \leq \bar W^e, \quad \forall \; t\in [0,T], \;   I, J \in \cI.
\end{equation}
In particular, $\forall t \in [0,T]$,
\begin{equation*}
    \lambda^e_t (I,Z_t) = \sum_{j=1}^{N_t} W^e(t, I, I_j)  \leq  \bar W^e N_t, \quad \text{and} \quad \Lambda^e_t (Z_t) \leq \bar W^e (N_t)^2.
\end{equation*}
\end{assumption}
Assumptions \ref{AssumptionIntensityPoisson},  \ref{AssumptionIntensity1} and \ref{AssumptionIntensity2} yield that events in the population occur with  the global event intensity~$\Lambda_t(Z_t)$ \eqref{eq:globalEvIntensity}, which is dominated by a polynomial function in the population size:
\begin{equation}
\label{eq:defbarLambda}
\Lambda_t(Z_t) \leq \bar \Lambda(N_t), \quad \text{with }  \bar \Lambda (n) = \sum_{e \in \mathcal{P}} \bar \mu^e + \sum_{e\in \mathcal{E}}\bar \lambda^e  n + \sum_{e \in \mathcal E_W} \bar W^e n^2.
\end{equation}
This bound is linear in the population size if there are no interactions, and quadratic if there at least  is an event including interactions.
This assumption is the key to the algorithms implemented in \IBMPopSim. Before presenting the simulation algorithm, we close this section with a rigorous definition of an IBM, based on the pathwise representation of its dynamics a Stochastic Differential Equation (SDE) driven by Poisson random measures.\\

\subsection{Pathwise representation}
Since the seminal paper of \cite{FouMel04}, it has been shown in many examples that a stochastic IBM dynamics can be defined  rigorously as the  unique solution of an SDE driven by Poisson measures, under reasonable non explosion conditions. In the following, we introduce a unified framework for the pathwise representation of the class of stochastic IBMs introduced above.  Some recalls on Poisson random measures are presented in the Appendix \ref{section::preliminaries}, and for more details on these
representations on particular examples, we refer to the abundant literature on the subject.

In the following we consider an individual-based stochastic population $(Z_t)_{t\in [0,T]}$, keeping  the notations introduced in Section \ref{sec::events} and \ref{sec::event_intensity} for the events and their intensities. In particular, the set of events types that define the population evolution is denoted by $\mathcal{P} \cup \mathcal{E} \cup \mathcal{E}_W \subset E$, with $\mathcal{P}$ the set of events types with Poisson intensity verifying 
assumption \ref{AssumptionIntensityPoisson}, $\mathcal{E}$ the set of events types with individual intensity and no interaction, verifying Assumption  \ref{AssumptionIntensity1}, and finally $\mathcal{E}_W$ the set of event types with interactions, verifying Assumption \ref{AssumptionIntensity2}. 

\paragraph{Non explosion criterion} First, one has to ensure that the number of events occurring in the population will  not explode in finite time, leading to an infinite simulation time.  Assumptions \ref{AssumptionIntensity1} and \ref{AssumptionIntensity2} are not sufficient to guarantee the non explosion of the event number, due to the potential explosion of the population size in the presence of interactions. An example is the case when only birth events occur,  with an intensity $\Lambda^b_t(Z_t) = C_b (N_t^a)^2$ ($W^b(t, I,J) =C_b$).  Then, the number of alive individuals  $(N_t^a)_{t\geq 0}$  is a well-known pure birth process of intensity function $g(n) = C_b n^2$ (intensity of moving from state $n$ to $n+1$). This process explodes in finite time, since $g$ does not verify the necessary and sufficient non explosion criterion for pure birth Markov processes: $\sum_{n=1}^\infty \frac{1}{g(n)} = \infty$ (see e.g. Theorem 2.2 in \cite{BanMel15}).  There is thus an explosion in finite time of  birth events.

This example shows that the important point for non explosion is to control the population size. We give below a general sufficient condition on birth and entry event intensities, in order for the population size to stay finite in finite time.  This ensures that the number of events does not explode in finite time. Informally, the idea is to control the intensities by a pure birth intensity function verifying the non-explosion criterion.
\begin{assumption}
\label{Assumption:nonExplosion}
Let $e=b$ or $en$,  a birth or entry event type. If the intensity at which the events of type $e$ occur in the population are not Poissonian, i.e. $e \in \mathcal{E} \cup \mathcal{E}_W$, then there exists a function $f^e : \N \to (0, +\infty)$, such that
\begin{equation}
\sum_{n=1}^{\infty} \frac{1}{nf^e(n)} = \infty,
\end{equation}
and  for all individual $I \in \cI$ and population measure  $Z = \sum_{k=1}^{n} \delta_{I_k}$ of size $n$,
\begin{equation}
\lambda^e_t (I, Z) \leq f^e(n), \; \forall \; 0\leq t \leq T.
\end{equation}
\end{assumption}
\begin{remark}
If $e \in \mathcal{E}$,  $\lambda_t^e(I,Z) = \lambda^e(t,I) \leq \bar{\lambda}^e$ by  the domination Assumption \ref{AssumptionIntensity2}. In this case,  Assumption \ref{Assumption:nonExplosion} is always verified with $f^e(n) = \bar{\lambda}^e$. 
\end{remark}
Assumption \ref{Assumption:nonExplosion}  yields that the global intensity $\Lambda_t^e(\cdot)$ of event $e$ is bounded by a function $g^e$ only depending on the population size:
\begin{equation*}
\Lambda_t^e (Z) \leq g^e(n) := nf^e(n), \quad \text{with }\sum_{n=1}^{\infty} \frac{1}{g^e(n)} = \infty.
\end{equation*}
If $e\in \mathcal{P}$ has a Poisson intensity, then $\Lambda_t^e(Z) =\mu^e_t$ always verifies the previous equation with $g^e(n) = \bar \mu^e$.
\paragraph*{}
Before introducing the IBM SDE, let us give an idea of the equation construction.  Between two successive events, the population composition $Z_t$ stays constant, since  the population process $(Z_t)_{t \geq 0}$ is a pure jump process. Furthermore, since each event type is characterized by an intensity function,  the jumps occurring in the population can be represented by restriction and projection of a Poisson measure defined on a larger state space. More precisely,  we introduce   a random Poisson measure $Q$ on $\mathbb R^+ \times \J \times \mathbb{R}^+$,  with $\J = \mathbb N \times(\mathcal E \cup \mathcal{E}_W)$. $Q$ is composed  of  random quadruplets $(\tau, k , e, \theta)$, where $\tau$ represents a potential event time for an individual $I_k$ and event type $e$. The last variable $\theta$ is used to accept/reject  this proposed event, depending on the event intensity. Hence, the Poisson measure is restricted to a certain random set and then projected on the space of interest $\R^+ \times \J$. If the event is accepted, then a jump $\phi^e(\tau,I_k)$ occurs. 

The proof of Theorem \ref{ThEqZ} is detailed in the Appendix \ref{ProofThPathwise}. Note that Equation \eqref{SDE_pop} is an SDE describing the evolution of the IBM, the intensity of the events in the right hand side of the equation depending on the population process $Z$ itself. The main idea of  the proof of  Theorem \ref{ThEqZ} is to use the non explosion property of Lemma \ref{lemma:nonExplosionSDE},  and to write the r.h.s of \eqref{SDE_pop} as a sum of simple equations between two successive events, solved by induction.\\
The proof of Lemma \ref{lemma:nonExplosionSDE}, detailed in  Appendix \ref{section:proofLemma}, is more technical and rely on    pathwise comparison result, generalizing those obtained in \cite{KaaElK20}. 
An alternative pathwise representation of the population process, inspired by the randomized Algorithm \ref{algo::rzndomized}, is given as well  in   Theorem \ref{ThEqZrandomized}.

\begin{theo}[Pathwise representation]
\label{ThEqZ}
Let $T\in \R^+$ and $\J = \mathbb N \times(\mathcal E \cup \mathcal{E}_W)$. \\
Let $Q$ be a random Poisson measure on $\mathbb R^+ \times \J \times \mathbb{R}^+$,  of intensity  $ \d t \delta_{\J}(\d k,\d e)  \mathbf{1}_{[0,\bar \lambda^e]} (\theta)\d \theta $, with $\delta_\J$ the counting measure on $\J$. Finally, let  $Q^{\mathcal P}$ be a random Poisson measure on $\mathbb R^+ \times \mathcal{P}  \times \mathbb{R}^+$,  of intensity  $ \d t \delta_{\cal P}(\d e)  \mathbf{1}_{[0,\bar \mu^e]} (\theta)\d \theta $, and $Z_0= \sum_{k=1}^{N_0} \delta_{I_k}$ an initial population. \\
Then, under Assumption  \ref{Assumption:nonExplosion}, there exists a unique  measure-valued population process $Z$, strong solution on the following SDE driven by the Poisson measure $Q$:
\begin{align}
\label{SDE_pop}
Z_t = Z_0 &+ \int_0^t \int_{ \J \times \mathbb R^+ }\phi^e (s , I_k)  \mathbf{1}_{\{k \leq N_{s^-}\} }\mathbf{1}_{\{\theta \leq \lambda_s^e(I_k, Z_{s^-})\}} Q (\d s ,\d k , \d e, \d \theta )\\
\nonumber &+   \int_0^t \int_{\mathcal{P} \times \mathbb R^+}  \phi^e(s, I_{s^-}) \mathbf{1}_{\{\theta \leq \mu^e(s) \}} Q^{\mathcal{P}} (\d s ,\d e,  \d \theta),  \qquad \forall  0 \leq t \leq T,
\end{align}
and where $I_{s^-}$ is an individual, chosen uniformly among alive individuals in the population $Z_{s^-}$.
\end{theo}
\begin{lemma}
\label{lemma:nonExplosionSDE}
 Let $Z$ be a solution of \eqref{SDE_pop} on $\R^+$,  with $(T_n)_{n\geq 0}$ its jump times, $T_0 = 0$.  If  Assumption \ref{Assumption:nonExplosion} is satisfied, then
 \begin{equation}
 \lim_{n \to \infty} T_n = \infty, \quad \P\text{-a.s.}
 \end{equation}
\end{lemma}
\section{Population simulation}
\label{sec::simulation}
%%%
We now present the main algorithm for simulating the evolution of an IBM over $[0,T]$.The algorithm implemented in \IBMPopSim allows the exact simulation of \eqref{SDE_pop},  based on an acceptance/reject algorithm for simulating random times called \emph{thinning}. The exact simulation of event times with this acceptance/reject procedure is closely related to the simulations of  inhomogeneous Poisson processes by the so-called thinning algorithm, often attributed to \cite{LewShe79}.   The simulation methods for  inhomogeneous Poisson processes can be adapted to IBMs, and we introduce in this section a general algorithm extending those by \cite{FouMel04} (see also \cite{FerTra09}, \cite{Ben10}).

The algorithm is based on  exponential ``candidate'' event times, chosen  with a (constant) intensity  which must be greater than the global event intensity $(\Lambda_t(Z_t))$ \eqref{eq:GlobalIntensity}.  Starting from time $t$, once a candidate event time $t + \bar T_\ell$ has been proposed, a candidate event type $e$ (birth, death,...) is chosen with a  probability $p^e$ depending on the event intensity bounds $\bar \mu^e$, $\bar \lambda^e$ and $\bar W^e$, as defined in Assumption \ref{AssumptionIntensity1} and \ref{AssumptionIntensity2}.  An individual $I$ is then drawn from the population. Finally, it remains to  accept or reject the candidate event with a  probability  $q^e(t,I,Z_t)$ depending on the true event intensity. If the candidate event time is accepted, then the event $e$ occurs at time $t + \bar T_\ell$ to  the individual $I$. The main idea of the algorithm implemented can be summarized as follows:
\begin{enumerate}
    \item Draw a candidate time $t + \bar T_\ell$ and candidate event type $e$.
    \item Draw a uniform variable $\theta \sim \mathcal{U}([0, 1])$ and individual $I$.
    \item \textbf{If} $\theta \leq q^e(t,I,Z_t)$ \textbf{then} event $e$ occur to individual $I$,
    \textbf{else} Do nothing and start again from $t + \bar T_\ell$.
\end{enumerate}

Before introducing the main algorithms in more details,  we recall briefly the thinning procedure for simulating inhomogeneous Poisson processes, as well as the links with pathwise representations. Some recalls on Poisson random measures are presented in Appendix \ref{section::preliminaries}.  For a more general presentation of thinning of a Poisson random measure, see \cite{Dev86, Cin11,Kal17}.

\subsection{Thinning of Poisson measure}
\label{sec::thinning}

 Let us start with  the simulation and pathwise representation of an inhomogeneous Poisson process  on $[0,T]$ with intensity  $(\Lambda(t))_{t\in [0,T]}$. The thinning procedure is based on the fundamental assumption that $\Lambda(t) \leq  \bar \Lambda$ is bounded on $[0,T]$. In this case, the inhomogeneous Poisson can be  obtained from an homogeneous Poisson process of intensity $\bar \Lambda$, which can be simulated easily (see Appendix \ref{section::preliminaries}). 

First, the Poisson process can be extended to a Marked Poisson measure $\bar Q:= \sum_{\ell \ge 1} \delta_{(\bar T_\ell, \bar \Theta_\ell)}$ on $(\mathbb{R}^+)^2$, defined as follow:
\begin{itemize}
\item The  jump times of $(\bar T_\ell)_{\ell \ge 1}$ of $\bar Q$ are the jump times of a Poisson process of intensity $\bar \Lambda$. 
\item  The  marks $(\bar \Theta_\ell)_{\ell \ge 1}$ are \emph{i.i.d.} random variables, uniformly distributed on $[0,\bar \Lambda]$.
\end{itemize}
By Proposition \ref{PropMarkedPoisson},  $\bar{Q}$ is a Poisson random measure with mean measure $$ \bar \mu(\d t, \d \theta): = \bar \Lambda \d t
    \frac{\mathbf{1}_{[0, \bar \Lambda]}(\theta)}{\bar \Lambda} \d \theta= \d t  \mathbf{1}_{[0, \bar \Lambda]}(\theta) \d \theta.$$
In particular, the average number of atoms $(\bar T_\ell, \bar \Theta_\ell)$ in $[0,t]\times [0,h]$ is
$$\E[Q([0,t]\times [0,h])]=\E[\sum_{\ell} \boldsymbol{1}_{[0,t]\times [0,h]} (\bar T_\ell, \bar \Theta_{\ell})]  = \int_{(\mathbb{R}^+)^2}  \bar \mu(\d t, \d \theta)  = t (\bar \Lambda \wedge h).$$
The thinning  is based on the restriction property for Poisson measure: for a measurable set $\Delta\subset \R^+\times \R^+$,  the restriction $Q^\Delta:= \boldsymbol{1}_{\Delta}\bar Q $ of $\bar Q$ to $\Delta$ (by taking only atoms in $\Delta$)  is also a  Poisson random measure  of mean measure $\mu^{\Delta}(\d t, \d \theta)  = \boldsymbol{1}_{\Delta}(t,\theta) \bar \mu(\d t, \d \theta).$

\begin{figure}[H]
    \centering
    \input{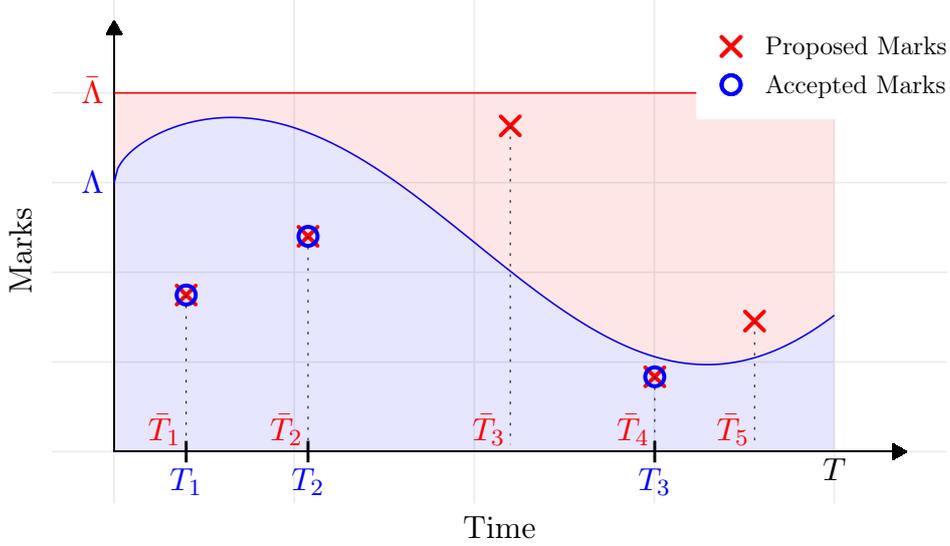}
    \caption{Realization of a Marked Poisson measure $\bar Q$ on $[0,T]$ with mean measure $\bar \mu(\d t, \d \theta) = \d t \mathbf{1}_{[0, \bar \Lambda]}(\theta) \d \theta$ (red crosses), and realization of the restriction $\bar Q^\Delta$ where $\Delta = \{ (t,\theta)\in [0,T]\times[0, \bar \Lambda], \theta \le \Lambda(t) \}$ (blue circles). The projection of $\bar Q^\Delta$ on first component is an inhomogeneous Poisson process on $[0,T]$ of intensity $(\Lambda(t))$ and jump times $(T_k)_{k \ge 1}$.}
    \label{plot:thinning}
\end{figure}

In order to obtain an inhomogeneous Poisson measure of intensity $(\Lambda(t))$, the ``good'' choice of $\Delta$ is the hypograph of $\Lambda$: $\Delta =\{ (t,\theta) \in [0,T]\times [0,\bar \Lambda] ; \; \theta \leq \Lambda(t)\}$ (see Figure \ref{plot:thinning}). Then, 
\begin{align*}
  &  Q^\Delta = \sum_{\ell \ge 1} \mathbf{1}_{\left\{\bar \Theta_\ell \le \Lambda(\bar T_\ell)\right\}} \delta_{(\bar T_\ell, \bar \Theta_\ell)},
 \end{align*}
and since $\Lambda(t) \leq \bar \Lambda$, on $[0,T]$: 
\begin{align*}
    \mu^{\Delta}(\d t, \d \theta) & = \boldsymbol{1}_{\{ \theta \leq \Lambda(t)\}}  \d t
\mathbf{1}_{[0, \bar \Lambda]}(\theta)\d \theta = \boldsymbol{1}_{\{\theta \leq \Lambda(t)\}} \d t \d \theta.
\end{align*}
Finally, the inhomogeneous Poisson process is obtained by the projection Proposition \ref{PropProjPoissonMeasure}, which states that the jump times of $Q^\Delta$ are the jump times of an inhomogeneous Poisson process of intensity $(\Lambda(t))$:
\begin{prop}
\label{PropThinning}
The counting process  $N^\Lambda$, projection of $Q^{\Delta}$ on the time component and defined by, 
\begin{equation}
\label{EqThinInhomoeneousPoisson}
N^{\Lambda}_t := Q^{\Delta}( [0,t] \times \R^+) = \int_0^t \int_{\R^+} \boldsymbol{1}_{\{\theta \leq \Lambda(s)\}} \bar Q(\d s, \d \theta) = \sum_{\ell \geq 1} \mathsf{1}_{\{ \bar T_\ell \leq t \}} \mathsf{1}_{\{\bar \Theta_\ell \leq  \Lambda(\bar T_\ell) \}},  \quad \forall t \in [0,T], 
\end{equation}
is an inhomogeneous Poisson process on $[0,T]$ of intensity  function $(\Lambda(t))_{t\in [0,T]}$.
The thinning Equation~\eqref{EqThinInhomoeneousPoisson} is  a pathwise representation of $N^\Lambda$ by \emph{restriction and projection} of the Poisson  measure $Q$ on $[0,T]$.
\end{prop}
The previous proposition yields a straightforward thinning algorithm to simulate the jump times $(T_k)_{k \ge 1}$ of an inhomogeneous Poisson process of intensity $\Lambda(t)$, by selecting jump times $\bar T_\ell$ such that $\bar \Theta_\ell \le \Lambda(\bar T_\ell)$. 

\paragraph{Multivariate Poisson process}
\label{sec:vector}
This can be extended to the simulation of multivariate inhomogeneous  Poisson processes, which is  an important example before tackling the simulation of an IBM.  

Let $(N^j)_{j \in \J}$ be a (inhomogeneous) multivariate Poisson process indexed by a finite set $\J$, such that $\forall j \in \J$,  the intensity $(\lambda_j(t))_{t\in [0,T]}$ of $N_j$ is bounded on $[0,T]$:
\begin{equation*}
    \sup_{t\in[0,T]} \lambda_j(t) \le \bar \lambda_j, \text{ and let }
    \bar \Lambda = \sum_{j \in \J} \bar \lambda_j.
\end{equation*}
Recall that such multivariate counting process can be rewritten as a Poisson random measure  $N= \sum_{k\geq 1} \delta_{(T_k,J_k)}$ on $\R^+\times \J$ (see e.g. Sec. 2 of Chapter 6 in~\cite{Cin11}), where  $T_k$ is the $k$th jump time of $\sum_{j\in \J} N^j$ and $J_k$ corresponds to the component of the the vector which jumps. In particular, $N^j_t = N([0,t]\times \{j\})$.

Once again the simulation of such process can be obtained from the simulation of a (homogeneous) multivariate Poisson process of intensity vector  $(\bar{\lambda}_j)_{j \in \J}$, extended into a Poisson measures by adding marks on $\R^+$.  Thus, we introduce the Marked Poisson measure $\bar Q = \sum \delta_{(\bar T_\ell, \bar J_\ell, \bar \Theta_\ell)}$ on $\R^+  \times \J \times \R^+$, such that: 
\begin{itemize}
\item  The jump times $(\bar T_\ell)$  of $\bar Q$ are the jump times of a Poisson measure of intensity $\bar \Lambda$.
\item The  variables $(\bar J_\ell)$ are \emph{i.i.d.} random  variables on $\J$, with  $\displaystyle p_j= \P(\bar J_1 = j)= \bar \lambda_j/\bar \Lambda$ and representing the component of the vector which jumps. 
\item The marks $(\bar \Theta_\ell)$  are independent variables with  $\bar \Theta_\ell$ a uniform random variable on $[0,\bar \lambda_{{\bar J}_\ell}]$, $\forall \ell \geq 1$. 
\end{itemize}
By Proposition \ref{PropMarkedPoisson} and \ref{PropProjPoissonMeasure}, each measure $ \bar Q_j (\d t, \d \theta) = \bar Q(\d t, \{j\}, \d \theta) = \sum_{\ell \geq 1} \mathsf{1}_{\{\bar J_\ell=j \}} \delta_{(\bar T_\ell, \bar \Theta_\ell)}$ is a marked Poisson measure of intensity 
$$\bar \mu_j ( \d t  ,\d \theta) = \bar{\Lambda}p_j \d t  \frac{\mathsf{1}_{\{\theta \leq \bar \lambda_{j}\}}(\theta)}{\bar \lambda_{j}} \d \theta = \d t  \mathsf{1}_{\{\theta \leq \bar \lambda_{j}\}}(\theta) \d \theta.$$
As a direct application of Proposition \ref{PropThinning}, the inhomogeneous multivariate Poisson process is obtained by restriction of each measures $\bar Q_j $ to $\Delta_j = \{ (t, \theta) \in [0,T] \times [0,\bar \lambda_j] ;\; \theta \leq \lambda_j(t) \}$ and projection:
\begin{prop}
\label{PropThinningVector}
The multivariate counting process $(N^j)_{j \in \J}$,  defined for all $j \in \J$ and  $ t \in [0,T]$ by thinning and projection of $\bar Q$:
\begin{equation}
N^j_t  := \int_0^t \int_{ \R^+} \boldsymbol{1}_{\{\theta \leq \lambda_j(s)\}} \bar{Q}(\d s, \{j\}, \d \theta) = \sum_{\ell \geq 1} \mathsf{1}_{\{ \bar T_\ell \leq t \}}\mathsf{1}_{\{\bar J_\ell = j\}} \mathsf{1}_{\{\bar \Theta_\ell \leq  \lambda_j (\bar T_\ell ) \}}, 
\end{equation}
is an inhomogeneous Poisson process  of intensity vector $(\lambda_j(t))_{j\in\J}$ on $[0,T]$.
\end{prop}
Proposition  \ref{PropThinningVector} yields the following simulation algorithm for multivariate Poisson processes:\\

\begin{algorithm}[H]
    \DontPrintSemicolon
    \SetKwInOut{Input}{Input}\SetKwInOut{Output}{Output}
    \Input{Functions $\lambda_j: [0,T] \to [0,\bar \lambda]$ and $\bar \lambda_j$, $\bar \Lambda = \sum_{j \in \J} \bar \lambda_j$.}
    \Output{Points $(T_k, J_k)$ of Poisson measure $N$ on $[0,T] \times \J$.}
    Initialization $T_0 \longleftarrow 0$, $\bar T_0 \longleftarrow 0$ \;
    \While{$T_k < T$}{
        \Repeat{accepted event $\bar \Theta_\ell \le \lambda_{\bar J_\ell}(\bar T_{\ell})$} {
            increment iterative variable $\ell \longleftarrow \ell+1$ \;
            compute next proposed time $\bar T_{\ell} \longleftarrow \bar T_{\ell-1} + S_\ell$  with $S_\ell \sim \mathcal{E}(\bar \Lambda)$\;
            draw $\bar J_\ell \sim Discrete\big(\big\{\frac{\bar \lambda_j}{\bar \Lambda}, j \in \J \big\} \big)_j)$ \emph{i.e.} $\mathbb{P}(\bar J_\ell = j) = \frac{\bar \lambda_j}{\bar \Lambda} $ \;
            draw $\bar \Theta_\ell \sim \mathcal{U}([0,\bar \lambda_{\bar J_\ell}])$\;
        }
        increment iterative variable $k \longleftarrow k+1$ \;
        record $(T_k, J_k) \longleftarrow (\bar T_{\ell}, \bar J_\ell)$ as accepted point \;
    }
\caption{Thinning algorithm for multivariate inhomogeneous Poisson processes.}
\label{AlgoThinning2}
\end{algorithm}
\begin{remark}
\label{RqAlternateThinningalgo}
The acceptance/rejection algorithm \ref{AlgoThinning2}  can be efficient when the functions $\lambda_j$ are of different order, and thus bounded by different $\bar \lambda_j$. However,  it is important to note that  the simulation of the discrete random variables $(\bar J_\ell)$ can be costly (compared to a uniform law) when $\J$ is large, for instance when an individual is drawn from a large population. In this case, an alternative is to choose the same bound $\bar \lambda_j= \bar \lambda$ for all $j \in \J$. 
Then the marks $(\bar J_\ell, \bar \Theta_\ell)$ are i.i.d uniform variables on $\J \times [0,\bar \lambda]$, faster to simulate.
\end{remark}

\subsection{Simulation algorithm}
\label{sec::simulation_algo}
Let us now come back to the simulation of the IBM  introduced in Section \ref{section::IBM}. For ease of notations, we assume that there are no event with Poisson intensity ($\mathcal{P} =\emptyset$), so that all events that occur are of type $e \in \mathcal{E} \cup \mathcal{E}_W$, with individual intensity $\lambda_t^e(I,Z_t)$  depending on the population composition $Z_t$ ($e \in \mathcal{E}_W$) or  not  ($e \in \mathcal{E}$),  as defined in \eqref{IndividualIntensity} and verifying either Assumption \ref{AssumptionIntensity1} or \ref{AssumptionIntensity2}. 
The global intensity~\eqref{eq:globalEvIntensity} at time $t \in [0,T]$ is thus
\begin{equation*}
\label{eq:defintensity}
    \Lambda_t(Z_t) = \sum_{e \in \mathcal{E}} \Big( \sum_{k=1}^{N_t} \lambda^e(t, I_k) \Big)
    + \sum_{e \in \mathcal{E}_W} \Big( \sum_{k=1}^{N_t} \sum_{j=1}^{N_t} W^e(t, I_k, I_j) \Big) \leq \bar \Lambda(N_t),
\end{equation*}
with $\bar \Lambda(n) = \big(\sum_{e \in \mathcal{E}} \bar \lambda^e \big) n + \big( \sum_{e \in \mathcal{E}_W} \bar W^e \big) n^2$.

One of the main difficulty is that the intensity of events is not deterministic as in the case of inhomogeneous Poisson processes, but a function $\Lambda_t(Z_t)$ of the population state, bounded by a function which also depends on the population size. However, the algorithm~\ref{AlgoThinning2} can be adapted to simulate the IBM. The construction is done by induction, by conditioning on the state of the population $Z_{T_k}$ at the $k$th event time $T_k$ ($T_0 = 0$).

We first present the construction of the first event at time $T_1$.
\paragraph{First event simulation} Before the first event time  (on $\{ t< T_1\}$),  the population composition is constant : $Z_t = Z_0 = \{ I_1, \dots, I_{N_0}\}$. For each type of event $e$ and individual $I_k$, $k \in \{1,\dots N_0\}$,  we denote by $N^{k,e}$ the counting process of intensity $\lambda_t^e (I_k,Z_t)$, counting the occurrences of the events of type  $e$ happening to the individual $I_k$. Then, the first event $T_1$ is the first jump time of the multivariate  counting vector $(N^{(k,e)})_{ (k,e) \in \J_0}$, with $\mathcal{J}_0 = \{1,\dots , N_0\}\times \big(\mathcal{E} \cup \mathcal{E}_W \big)$.

Since the population composition is constant before the first event time,   each counting process  $N^{j}$ coincides on $[0,T_1[$ with an inhomogeneous  Poisson process, of intensity $\lambda_t^e (I_k,Z_0)$.  Thus (conditionally to $Z_0$), $T_1$ is also the first jump time of  an inhomogeneous multivariate Poisson process $N^0 = (N^{0,j})_{j \in \mathcal{J}_0}$ of intensity  function  $(\lambda_j)_{j\in \mathcal J_0}$, defined for all $j = (k,e) \in \J_0$ by:
\[
\lambda_j(t) = \lambda^e_t(I_k,Z_0) \le \bar \lambda^e_0 \quad \text{with} \quad \bar \lambda^e_0 = \bar \lambda^e \mathbf{1}_{e \in \mathcal{E}} + \bar W^e N_0 \mathbf{1}_{e \in \mathcal{E}_W},
\]
by Assumptions~\ref{AssumptionIntensity1} and~\ref{AssumptionIntensity2}. 
% By Proposition \ref{prop:OpePoisson}, this process can be represented by the Marked Poisson measure  $Q^0(\d t , \d k, \d e )$ on $\mathbb R^+ \times \mathcal{J}_0$, of mean measure
%\begin{equation}
%\label{DefIntensityQ0}
%\mu^0(\d t, \d k, \d e ) = \lambda^e_t ( I_k,Z_0)\d t \delta_{\mathcal J_0}(\d k, \d e) , %= \Lambda_t(Z_0) \d t \frac{\lambda^e_{t} ( I_k,Z_0)}{\Lambda_t(Z_0)}\delta_{\mathcal J_0}(\d k, \d e),
%\end{equation}
%with $\delta_{\mathcal J_0}(\d k, \d e)$ the counting measure on $\J_0$.
In particular,  the  jump times of $N^0$  occur at the intensity 
$$\Lambda(t) =\sum_{j \in \mathcal{J}_0} \lambda_j(t)  =\sum_{e \in \mathcal{E} \cup \mathcal{E}_W} \sum_{k=1}^{N_0}  \lambda^e_t(I_k,Z_0) \leq \bar \Lambda(N_0)=N_0 \sum_{e \in \mathcal{E} \cup \mathcal{E}_W} \bar \lambda^e_0.$$
By Proposition \ref{PropThinningVector}, $N^0$ can be obtained by thinning of the marked Poisson measure $\bar Q^0 = \sum_{\ell \geq 1} \delta_{(\bar T_\ell , (\bar{K}_\ell, \bar E_\ell), \bar \Theta_\ell)}$ on $\R^+\times\J_0 \times \R^+$,  with:
\begin{itemize}
\item $(\bar T_\ell)_{\ell \in \N^*}$ the jump times of a Poisson process of rate $\bar \Lambda(N_0)$.
\item $(\bar{K}_\ell, \bar E_\ell)_{\ell \in \N^*}$  discrete \emph{i.i.d.} random variables  on $\J_0 = \{1,\dots , N_0\}\times \big(\mathcal{E} \cup \mathcal{E}_W \big)$, with $K_\ell$ representing the index of the chosen individual and $E_\ell$ the event type for the proposed event, such that: 
\[
\mathbb{P}( \bar  K_1 = k, \bar E_1 = e) =  \frac{\bar \lambda^e_0 }{\bar \Lambda(N_0)}
= \frac{1}{N_0} \frac{\bar \lambda^e_0 N_0}{\bar \Lambda(N_0)},
\]
%Note that $\frac{\bar \lambda^e_0 N_0}{\bar \Lambda(N_0)} = \frac{\bar \lambda^e \mathbf{1}_{e \in \mathcal{E}} + \bar W^e N_0 \mathbf{1}_{e \in \mathcal{E}_W}}{\sum_{e \in \mathcal{E}} \bar \lambda^e + \sum_{e \in \mathcal{E}_W} \bar W^e N_0}$.
i.e. $(\bar K_1, \bar E_1)$ are distributed as independent random  variables where $\bar K_1 \sim \mathcal U(\{1,\dots, N_0\})$ and $\bar E_1$ such that
\begin{equation*}
p_e := \mathbb{P}( \bar E_1 = e)
    = \frac{\bar \lambda^e_0 N_0}{\bar \Lambda(N_0)}.
\end{equation*}
\item $(\bar \Theta_\ell)_{\ell \in \N^*}$ are independent  uniform random variables, with $ \bar \Theta_\ell \sim \mathcal{U}([0,\bar \lambda^{\bar E_\ell}]).$
\end{itemize}
% of mean measure:
%\begin{equation*}
%    \bar \mu^0(\d t, \d k, \d e , \d \theta ) = \d t \delta_{\J_0}(\d k, \d e) \mathbf{1}_{[0,\bar \lambda^e_0]}(\theta) \d \theta  = \bar \Lambda(N_0) \d t
%    \frac{\bar \lambda^e_0}{\bar \Lambda(N_0)} \delta_{\J_0}(\d k, \d e) \frac{1}{\bar \lambda^e_0} \mathbf{1}_{[0,\bar \lambda^e_0]}(\theta) \d \theta.
%\end{equation*}
%
Since the first event is the first jump of $N^0$, by Proposition \ref{PropThinningVector} and Algorithm \ref{AlgoThinning2}, the first event time $T_1$ is the first jump time $\bar T_\ell$  of $\bar Q^0$  such that  $\bar \Theta_\ell \leq \lambda^{\bar E_\ell}_{\bar T_\ell}(I_{\bar K_\ell}, Z_0)$.

 At $T_1 =\bar T_{\ell}$,  the event $\bar E_\ell$ occurs to the individual $I_{\bar K_\ell} = (\tau^b, \infty, x)$.  For instance, if $\bar E_\ell =d$, a death/exit event occurs, so that $Z_{T_1} = Z_{0} + \delta_{(\tau^b, T_1, x)} - \delta_{I_{\bar K_\ell}}$ and $N_{T_1} = N_{0}$. If $\bar E_\ell =b$ or $en$, a birth or entry event occurs, so that $N_{T_1} = N_{0} + 1$, and a new individual $I_{N_0+1}$ is added to the population, chosen as described in Table \ref{TableEvAction}. Finally, if $\bar E_\ell=s$, a swap event occurs, the  population size stays constant and  $I_{\bar K_\ell}$ is replaced by an individual  $I_{\bar K_\ell}'$, chosen as described in Table \ref{TableEvAction}.

 The steps for simulating the first event in the population can be iterated in order to simulate the population. At the $k$th step, the same procedure is repeated to simulate the $k$th event, starting from a population $Z_{T_{k-1}}$ of size $N_{T_{k-1}}$. \\

\begin{algorithm}[H]
    \DontPrintSemicolon
    \SetKwInOut{Input}{Input}\SetKwInOut{Output}{Output}
    \Input{Initial population $Z_0$, horizon $T > 0$, and events described by \\
        Intensity functions and bounds $(\lambda^e, \bar \lambda^e)$ for $e \in \mathcal{E}$ and $(W^e, \bar W^e)$ for $e \in \mathcal{E}_W$ \\
    Event action functions $\phi^e(t, I)$ for $e \in \mathcal{E} \cup \mathcal{E}_W$ (see Table \ref{TableEvAction})}
    \Output{Population $Z_T$}
    Initialization $T_0 \longleftarrow 0$, $\bar T_0 \longleftarrow 0$ \;
    \While{$T_k < T$}{
        \Repeat{accepted event $\bar \Theta_\ell \le \lambda^{\bar E_\ell}_{\bar T_\ell}(I_{\bar K_\ell},Z_{T_{k}})$} {
            increment iterative variable $\ell \longleftarrow \ell+1$ \;
            compute next proposed time $\bar T_{\ell} \longleftarrow \bar T_{\ell-1} + \mathcal{E}\big(\bar \Lambda(N_{T_k}) \big)$ \;
            draw a proposed event $\bar E_\ell \sim Discrete(p_e)$ with $p_e = \frac{\bar \lambda^e \mathbf{1}_{e \in \mathcal{E}} + \bar W^e N_{T_k} \mathbf{1}_{e \in \mathcal{E}_W}}{\sum_{e \in \mathcal{E}} \bar \lambda^e + \sum_{e \in \mathcal{E}_W} \bar W^e N_{T_k}}$ \;
            draw an individual index $\bar K_\ell \sim \mathcal{U}(\{1,\dots,N_{T_k}\})$ \;
            draw $\bar \Theta_\ell \sim \mathcal{U}([0, \bar \lambda^{\bar E_\ell}])$ if $\bar E_\ell \in \mathcal{E}$ or $\bar \Theta_\ell \sim \mathcal{U}([0, \bar W^{\bar E_\ell} N_{T_k}])$ if $\bar E_\ell \in \mathcal{E}_W$  \;
        }
        increment iterative variable $k \longleftarrow k+1$ \;
        record $(T_k, E_k, I_k) \longleftarrow (\bar T_{\ell}, \bar E_\ell,  I_{\bar K_\ell})$ as accepted time, event and individual \;
        update the population $Z_{T_k} = Z_{T_{k-1}} + \phi^{E_k}(T_k, I_k)$
    }
\caption{IBM simulation algorithm (without events of Poissonian intensity)}
\label{algo:PopNointeraction2}
\end{algorithm}
The proof of Theorem \ref{theorem:algoNoInteraction} is detailed in the Appendix \ref{proof:algonointeraction}. 
\begin{theo}
\label{theorem:algoNoInteraction}
 Algorithm \ref{algo:PopNointeraction2}  are exact simulations of  Equation \eqref{SDE_pop}'s solution.
\end{theo}
\begin{remark}
\label{remark::removaldeadIndividuals}
The population  $Z_{T_k}$ includes dead/exited individuals before the event time $T_k$. Thus, $N_{T_k} > N_{T_k}^a$ is greater than the number of alive individuals at time $T_k$. When a dead individual $I_{\bar K_l}$ is drawn from the population during the rejection/acceptance phase of the algorithm, the proposed event $ (\bar T_{\ell}, \bar E_\ell,  I_{\bar K_\ell})$ is automatically rejected since the event intensity is $\lambda^{\bar E_\ell}_{T_\ell}(I_{\bar K_\ell}, Z_{T_k}) = 0$ (nothing can happen to a dead individual). This can slow down the algorithm, especially when the proportion of dead/exited individuals in the population increases. However,  the computational cost of keeping dead/exited  individuals in the population is much lower than the cost of removing an individual from the population at each death/exit event, which is linear in the population size.

Actually, dead/exited individuals are regularly removed from the population in the \IBMPopSim algorithm, in order to optimize the trade-off between having to many dead individuals and removing dead individuals from the population too often. The frequency at which dead individuals are  ``removed from the population'' can be chosen by the user, as an optional argument of the main function \texttt{popsim} (see details in Section \ref{simulation}).
\end{remark}
\begin{remark} In practice, the bounds $\bar \lambda^e$ and $\bar W^e$ should be chosen as sharp as possible. It is easy to see that conditionally to  $\{\bar E_\ell = e, \bar T_\ell = t, \bar K_\ell = l \}$ the probability of accepting the event  is, depending if there are interactions,
    \begin{equation*}
 \mathbb{P}\big(\bar \Theta_\ell \le \lambda^e_t(I_l,Z_{T_k})| \mathcal{F}_{T_k}\big)
        = \frac{\lambda^e(t, I_l)}{\bar \lambda^e} \mathbf{1}_{e \in \mathcal{E}} +  \frac{\sum_{j=1}^{N_{T_k}} W^e(t, I_l, I_j)}{\bar W^e N_{T_k}} \mathbf{1}_{e \in \mathcal{E}_W}.
    \end{equation*}
The sharper the bounds $\bar \lambda^e$ and $\bar W^e$ are, the higher is the acceptance rate. \\
For even sharper bounds, an alternative is to define bounds $\bar \lambda^e(I_l)$ and $\bar W^e(I_l)$  depending on the individuals' characteristics. However,  the algorithm is modified and the individual $I_l$ is not chosen uniformly in the population anymore. Due to the population size, this is way more costly than choosing uniform bounds,  as explained in Remark \ref{RqAlternateThinningalgo}.
\end{remark}

\subsection{Simulation algorithm with randomization}
\label{sec::simulation_algo_randomized}
Let $e \in \cal E_W$ be an event with interactions. In order to evaluate the individual intensity $\lambda^e_t (I,Z_t) = \sum_{j=1}^{N_t} W^e(t, I,I_j)$ one must compute $W^e(t, I_l, I_j) $ for all individuals in the population. This step can be computationally costly, especially for large populations.
%In the case of an event with interactions,  $e\in \mathcal E_W$, the individual intensity $\lambda^e_t (I,Z_t) = \sum_{j=1}^{N_t} W^e(t, I,I_j)$ has a sublinear growth. The larger is the population, the higher  is the intensity of the event. The simulation time is therefore much greater in the case of the presence of events with interaction.
%Moreover,  the evaluation of the individual intensity function $\lambda^e_t(I,Z)$ is linear in the size of the population. Indeed, in the previous algorithm, at the $k$th step  (last event time $T_k$), and conditionally to  $\{ \bar E_\ell = e \in \mathcal{E}_W, \bar T_\ell = t, \bar K_\ell = l\}$ the probability of accepting the event is given by the drawing of $\bar \Theta_\ell \sim \mathcal{U}([0, \bar W^e N_{T_k}])$ and the global summation $\lambda_t^e(I_l, Z_{T_k})$ \emph{i.e.}
%\begin{equation*}
%\mathbb{P}\big(\bar \Theta_\ell \le \lambda^e_t(I_l,Z_{T_k}) | \mathcal{F}_{T_k} \big)
%    = \mathbb{P}\Big(\bar \Theta_\ell \le \sum_{j=1}^{N_{T_k}} W^e(t, I_l, I_j) | \mathcal{F}_{T_k}  \Big) = \frac{\sum_{j=1}^{N_{T_k}} W^e(t, I_l, I_j)}{\bar W^e N_{T_k}}.
%\end{equation*}
%To evaluate $\lambda^e_t(I_l,Z_{T_k})$, one must compute $W^e(t, I_l, I_j) $ for all individuals in the population.
One way to avoid this summation is to use randomization (see also~\cite{FouMel04} in a model without age). The randomization consists in replacing  the summation by an evaluation of the interaction function $W^e$ using an individual $J$  drawn uniformly from the population.

More precisely, if $J \sim \mathcal{U}(\{1, \dots, N_{T_k}\})$ is independent of $\bar \Theta_\ell$,  we have
\begin{equation}
\label{eq:randomizedevent}
\mathbb{P}\Big(\bar \Theta_\ell \le \sum_{j=1}^{N_{T_k}} W^e(t, I_l, I_j) | \mathcal{F}_{T_k} \Big)
    = \mathbb{P}\big(\bar \Theta_\ell \le N_{T_k} W^e(t, I_l, I_J) | \mathcal{F}_{T_k}\big).
\end{equation}
Equivalently, we can write this probability as $\mathbb{P}\big(\tilde \Theta_\ell \le W^e(t, I_l, I_J) \big)$ where $\tilde \Theta_\ell  = \frac{\bar \Theta_\ell}{N_{T_k}}\sim \mathcal{U}([0, \bar W^e])$ is independent of $J \sim \mathcal{U}(\{1, \dots, N_{T_k}\})$.

\begin{remark}
The efficiency of the randomization procedure increases with the population homogeneity. If the function $W^e$ varies little according to the individuals in the population, the randomization approach is very efficient in practice, especially when the population is large.
\end{remark}

We now present the main algorithm implemented in the \texttt{popsim} function of the \IBMPopSim package in the case where events arrive with individual intensities, but also with interactions (using randomization) and Poisson intensities. In this general case, $\bar \Lambda(n)$ is defined by \eqref{eq:defbarLambda}. \\

\begin{algorithm}[H]
\label{algo::rzndomized}
    \DontPrintSemicolon
    \SetKwInOut{Input}{Input}\SetKwInOut{Output}{Output}
    \Input{Initial population $Z_0$, horizon $T > 0$, and events described by \\
        Intensity functions and bounds $(\lambda^e, \bar \lambda^e)$ for $e \in \mathcal{E}$, $(W^e, \bar W^e)$ for $e \in \mathcal{E}_W$ \\
        and $(\mu^e, \bar \mu^e)$ for $e \in \mathcal{P}$  \\
    Event action functions $\phi^e(t, I)$ for $e \in \mathcal{E} \cup \mathcal{E}_W \cup \mathcal{P}$ (Table \ref{TableEvAction})}
    \Output{Population $Z_T$}
    Initialization $T_0 \longleftarrow 0$, $\bar T_0 \longleftarrow 0$ \;
    \While{$T_k < T$}{
        \Repeat{accepted}  {
            increment iterative variable $\ell \longleftarrow \ell+1$ \;
            compute next proposed time $\bar T_{\ell} \longleftarrow \bar T_{\ell-1} + \mathcal{E}\big(\bar \Lambda(N_{T_k}) \big)$ \;
            draw an individual index $\bar K_\ell \sim \mathcal{U}(\{1,\dots,N_{T_k}\})$ \;
            draw a proposed event $\bar E_\ell \sim Disc(p_e)$ with $p_e = \frac{\bar \mu^e \mathbf{1}_{e \in \mathcal{P}} + \bar \lambda^e  N_{T_k}\mathbf{1}_{e \in \mathcal{E}} + \bar W^e (N_{T_k})^2 \mathbf{1}_{e \in \mathcal{E}_W}}{\bar \Lambda(N_{T_k})}$ \;
            \If{$\bar E_\ell \in \mathcal{E}$ (without interaction)}{
                draw $\bar \Theta_\ell \sim \mathcal{U}\big([0, \bar \lambda^{\bar E_\ell}] \big)$ \;
                \emph{accepted} $\longleftarrow \bar \Theta_\ell \le \lambda^{\bar E_\ell}(\bar T_\ell, I_{\bar K_\ell})$ \;
            }
            \If{$\bar E_\ell \in \mathcal{E}_W$ (with interaction)}{
                draw $(\bar \Theta_\ell, J_\ell) \sim  \mathcal{U}\big([0, \bar W^{\bar E_\ell}] \times \{1, \dots, N_{T_k}\} \big)$ \;
                \emph{accepted} $\longleftarrow \bar \Theta_\ell \le W^{\bar E_\ell}(\bar T_\ell, I_{\bar K_\ell}, I_{J_\ell})$ \;
            }
            \If{$\bar E_\ell \in \mathcal{P}$ (Poissonian intensity)}{
                draw $\bar \Theta_\ell \sim \mathcal{U}\big([0, \bar \mu^{\bar E_\ell}] \big)$ \;
                \emph{accepted} $\longleftarrow \bar \Theta_\ell \le \mu^{\bar E_\ell}(\bar T_\ell)$ \;
            }
        }
        increment iterative variable $k \longleftarrow k+1$ \;
        record $(T_k, E_k, I_k) \longleftarrow (\bar T_{\ell}, \bar E_\ell, \bar I_{\bar K_\ell})$ as accepted time, event and individual \;
        update the population $Z_{T_k} = Z_{T_{k-1}} + \phi^{E_k}(T_k, I_k)$
    }
\caption{Randomized IBM simulation algorithm}
\label{algo:Popinteraction2}
\end{algorithm}
\begin{prop}
\label{prop:algorandomized}
The population processes $(Z_t)_{t\in [0,T]}$ simulated by the Algorithm \ref{algo:PopNointeraction2} and \ref{algo:Popinteraction2} have the same law.
\end{prop}
\begin{proof}
The only difference between Algorithm \ref{algo:PopNointeraction2} and \ref{algo:Popinteraction2} is in the acceptance/rejection step of proposed events, in the presence of interactions. In Algorithm \ref{algo:Popinteraction2}, a proposed event $(\bar T_\ell, \bar E_\ell, \bar K_\ell)$, with $\bar E_l \in \mathcal{E}_W$ an event with interaction, is accepted as a true event in the population if
\begin{equation*}
\bar \Theta_\ell \le W^{\bar E_\ell}(\bar T_\ell, I_{\bar K_\ell}, I_{\bar J_\ell}), \text{ with } (\bar \Theta_\ell, \bar J_\ell) \sim  \mathcal{U}\big([0, \bar W^{\bar E_\ell}] \times \{1, \dots, N_{T_k}\} \big).
\end{equation*}
By \eqref{eq:randomizedevent}, the probability of accepting this event is the same than in Algorithm \ref{algo:PopNointeraction2}, which achieves the proof.
\end{proof}

\begin{cor}
\label{theorem:algoInteraction}
 Algorithm \ref{algo:Popinteraction2}  is an exact simulation of  Equation \eqref{SDE_pop}'s solution.
\end{cor}

\section{Model creation and simulation with IBMPopSim}
\label{sec::package}
The use of the \IBMPopSim package is mainly done in two steps: a first model creation followed by the simulation of the population evolution. 
The creation of a model is itself based on two steps: the description of the population $Z_t$, as introduced in Section~\ref{sec::population}, and 
the description of the events types, along with their associated intensities, as detailed in Sections~\ref{sec::events} and~\ref{sec::event_intensity}.
A model is compiled by calling the \r{mk\_model} function, which internally uses a template mechanism to generate automatically the source code describing the model, which is subsequently compiled using the \Rcpp package to produce the object code.

After the compilation of the model, the simulations are launched by calling the \r{popsim} function. This function depends on the previously compiled model and simulates a random trajectory of the population evolution based on an initial population and on parameter values, which can change from a call to another.

In this section, we take a closer look at each component of a model in \IBMPopSim. We also refer to the \href{https://daphnegiorgi.github.io/IBMPopSim/}{IBMPopSim website} and to the \texttt{vignettes} of the package for more details on the package and various examples of model creation. 

\subsection{Population}

A population $Z$ is represented by an object of class \texttt{population} containing a data frame where each row corresponds to an individual $I=(\tau^b, \tau^d, x)$, and which has at least two columns, \texttt{birth} and \texttt{death}, corresponding to the birth date $\tau^b$ and death/exit date $\tau^d$ ($\tau^d$ is set to \texttt{NA} for alive individuals).
The data frame can contain more than two columns if individuals are described by additional characteristics $x= (x_1,\dots x_n)$.

\paragraph{Entry and exit events} If  entry events  can occur in the population, the population shall contain a characteristic named \texttt{entry}. This can be done by setting the flag \texttt{entry=TRUE} in the \texttt{population} function, or by calling the  \texttt{add\_characteristic} function on an existing population.  During the simulation, the date at which an individual enters the population is automatically recorded in the variable \texttt{I.entry}.\\
If exit events  can occur, the population shall contain a characteristic named \texttt{out}. This can be done by setting the flag \texttt{out=TRUE} in the \texttt{population} function, or by calling the  \texttt{add\_characteristic} function. When an individual \texttt{I} exits the population during the simulation, \texttt{I.out} is set to \texttt{TRUE} and its exit time is recorded as a ``death'' date.

In the example below, individuals are described by their birth and death dates, as well a Boolean characteristics called male, and the \texttt{entry} characteristic. For instance, the first individual is a female whose age at $t_0=0$ is $107$ and who was originally in the population.

\begin{knitrout}
\definecolor{shadecolor}{rgb}{0.969, 0.969, 0.969}\color{fgcolor}\begin{kframe}
\begin{alltt}
\hlstd{pop_init} \hlkwb{<-} \hlkwd{population}\hlstd{(EW_pop_14}\hlopt{$}\hlstd{sample,}\hlkwc{entry}\hlstd{=}\hlnum{TRUE}\hlstd{)}
\hlkwd{str}\hlstd{(pop_init)}
\hlcom{## Classes 'population' and 'data.frame':	100000 obs. of  4 variables:}
\hlcom{##  birth: num  -107 -107 -105 -104 -104 ...}
\hlcom{##  death: num  NA NA NA NA NA NA NA NA NA NA ...}
\hlcom{##  male : logi  FALSE FALSE TRUE FALSE FALSE FALSE ...}
\hlcom{##  entry: logi  NA NA NA NA NA NA ...}
\end{alltt}
\end{kframe}
\end{knitrout}

\paragraph{Individual}
In the \cpp model which is automatically generated and compiled, an individual \texttt{I} is an object of an internal class containing some attributes (\texttt{birth\_date}, \texttt{death\_date} and the characteristics, here \texttt{male}), and some methods including:
    \begin{itemize}
        \item \texttt{I.age(t)}: a \texttt{const} method returning the age of an individual \texttt{I} at time \texttt{t},
        \item \texttt{I.set\_age(a, t)}: a method to set the age \texttt{a} at time \texttt{t} of an individual \texttt{I} (set \texttt{birth\_date} at \texttt{t-a}),
        \item \texttt{I.is\_dead(t)}: a \texttt{const} method returning \texttt{true} if the individual \texttt{I} is dead at time \texttt{t}.
	\end{itemize}

\begin{remark}[Characteristics type]
A characteristic $x_i$ must be of atomic type: logical, integer, double or character. The function \texttt{get\_characteristic} allows to easily get characteristics names and their types  from a population data frame.  We draw the attention to the fact that some names for characteristics are forbidden, or reserved to specific cases : this is the case for \texttt{birth, death, entry, out, id}. 
\end{remark}

\subsection{Events}\label{sec::package_events}

The most important step of the model creation is the events creation.  The call to the function creating an event  is of form

\begin{knitrout}
\definecolor{shadecolor}{rgb}{0.969, 0.969, 0.969}\color{fgcolor}\begin{kframe}
\begin{alltt}
\hlkwd{mk_event_CLASS}\hlstd{(}\hlkwc{type} \hlstd{=} \hlstr{"TYPE"}\hlstd{,} \hlkwc{name} \hlstd{=}\hlstr{"NAME"}\hlstd{, ...)}
\end{alltt}
\end{kframe}
\end{knitrout}
\noindent where \texttt{CLASS} is replaced by the class of the event intensity, described in Section~\ref{sec::event_intensity}, and \texttt{type} corresponds to the event type, described in Section~\ref{sec::events}. Tables~\ref{tab::intensity_classes} and~\ref{tab::event_types}  summarize the different possible choices for intensity classes and types of event. The optional argument \texttt{name} gives a name to the event. If not specified, the name of the event is its type, for instance
\texttt{death}. However, a name must be specified if the model is
composed of several events with the same type (for instance when there are multiple death events corresponding to different causes of death).  The other arguments depend on the intensity class and on the event type. 

\begin{minipage}{.6\textwidth} %
\begin{longtable}[]{@{}lll@{}}
\caption{Intensity classes}
\label{tab::intensity_classes}\tabularnewline
\toprule
Intensity class & Set & \texttt{CLASS}  \\
\midrule
\endfirsthead
\toprule
Intensity class & Set &  \texttt{Class} \\
\midrule
\endhead
Individual & $\mathcal{E}$ &  \texttt{individual} \\
Interaction & $\mathcal{E}_W$ & \texttt{interaction} \\
Poisson & $\mathcal{P}$ &  \texttt{poisson} \\
Inhomogeneous Poisson & $\mathcal{P}$ &  \texttt{inhomogeneous\_poisson} \\
\bottomrule
\end{longtable}
\end{minipage}
\begin{minipage}{.4\textwidth} %
\begin{longtable}[]{@{}ll@{}}
\caption{Event types}
\label{tab::event_types}\tabularnewline
\toprule
Event type & \texttt{TYPE} \\
\midrule
\endfirsthead
\toprule
Event type & \texttt{TYPE} \\
\midrule
\endhead
Birth & \texttt{birth} \\
Death & \texttt{death} \\
Entry & \texttt{entry} \\
Exit & \texttt{exit} \\
Swap & \texttt{swap} \\
\bottomrule
\end{longtable}

\end{minipage} 

\vspace{0.5cm}

The intensity function and the kernel of an event are defined through
arguments of the function \texttt{mk\_event\_CLASS}. These arguments are
strings composed of few lines of code. Since the model is compiled
using \Rcpp, the code should be written in \cpp. However, thanks to the
functions/variables of the package, even the
non-experienced \cpp user can define a model quite easily. To facilitate the implementation,  the user can also define a list of \textbf{model parameters}, which can be used in the event and intensity definitions.  These parameters are stored in a named list and  can be of various types: atomic type, 
numeric vector or matrix,
predefined function of one variable (\texttt{stepfun}, \texttt{linfun}, \texttt{gompertz},
  \texttt{weibull}, \texttt{piecewise\_x}),
piecewise functions of two variables (\texttt{piecewise\_xy}). We refer to the \texttt{vignette(\textquotesingle{}IBMPopSim\_cpp\textquotesingle{})}  for more details on parameters types and basic \cpp tools. Another advantage of the model parameters is that their value can be modified from a simulation to another without changing the model.

\subsubsection{Intensities}
In \IBMPopSim, the intensity of an event can belong to three classes (see Section~\ref{sec::event_intensity}): individual intensities without interaction between individuals, corresponding to events $e\in\mathcal{E}$, individual intensities with interaction, corresponding to events $e\in\mathcal{E}_W$, and Poisson intensities (homogeneous and inhomogeneous), corresponding to events $e\in\mathcal{P}$.

\paragraph{Event creation with individual intensity}
\label{event-creation-with-individual-intensity}

An event \(e\in \mathcal{E}\) (see ~\eqref{eq:intensityNointeraction}) has an intensity of the form \(\lambda^e(t, I)\)
which depends only on the individual \texttt{I} and 
time. Events with such intensity are created using the function

\begin{knitrout}
\definecolor{shadecolor}{rgb}{0.969, 0.969, 0.969}\color{fgcolor}\begin{kframe}
\begin{alltt}
\hlkwd{mk_event_individual}\hlstd{(}\hlkwc{type} \hlstd{=} \hlstr{"TYPE"}\hlstd{,}
                    \hlkwc{name} \hlstd{=} \hlstr{"NAME"}\hlstd{,}
                    \hlkwc{intensity_code} \hlstd{=} \hlstr{"INTENSITY"}\hlstd{, ...)}
\end{alltt}
\end{kframe}
\end{knitrout}
\noindent The \texttt{intensity\_code} argument is a character string containing
few lines of \cpp code describing the intensity function \(\lambda^e(t, I)\). 
The intensity value has to be stored in a variable called
\texttt{result} and the available
variables for the intensity code are given in Table~\ref{tab::intensity-variables}.

For instance, the intensity code below  corresponds to an individual death intensity $\lambda^d(t,I)$ equal to
\(d_1(a(I,t)) = \alpha_1 \exp(\beta_1 a(I,t))\) for males and
\(d_2(a(I,t)) = \alpha_2 \exp(\beta_2 a(I,t))\) for females, where $a(I,t)=t-\tau^b$ is the age of the individual $I=(\tau^b, \tau^d,x)$ at time $t$. In this case, the
intensity function depends on the individuals' age, gender, and on the
model parameters \(\alpha = (\alpha_1, \alpha_2)\) and
\(\beta = (\beta_1, \beta_2)\).

\begin{knitrout}
\definecolor{shadecolor}{rgb}{0.969, 0.969, 0.969}\color{fgcolor}\begin{kframe}
\begin{alltt}
\hlstd{death_intensity}  \hlkwb{<-} \hlstr{"if (I.male)
                       result = alpha_1*exp(beta_1*I.age(t));
                     else
                       result = alpha_2*exp(beta_2*I.age(t));"}
\end{alltt}
\end{kframe}
\end{knitrout}

\paragraph{Event creation with interaction intensity}\label{event-creation-with-interaction-intensity}

An event $e\in \mathcal{E}_W$ is an event
which occurs to an individual at a frequency which is the result of
interactions with other members of the population (see Equation~\eqref{eq:intensityInteraction}), and which can be written as $\lambda^e_t(I, Z_t)=\sum_{J\in Z_t} W^e(t, I, J)$  where $W^e(t, I, J)$ is the intensity of the interaction between individual $I$ and individual $J$.

An event $e\in \mathcal{E}_W$ with such intensity is created by
calling the function

\begin{knitrout}
\definecolor{shadecolor}{rgb}{0.969, 0.969, 0.969}\color{fgcolor}\begin{kframe}
\begin{alltt}
\hlkwd{mk_event_interaction}\hlstd{(}\hlkwc{type} \hlstd{=} \hlstr{"TYPE"}\hlstd{,}
                     \hlkwc{name} \hlstd{=} \hlstr{"NAME"}\hlstd{,}
                     \hlkwc{interaction_code} \hlstd{=} \hlstr{"INTERACTION_CODE"}\hlstd{,}
                     \hlkwc{interaction_type} \hlstd{=} \hlstr{"random"}\hlstd{, ...)}
\end{alltt}
\end{kframe}
\end{knitrout}
\noindent The \texttt{interaction\_code} argument contains few lines of \cpp code
describing the interaction function \(W^e(t, I, J)\). The interaction function
value has to be stored in a variable called \texttt{result} and the
available variables for the intensity code are given in Table~\ref{tab::intensity-variables}. For example, if we set

\begin{knitrout}
\definecolor{shadecolor}{rgb}{0.969, 0.969, 0.969}\color{fgcolor}\begin{kframe}
\begin{alltt}
\hlstd{death_interaction_code} \hlkwb{<-} \hlstr{"result = max(J.size -I.size,0);"}
\end{alltt}
\end{kframe}
\end{knitrout}
\noindent the death intensity of an individual \texttt{I} is
the result of the competition between individuals, depending on a
characteristic named \texttt{size}, as defined in \eqref{ex:interaction}.

The argument \texttt{interaction\_type}, set by default at
\texttt{random}, is the algorithm choice for simulating the model. When
\texttt{interaction\_type=full}, the simulation follows Algorithm~\ref{algo:PopNointeraction2}, while when
\texttt{interaction\_type=random} it follows Algorithm~\ref{algo:Popinteraction2}.
In most cases, the \texttt{random} algorithm is much faster than the
\texttt{full} algorithm, as we illustrate for instance in Section~\ref{section:ExempleInteraction}, 
where we observe the gain of a factor of 40 between the two algorithms, on a set of standard parameters. 
This allows in particular to explore parameter sets that give larger population sizes, without reaching computation times that explode.

\begin{longtable}[]{@{}ll@{}}
\caption{\cpp variables available for intensity code}
\label{tab::intensity-variables}
\tabularnewline
\toprule
Variable & Description \\
\midrule
\endfirsthead
\toprule
Variable & Description \\
\midrule
\endhead
\texttt{I} & Current individual \\
\texttt{J} & Another individual in the population (only for interaction)\\
\texttt{t} & Current time \\
Model parameters & Depends on the model \\
\bottomrule
\end{longtable}

\paragraph{Events creation with Poisson and Inhomogeneous Poisson
intensity}\label{events-creation-with-poisson-and-inhomogeneous-poisson-intensity}

For events \(e\in\mathcal{P}\) with an intensity \(\mu^e(t)\) which does not
depend on the population, the event intensity is of class \texttt{inhomogeneous\_poisson} or \texttt{poisson}
depending on whether or not the intensity depends on time (in the second case the intensity is constant).

For Poisson (constant) intensities the events are created with the
function

\begin{knitrout}
\definecolor{shadecolor}{rgb}{0.969, 0.969, 0.969}\color{fgcolor}\begin{kframe}
\begin{alltt}
\hlkwd{mk_event_poisson}\hlstd{(}\hlkwc{type} \hlstd{=} \hlstr{"TYPE"}\hlstd{,}
                 \hlkwc{name} \hlstd{=} \hlstr{"NAME"}\hlstd{,}
                 \hlkwc{intensity} \hlstd{=} \hlstr{"CONSTANT"}\hlstd{, ...)}
\end{alltt}
\end{kframe}
\end{knitrout}

The following example creates a death event, where individuals die
at a constant intensity \texttt{lambda} (which has to be in the list of
model parameters):

\begin{knitrout}
\definecolor{shadecolor}{rgb}{0.969, 0.969, 0.969}\color{fgcolor}\begin{kframe}
\begin{alltt}
\hlkwd{mk_event_poisson}\hlstd{(}\hlkwc{type} \hlstd{=} \hlstr{"death"}\hlstd{,}
                 \hlkwc{intensity} \hlstd{=} \hlstr{"lambda"}\hlstd{)}
\end{alltt}
\end{kframe}
\end{knitrout}

When the intensity $(\mu^e(t))$ depends on
time, the event can be created similarly by
using the function

\begin{knitrout}
\definecolor{shadecolor}{rgb}{0.969, 0.969, 0.969}\color{fgcolor}\begin{kframe}
\begin{alltt}
\hlkwd{mk_event_inhomogeneous_poisson}\hlstd{(}\hlkwc{type} \hlstd{=} \hlstr{"TYPE"}\hlstd{,}
                               \hlkwc{name} \hlstd{=} \hlstr{"NAME"}\hlstd{,}
                               \hlkwc{intensity_code} \hlstd{=} \hlstr{"INTENSITY"}\hlstd{, ...)}
\end{alltt}
\end{kframe}
\end{knitrout}

\subsubsection{Event kernel code}\label{sub_par::event-kernel-code}

When an event occurs, the events kernels $k^e$ specify how the event modifies the population. The events kernels are defined in  the \texttt{kernel\_code} parameter of the \texttt{mk\_event\_CLASS(type = "TYPE", name ="NAME", ...)} function. The \texttt{kernel\_code}  is \texttt{NULL} by default and doesn't have to be specified for death, exit events and birth events, but  mandatory for entry and swap events.  Recall that the \texttt{kernel\_code} argument is a string composed of a few lines of \cpp code, characterizing the individual characteristics following the event. Table~\ref{tab::events_variables} summarizes the list of available variables that can be used in the \texttt{kernel\_code}.

\begin{itemize}
\item \textbf{Death/Exit event}  If the user defines a  death event, the death date of the current individual \texttt{I}  is set automatically to the current time \texttt{t}.
Similarly, when  an individual \texttt{I} exits the population,\texttt{I.out} is set automatically to \texttt{TRUE} and his exit time is recorded as a "death" date. For these events types, the \texttt{kernel\_code} doesn't have to be specified by the user.
\item \textbf{Birth event} The default generated event kernel is that an individual \texttt{I} gives birth to a new individual \texttt{newI} of age 0 at the current time \texttt{t}, with same characteristics than the parent  \texttt{I}. If no kernel is specified, the default generated \cpp code for a birth event is:

\begin{knitrout}
\definecolor{shadecolor}{rgb}{0.969, 0.969, 0.969}\color{fgcolor}\begin{kframe}
\begin{alltt}
individual newI = I;
newI.birth_date = t;
\hlkwd{pop.add}(newI);
\end{alltt}
\end{kframe}
\end{knitrout}

The user can modify the birth kernel, by specify the argument  \texttt{kernel\_code} of \texttt{mk\_event\_CLASS}. In this case, the generated code is

\begin{knitrout}
\definecolor{shadecolor}{rgb}{0.969, 0.969, 0.969}\color{fgcolor}\begin{kframe}
\begin{alltt}
individual newI = I;
newI.birth_date = t;
_KERNEL_CODE_
\hlkwd{pop.add}(newI);
\end{alltt}
\end{kframe}
\end{knitrout}

where \texttt{\_KERNEL\_CODE\_} is replaced by the content of the  \texttt{kernel\_code}   argument.

\item \textbf{Entry event} When  an individual \texttt{I} enters the population, \texttt{I.entry} is set automatically as the date at which the individual enters the population. When an entry  occurs the individual entering the population is not of age $0$. In this case, the user must specify the \texttt{kernel\_code}   argument indicating how the age and characteristics  of the new individual are chosen. For instance, the code below creates an event of type \texttt{entry}, named  \texttt{ev\_example}, where individuals  enter the population at a Poisson constant intensity.  When an individual \texttt{newI} enters the population at time \texttt{t}, his age is chosen as a normally distributed random variable, with mean 20 and variance 4. 

\begin{knitrout}
\definecolor{shadecolor}{rgb}{0.969, 0.969, 0.969}\color{fgcolor}\begin{kframe}
\begin{alltt}
\hlkwd{mk_event_poisson}\hlstd{(}\hlkwc{type} \hlstd{=} \hlstr{"entry"}\hlstd{,} \hlkwc{name} \hlstd{=} \hlstr{"ev_example"}\hlstd{,}
                                        \hlkwc{intensity} \hlstd{=} \hlstr{"lambda"}\hlstd{,}
                 \hlkwc{kernel_code} \hlstd{=} \hlstr{"double a_I= max(CNorm(20,2),0);
                               						   newI.set_age(a_I,t);"}\hlstd{)}
\end{alltt}
\end{kframe}
\end{knitrout}

\item \textbf{Swap event} The user must specify the \texttt{kernel\_code}  argument indicating how the  characteristics  of an individual are modified following a swap.             
\end{itemize}

\begin{longtable}[]{@{}ll@{}}
\caption{\cpp variables available for events kernel
code}
\label{tab::events_variables}\tabularnewline
\toprule
Variable & Description \\
\midrule
\endfirsthead
\toprule
Variable & Description \\
\midrule
\endhead
\texttt{I} & Current individual \\
\texttt{t} & Current time \\
\texttt{pop} & Current population (vector) \\
\texttt{newI} & \begin{tabular}{@{}l@{}} New individual.  By default for birth events \texttt{newI\ =\ I} with \texttt{newI.birth\ =\ t}. \\ Available only for birth and entry events. \end{tabular} \\
Model parameters & Depends on the model \\
\bottomrule
\end{longtable}

\begin{remark}
When there are several events of the same type, the user can identify which events
generated a particular event by adding a characteristic to the
population recording the event name/id when it occurs. 
See e.g.~\texttt{vignette(\textquotesingle{}IBMPopSim\_human\_pop\textquotesingle{})}
for an example with different causes of death.
\end{remark}

\subsection{Model creation}\label{Modelcreation}
\label{sec::model_creation}

Once the population, the events, and model parameters are defined, the IBM model is created using the function
\texttt{mk\_model}. 

\begin{knitrout}
\definecolor{shadecolor}{rgb}{0.969, 0.969, 0.969}\color{fgcolor}\begin{kframe}
\begin{alltt}
\hlstd{model} \hlkwb{<-} \hlkwd{mk_model}\hlstd{(}\hlkwc{characteristics} \hlstd{=} \hlkwd{get_characteristics}\hlstd{(pop_init),}
                  \hlkwc{event} \hlstd{= events_list,}
                  \hlkwc{parameters} \hlstd{= model_params)}
\end{alltt}
\end{kframe}
\end{knitrout}
\noindent During this step which can take a few seconds, the model is created and
compiled using the \Rcpp package. The model
structure in \IBMPopSim is that the model depends only on the population
characteristics' and parameters names and types, rather than their
values. This means that once the model has been created, various
simulations can be done with different initial populations and different
parameters values.

\paragraph{Example}  Here is an example of model with a population structured by age and gender, with birth and death events. The death intensity of an individual of age $a$ is
$d(a) = \alpha \exp(\beta a),$ and females between 15 and 40 can give birth with birth intensity  $ b(a) = \bar \lambda^b \mathbf{1}_{[15,40]}.$ The newborn is a male with probability $p_{male}$.

\begin{knitrout}
\definecolor{shadecolor}{rgb}{0.969, 0.969, 0.969}\color{fgcolor}\begin{kframe}
\begin{alltt}
\hlstd{params} \hlkwb{<-} \hlkwd{list}\hlstd{(}\hlstr{"p_male"}\hlstd{=} \hlnum{0.51}\hlstd{,}
               \hlstr{"birth_rate"} \hlstd{=} \hlkwd{stepfun}\hlstd{(}\hlkwd{c}\hlstd{(}\hlnum{15}\hlstd{,}\hlnum{40}\hlstd{),}\hlkwd{c}\hlstd{(}\hlnum{0}\hlstd{,}\hlnum{0.05}\hlstd{,}\hlnum{0}\hlstd{)),}
               \hlstr{"death_rate"} \hlstd{=} \hlkwd{gompertz}\hlstd{(}\hlnum{0.008}\hlstd{,}\hlnum{0.02}\hlstd{))}


\hlstd{death_event} \hlkwb{<-} \hlkwd{mk_event_individual}\hlstd{(}\hlkwc{type} \hlstd{=} \hlstr{"death"}\hlstd{,} \hlkwc{name}\hlstd{=} \hlstr{"my_death_event"}\hlstd{,}
                  \hlkwc{intensity_code} \hlstd{=} \hlstr{"result = death_rate(age(I,t));"}\hlstd{)}

\hlstd{birth_event} \hlkwb{<-} \hlkwd{mk_event_individual}\hlstd{(} \hlkwc{type} \hlstd{=} \hlstr{"birth"}\hlstd{,}
                  \hlkwc{intensity_code} \hlstd{=} \hlstr{"if (I.male)
                                        result = 0;
                                    else
                                        result=birth_rate(age(I,t));"}\hlstd{,}
                  \hlkwc{kernel_code} \hlstd{=} \hlstr{"newI.male = CUnif(0, 1) < p_male;"}\hlstd{)}
\hlstd{pop} \hlkwb{<-} \hlkwd{population}\hlstd{(EW_pop_14}\hlopt{$}\hlstd{sample)}

\hlstd{model} \hlkwb{<-} \hlkwd{mk_model}\hlstd{(}\hlkwc{characteristics} \hlstd{=} \hlkwd{get_characteristics}\hlstd{(pop),}
                  \hlkwc{events} \hlstd{=} \hlkwd{list}\hlstd{(death_event,birth_event),}
                  \hlkwc{parameters} \hlstd{= params)}
\end{alltt}
\end{kframe}
\end{knitrout}

\subsection{Simulation}\label{simulation}

The simulation of the IBM is based on the algorithms presented in Sections~\ref{sec::simulation_algo} and~\ref{sec::simulation_algo_randomized}. The user has first  to
specify bounds for the intensity or interaction functions of each event type. The random evolution of the population
can then be simulated over a period of time \([0,T]\) by calling the function \texttt{popsim}

\begin{knitrout}
\definecolor{shadecolor}{rgb}{0.969, 0.969, 0.969}\color{fgcolor}\begin{kframe}
\begin{alltt}
\hlstd{sim_out} \hlkwb{->} \hlkwd{popsim}\hlstd{(model, pop_init, events_bounds, parameters,}
                  \hlkwc{age_max}\hlstd{=}\hlnum{Inf}\hlstd{, time,}
                  \hlkwc{multithreading}\hlstd{=}\hlnum{FALSE}\hlstd{,} \hlkwc{num_threads}\hlstd{=}\hlkwa{NULL}\hlstd{,}
                  \hlkwc{clean_step}\hlstd{=}\hlkwa{NULL}\hlstd{,} \hlkwc{clean_ratio}\hlstd{=}\hlnum{0.1}\hlstd{,} \hlkwc{seed}\hlstd{=}\hlkwa{NULL}\hlstd{)}
\end{alltt}
\end{kframe}
\end{knitrout}

\paragraph{Events bounds} Since the IBM simulation algorithm is based on an acceptance-rejection method for simulating random times, the user has to specify bounds for the intensity (or interaction) functions of each event (see Assumptions \ref{AssumptionIntensity1} and \ref{AssumptionIntensity2}). These bounds should be stored  in a named vector, where for event $e$, the name corresponding to the event bound  $\bar{\mu}^e$, $\bar{\lambda}^e$ or  $\bar{W}^e$ is the event \texttt{name} defined during the event creation step.

 In the model example built in the previous section the intensity bound for birth events is $\bar\lambda_b$. Since the death intensity function is not bounded, the user will have to specify a maximum age $a_{max}$ in \texttt{popsim} (all individuals above $a_{max}$ die automatically). Then, the bound for death events is $ \bar \lambda_d = \alpha\exp(\beta a_{max}).$ In the example, the death event has been named  \texttt{my\_death\_event}. No name has been specified for the birth event which thus has the default name \texttt{birth}. Then,

\begin{knitrout}
\definecolor{shadecolor}{rgb}{0.969, 0.969, 0.969}\color{fgcolor}\begin{kframe}
\begin{alltt}
\hlstd{a_max} \hlkwb{<-} \hlnum{120} \hlcom{# maximum age}
\hlstd{events_bounds} \hlkwb{<-} \hlkwd{c}\hlstd{(}\hlstr{"my_death_event"} \hlstd{= params}\hlopt{$}\hlkwd{death_rate}\hlstd{(a_max),}
                   \hlstr{"birth"} \hlstd{=} \hlkwd{max}\hlstd{(params}\hlopt{$}\hlstd{birth_rate))}
\end{alltt}
\end{kframe}
\end{knitrout}

Once the model and events bounds have been defined, a random trajectory of the population can be simulated by calling

\begin{knitrout}
\definecolor{shadecolor}{rgb}{0.969, 0.969, 0.969}\color{fgcolor}\begin{kframe}
\begin{alltt}
\hlstd{sim_out} \hlkwb{<-} \hlkwd{popsim}\hlstd{(model, pop, events_bounds, params,}
                  \hlkwc{age_max} \hlstd{= a_max,} \hlkwc{time} \hlstd{=} \hlnum{30}\hlstd{)}
\end{alltt}
\end{kframe}
\end{knitrout}

\paragraph{Optional parameters} If there
are no events with intensity of class \texttt{interaction}, then the
simulation can be parallelized easily by setting the optional parameter
\texttt{multithreading} (\texttt{FALSE} by default) to \texttt{TRUE}.
By default, the number of threads is the number of concurrent threads
supported by the available hardware implementation. The number of threads
can be set manually with the optional argument \texttt{num\_threads}. 
By default, when the proportion of dead individuals in the population exceeds $10\%$,  dead individuals are removed from the current population used in the algorithm (see Remark \ref{remark::removaldeadIndividuals}). The user  can modify this ratio using the optional argument  \texttt{clean\_ratio}, or by removing dead individuals from the population with a certain frequency, given by the  \texttt{clean\_step} argument. Finally, the user can also define the seed of the random number generator stored in the argument \texttt{seed}.

\paragraph{Outputs and treatment of swap events} The output of the \texttt{popsim} function is a list containing three elements: a data frame \texttt{population} containing the output population $Z_T$ (or a list of populations $(Z_{t_1}, \dots Z_{t_n})$ if \texttt{time} is a vector of times), a numeric vector \texttt{logs} of variables related to the simulation algorithm (including the simulation time and number of proposed/accepted events), and the list \texttt{arguments} of the simulation inputs, including the initial population, parameters and event bounds used for the simulation.

When there are no swap events (individuals don't change of
characteristics), the evolution of the population over the period
\([0,T]\) is recorded in a single data frame
\texttt{sim\_out\$population} where each line contains the information of an individual
who lived in the population over the period \([0,T]\) (see Remark \ref{remark::popfinale}). 

When there are swap events (individuals can change of characteristics), recording
the dates of swap events and changes of characteristics following
each swap event and for each individual in the
population is a memory intensive and computationally costly
process. To maintain efficient simulations in the presence of swap
events, the argument \texttt{time} of
\texttt{popsim} can be defined as a vector of dates \((t_0,\dots, t_n)\). In this
case, \texttt{popsim} returns in the object \texttt{population} a list
of \(n\) populations representing the population at time
\(t_1,\dots t_n\), simulated from the initial time \(t_0\). For
\(i=1\dots n\), the \(i\)th data frame is the population $Z_{t_i}$, i.e. individuals who lived
in the population during the period \([t_0,t_i]\), with their
characteristics at time \(t_i\).

It is also possible to isolate the individuals' life course, by adding an \texttt{id} column to the population, which can be done by setting \texttt{id=TRUE} in the population construction, or by calling the \texttt{add\_characteristic} function to an existing population, in order to identify each individual with a unique integer. 

Base functions to study the simulation outputs are provided in the package. For instance, the population age pyramid can computed at a give time, as well as death and exposure tables. Several illustrations of the outputs functions are given in the example Sections \ref{SectionInsurancePortofio} and \ref{section:ExempleInteraction}.

\section{Insurance portfolio}
\label{SectionInsurancePortofio}
This section provides an example of how to use the \IBMPopSim package to simulate a heterogeneous life insurance portfolio (see also~\texttt{vignette(\textquotesingle{}IBMPopSim\_insurance\_portfolio\textquotesingle{})}).

We consider an insurance portfolio consisting of  male policyholders,  of age greater than 65. These policyholders are characterized by their age, assumed to be less than $a_{max} = 110$,  and risk class $x\in \mathcal X =\{1,2\}$.

\textbf{Entries in the portfolio} New policyholders enter the population at a constant Poisson rate $\mu^{en}=\lambda$, which means that on average, $\lambda$ individuals enter the portfolio each year. A new individual enters the population at an age a that is uniformly distributed between 65 and 70, and is in risk class 1 with probability $p$.

\textbf{Death events} A baseline age and time specific death rate is first calibrated on ``England and Wales (EW)'' males mortality historic data\footnote{source: Human Mortality Database~\url{https://www.mortality.org/}}, and projected for 30 years using the Lee-Carter model with the package \texttt{StMoMo} (see~\cite{stmomo}). The forecasted baseline death intensity is  denoted by $d(t,a)$, defined by:
\begin{equation}
\label{eq:insurance-baseline}
d(t,a) = \sum_{k=0}^{29}\mathbf{1}_{\{k\leq t < k+1\}} d_k(a), \quad \forall \; t\in [0,30] \text{ and } a \in [65, a_{max}], 
\end{equation}
with $d_k(a)$ the point estimate of the forecasted mortality rate for age $a$ and year $k$.\\
Individuals in risk class 1 are assumed to have mortality rates that are 20\% higher than the baseline mortality (for instance, the risk class could refer to smokers), while individuals in risk class 2 are assumed to have mortality rates that are 20\% lower than the baseline (non smokers). The death intensity of an individual $I= (\tau_b, \infty, x)$, of age $a(I,t) = t - \tau_b$ at time $t$ and in risk class $x \in \{1, 2\}$ is thus the function 
\begin{equation}
\label{eq:insurance-deathrates}
\lambda^d(t,I) = \alpha_x d(t,a(I,t)), \quad \alpha_1 = 1.2, \quad \alpha_2 = 0.8.
\end{equation}
In particular, the death intensity verifies Assumption  \ref{AssumptionIntensity1} since: 
\begin{equation}
\label{eq:insurance-bound-deathrates}
\lambda^d(t,I) \leq \bar d : = \alpha_1 \sup_{t \in [0,30]} d(t,a_{max}). 
\end{equation}

\textbf{Exits from the portfolio} Individuals exit the portfolio at a constant (individual) rate \(\lambda^{ex}(t,I) = \mu^{i}\) only depending on their risk class 
$i\in \{1,2\}$.
\subsection{Population}\label{insurance-population}

We start with an initial population of $30\,000$ males of age 65, distributed uniformly in each risk class. The population data frame has thus the two (mandatory) columns \texttt{birth} (here the initial time is \(t_0=0\)) and \texttt{death} (\texttt{NA} if alive), and  an additional column \texttt{risk\_cls} corresponding to the policyholders risk class. Since there are entry and exit events, the \texttt{entry} and \texttt{out} flags of the population constructor are set to \texttt{TRUE}.

\begin{knitrout}
\definecolor{shadecolor}{rgb}{0.969, 0.969, 0.969}\color{fgcolor}\begin{kframe}
\begin{alltt}
\hlstd{N} \hlkwb{<-} \hlnum{30000}
\hlstd{pop_df} \hlkwb{<-} \hlkwd{data.frame}\hlstd{(}\hlstr{"birth"} \hlstd{=} \hlkwd{rep}\hlstd{(}\hlopt{-}\hlnum{65}\hlstd{,N),} \hlstr{"death"} \hlstd{=} \hlkwd{rep}\hlstd{(}\hlnum{NA}\hlstd{,N),}
                     \hlstr{"risk_cls"} \hlstd{=} \hlkwd{rep}\hlstd{(}\hlnum{1}\hlopt{:}\hlnum{2}\hlstd{,}\hlkwc{each}\hlstd{=N}\hlopt{/}\hlnum{2}\hlstd{))}
\hlstd{pop_init} \hlkwb{<-} \hlkwd{population}\hlstd{(pop_df,} \hlkwc{entry}\hlstd{=}\hlnum{TRUE}\hlstd{,} \hlkwc{out}\hlstd{=}\hlnum{TRUE}\hlstd{)}
\end{alltt}
\end{kframe}
\end{knitrout}

\subsection{Events}\label{insurance-events}

\paragraph{Entry events}
% If an individual enters the population at time $t$, his \texttt{entry} characteristic is automatically set up to  be equal to $t$.
The age of the new individual is determined by the \texttt{kernel\_code} argument in the \texttt{mk\_event\_poisson} function.

\begin{knitrout}
\definecolor{shadecolor}{rgb}{0.969, 0.969, 0.969}\color{fgcolor}\begin{kframe}
\begin{alltt}
\hlstd{entry_params} \hlkwb{<-} \hlkwd{list}\hlstd{(}\hlstr{"lambda"} \hlstd{=} \hlnum{30000}\hlstd{,} \hlstr{"p"} \hlstd{=} \hlnum{0.5}\hlstd{)}
\hlstd{entry_event} \hlkwb{<-} \hlkwd{mk_event_poisson}\hlstd{(}
    \hlkwc{type} \hlstd{=} \hlstr{"entry"}\hlstd{,}
    \hlkwc{intensity} \hlstd{=} \hlstr{"lambda"}\hlstd{,}
    \hlkwc{kernel_code} \hlstd{=} \hlstr{"if (CUnif() < p) newI.risk_cls =1;
                   else newI.risk_cls= 2;
                   double a = CUnif(65, 70);
                   newI.set_age(a, t);"}\hlstd{)}
\end{alltt}
\end{kframe}
\end{knitrout}
\noindent Note that the variables \texttt{newI} and \texttt{t}, as well as the function \texttt{CUnif()}, are implicitly defined and usable in the \texttt{kernel\_code}. The field \texttt{risk\_cls} comes from the names of characteristics of individuals in the population. The names \texttt{lambda} and \texttt{p} are parameter names that will be specified in the \RR named list \texttt{params}.

Here we use a constant $\lambda$ as the event intensity, but we could also use a rate $\lambda(t)$ that depends on time, using the function \texttt{mk\_event\_poisson\_inhomogeneous}.

\paragraph{Death and exit events}
The baseline death intensity defined in~\eqref{eq:insurance-baseline} and obtained with the package \texttt{StMoMo} is stored in the variable \texttt{death\_male}. 
\begin{knitrout}
\definecolor{shadecolor}{rgb}{0.969, 0.969, 0.969}\color{fgcolor}\begin{kframe}
\begin{alltt}
\hlstd{EWStMoMoMale} \hlkwb{<-} \hlkwd{StMoMoData}\hlstd{(EWdata_hmd,} \hlkwc{series} \hlstd{=} \hlstr{"male"}\hlstd{)}
\hlstd{LC} \hlkwb{<-} \hlkwd{lc}\hlstd{()}
\hlstd{ages.fit} \hlkwb{<-} \hlnum{65}\hlopt{:}\hlnum{100}
\hlstd{years.fit} \hlkwb{<-} \hlnum{1950}\hlopt{:}\hlnum{2016}
\hlstd{LCfitMale} \hlkwb{<-} \hlkwd{fit}\hlstd{(LC,} \hlkwc{data} \hlstd{= EWStMoMoMale,} \hlkwc{ages.fit} \hlstd{= ages.fit,}
                 \hlkwc{years.fit} \hlstd{= years.fit)}
\hlstd{t} \hlkwb{<-} \hlnum{30}
\hlstd{LCforecastMale} \hlkwb{<-} \hlkwd{forecast}\hlstd{(LCfitMale,} \hlkwc{h} \hlstd{= t)}
\hlstd{d_k} \hlkwb{<-} \hlkwd{apply}\hlstd{(LCforecastMale}\hlopt{$}\hlstd{rates,} \hlnum{2}\hlstd{,} \hlkwa{function}\hlstd{(}\hlkwc{x}\hlstd{)} \hlkwd{stepfun}\hlstd{(}\hlnum{66}\hlopt{:}\hlnum{100}\hlstd{, x))}
\hlstd{breaks} \hlkwb{<-} \hlnum{1}\hlopt{:}\hlnum{29}
\hlstd{death_male} \hlkwb{<-} \hlkwd{piecewise_xy}\hlstd{(breaks,d_k)}
\end{alltt}
\end{kframe}
\end{knitrout}

The death and exit intensities are of class \texttt{individual} (see Table \ref{tab::intensity_classes} ). Hence, the death and exit events are created with the \texttt{mk\_event\_individual} function.
\begin{knitrout}
\definecolor{shadecolor}{rgb}{0.969, 0.969, 0.969}\color{fgcolor}\begin{kframe}
\begin{alltt}
\hlstd{death_params} \hlkwb{<-} \hlkwd{list}\hlstd{(}\hlstr{"death_male"} \hlstd{= death_male,} \hlstr{"alpha"} \hlstd{=} \hlkwd{c}\hlstd{(}\hlnum{1.2}\hlstd{,} \hlnum{0.8}\hlstd{))}
\hlstd{death_event} \hlkwb{<-} \hlkwd{mk_event_individual}\hlstd{(}
    \hlkwc{type} \hlstd{=} \hlstr{"death"}\hlstd{,}
    \hlkwc{intensity_code} \hlstd{=} \hlstr{"result = alpha[I.risk_cls-1] * death_male(t, I.age(t));"}\hlstd{)}
\end{alltt}
\end{kframe}
\end{knitrout}

%The \texttt{out} value  in the initial population is set to \texttt{FALSE} by default. When an individual leaves the population, his characteristic \texttt{out} is set to \texttt{TRUE} and the date at which he exited the population is recorded in the column \texttt{death}.

\begin{knitrout}
\definecolor{shadecolor}{rgb}{0.969, 0.969, 0.969}\color{fgcolor}\begin{kframe}
\begin{alltt}
\hlstd{exit_params} \hlkwb{=} \hlkwd{list}\hlstd{(}\hlstr{"mu"} \hlstd{=} \hlkwd{c}\hlstd{(}\hlnum{0.001}\hlstd{,} \hlnum{0.06}\hlstd{))}
\hlstd{exit_event} \hlkwb{<-} \hlkwd{mk_event_individual}\hlstd{(}
    \hlkwc{type} \hlstd{=} \hlstr{"exit"}\hlstd{,}
    \hlkwc{intensity_code} \hlstd{=} \hlstr{"result = mu[I.risk_cls-1]; "}\hlstd{)}
\end{alltt}
\end{kframe}
\end{knitrout}

\subsection{Model creation and simulation}\label{insurance-simulation}

The model is created from all the previously defined building blocks, by calling the \texttt{mk\_model}. 

\begin{knitrout}
\definecolor{shadecolor}{rgb}{0.969, 0.969, 0.969}\color{fgcolor}\begin{kframe}
\begin{alltt}
\hlstd{model} \hlkwb{<-} \hlkwd{mk_model}\hlstd{(}
    \hlkwc{characteristics} \hlstd{=} \hlkwd{get_characteristics}\hlstd{(pop_init),}
    \hlkwc{events} \hlstd{=} \hlkwd{list}\hlstd{(entry_event, death_event, exit_event),}
    \hlkwc{parameters} \hlstd{=} \hlkwd{c}\hlstd{(entry_params, death_params, exit_params))}
\end{alltt}
\end{kframe}
\end{knitrout}

Once the model is compiled, it can be used with different parameters and run simulations for various scenarios. Similarly, the initial population (here \texttt{pop\_df}) can be modified without rerunning the \texttt{mk\_model} function.  The bounds for entry events is simply the intensity $\lambda$. For death events, the bound is given by $\bar{d}$ defined \eqref{eq:insurance-bound-deathrates}, which is stored in the \texttt{death\_max} variable. 

\begin{knitrout}
\definecolor{shadecolor}{rgb}{0.969, 0.969, 0.969}\color{fgcolor}\begin{kframe}
\begin{alltt}
\hlstd{bounds} \hlkwb{<-} \hlkwd{c}\hlstd{(}\hlstr{"entry"} \hlstd{= entry_params}\hlopt{$}\hlstd{lambda,}
            \hlstr{"death"} \hlstd{= death_max,}
            \hlstr{"exit"} \hlstd{=} \hlkwd{max}\hlstd{(exit_params}\hlopt{$}\hlstd{mu))}
\hlstd{sim_out} \hlkwb{<-} \hlkwd{popsim}\hlstd{(}
    \hlkwc{model} \hlstd{= model,}
    \hlkwc{initial_population} \hlstd{= pop_init,}
    \hlkwc{events_bounds} \hlstd{= bounds,}
    \hlkwc{parameters} \hlstd{=} \hlkwd{c}\hlstd{(entry_params, death_params, exit_params),}
    \hlkwc{time} \hlstd{=} \hlnum{30}\hlstd{,}
    \hlkwc{age_max} \hlstd{=} \hlnum{110}\hlstd{,}
    \hlkwc{multithreading} \hlstd{=} \hlnum{TRUE}\hlstd{)}
\end{alltt}
\end{kframe}
\end{knitrout}

\subsection{Outputs}

The data frame \texttt{sim\_out\$population} consists of all individuals present in the portfolio during the period of $[0, 30]$, including the individuals in the initial population and those who entered the portfolio. Each row represents an individual, with their date of birth, date of death (\texttt{NA} if still alive at the end of the simulation), risk class, and characteristics \texttt{entry} and  \texttt{out}. Recall that if an individual enters the population at time $t$, his \texttt{entry} characteristic is automatically set up to  be equal to $t$. The characteristics \texttt{out} is set to \texttt{TRUE} for individuals who left the portfolio due to an exit event. 

In this example, the simulation time over 30 years, starting from an initial population of 30 000 individuals is of $2\times 10^{-4}$ seconds, for an acceptance rate of proposed event of approximately 25\%. At the end of the simulation, the number of alive individuals is approximately 430 000. 

\begin{knitrout}
\definecolor{shadecolor}{rgb}{0.969, 0.969, 0.969}\color{fgcolor}\begin{kframe}
\begin{alltt}
\hlkwd{dim}\hlstd{(}\hlkwd{population_alive}\hlstd{(sim_out}\hlopt{$}\hlstd{population,}\hlkwc{t} \hlstd{=} \hlnum{30}\hlstd{))}
\hlcom{## 428517      5}

\hlstd{sim_out}\hlopt{$}\hlstd{logs[[}\hlstr{"duration_ns"}\hlstd{]]}\hlopt{/}\hlnum{1e9}
\hlcom{## 0.000106429}
\end{alltt}
\end{kframe}
\end{knitrout}

\begin{figure}[!ht]
    \begin{subfigure}[b]{0.48\textwidth}
    \includegraphics[width=\textwidth]{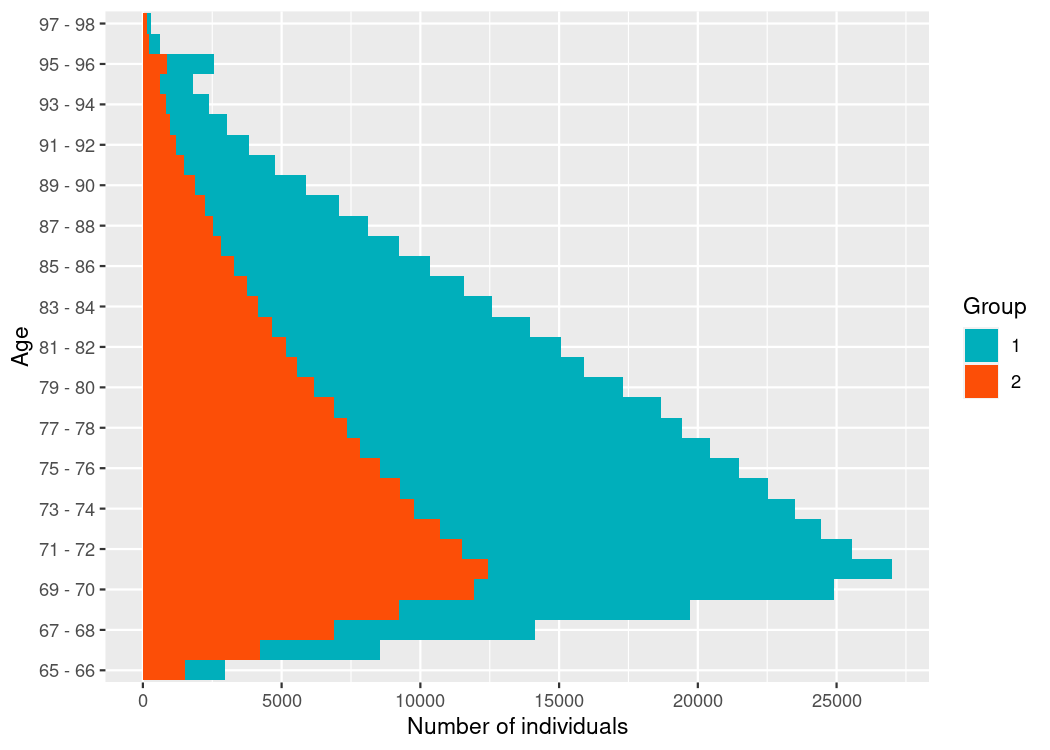}
    \caption{Portfolio age pyramid at $t=30$.}
    \end{subfigure}
    \hfill
    \begin{subfigure}[b]{0.48\textwidth}
        \includegraphics[width=\textwidth]{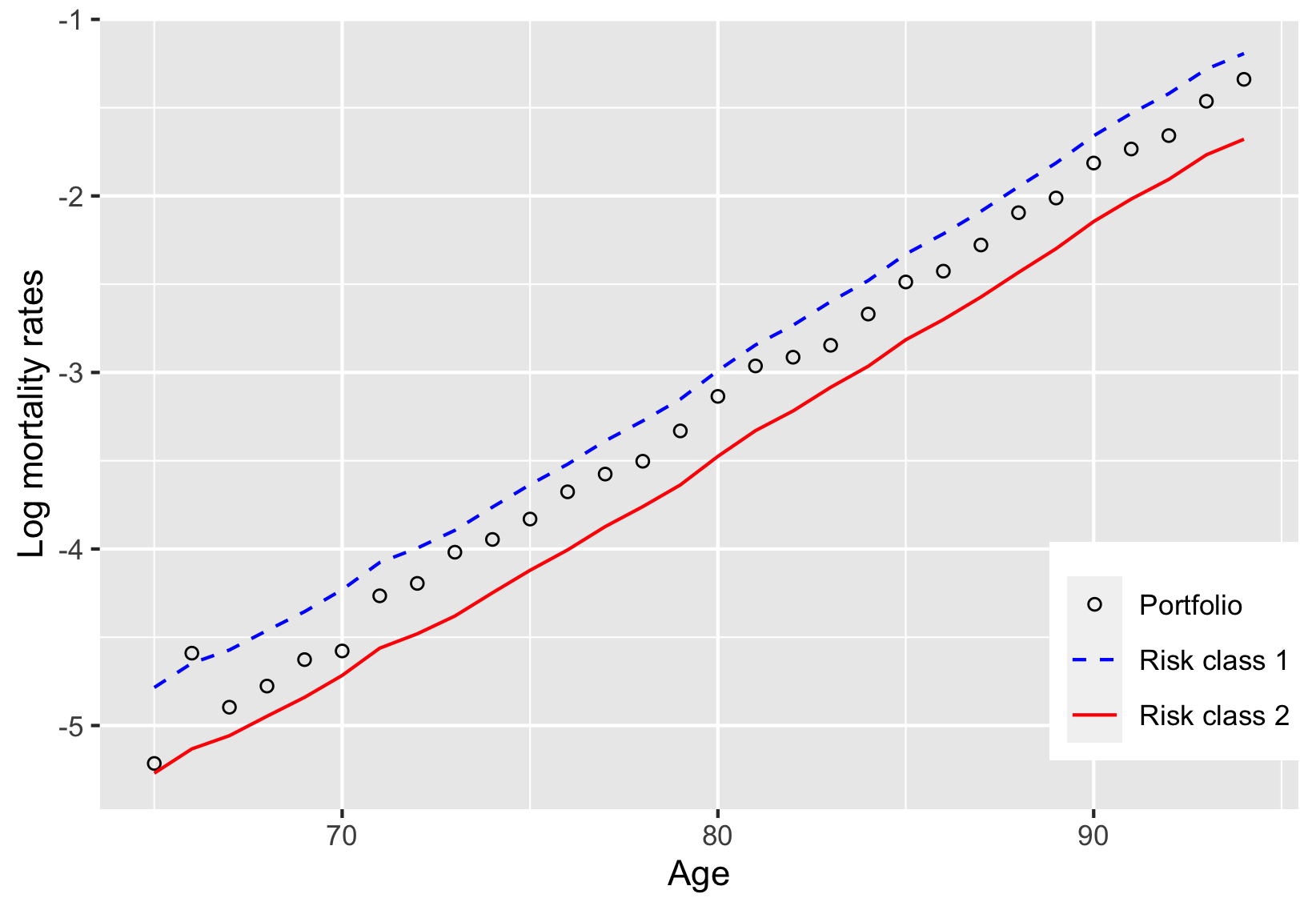}
        \caption{Portfolio central death rates at $t=30$ (black).}
    \end{subfigure}
    \caption{Information obtained from a simulation of the portfolio evolving over 30 years with individuals in risk class 1 (blue) and 2 (red).}
    \label{fig:insur}
\end{figure}

Initially in the portfolio (at $t=0$), there is the same number of 65 years old policyholders in each risk class. However, policyholders in the risk class 2 with lower mortality rates leave the portfolio at higher rate than policyholders in the risk class 1 : $\mu^2 > \mu^1$. Therefore, the heterogeneous portfolio composition changes with time, including more and more individuals in risk class 1 with higher mortality rates, but with variations across age classes.  To illustrate the composition of the total population at the end of the simulation ($t=30$), we present in Figure~\ref{fig:insur}(a) the age pyramid  of the final composition of the portfolio obtained with the \texttt{age\_pyramid} and \texttt{plot} function of the \texttt{pyramid} class.

\begin{knitrout}
\definecolor{shadecolor}{rgb}{0.969, 0.969, 0.969}\color{fgcolor}\begin{kframe}
\begin{alltt}
\hlstd{age_grp} \hlkwb{<-} \hlnum{65}\hlopt{:}\hlnum{95}
\hlstd{pyr} \hlkwb{=} \hlkwd{age_pyramid}\hlstd{(sim_out}\hlopt{$}\hlstd{population,} \hlkwc{time} \hlstd{=} \hlnum{30}\hlstd{,} \hlkwc{ages}\hlstd{=age_grp)}
\hlkwd{colnames}\hlstd{(pyr)[}\hlnum{2}\hlstd{]}\hlkwb{<-} \hlstr{"group_name"}
\hlstd{pyr}\hlopt{$}\hlstd{group_name} \hlkwb{<-} \hlkwd{as.character}\hlstd{(pyr}\hlopt{$}\hlstd{group_name)}
\hlkwd{plot}\hlstd{(pyr,}\hlkwc{colors} \hlstd{=}  \hlkwd{c}\hlstd{(}\hlstr{"1"}\hlstd{=}\hlstr{"#00AFBB"}\hlstd{,}\hlstr{"2"}\hlstd{=}\hlstr{"#FC4E07"}\hlstd{) ,}
         \hlkwc{age_breaks} \hlstd{=} \hlkwd{as.integer}\hlstd{(}\hlkwd{seq}\hlstd{(}\hlnum{1}\hlstd{,}\hlkwd{length}\hlstd{(age_grp)}\hlopt{-}\hlnum{1}\hlstd{,}\hlkwc{by}\hlstd{=}\hlnum{2}\hlstd{)))}
\end{alltt}
\end{kframe}
\end{knitrout}

\IBMPopSim  also allows the fast computation of exact life tables from truncated and censored individual data (due to entry and exit events), using the functions \texttt{death\_table} and \texttt{exposure\_table}. These function are particularly efficient, since the computations are made using the \texttt{Rccp} library. 

\begin{knitrout}
\definecolor{shadecolor}{rgb}{0.969, 0.969, 0.969}\color{fgcolor}\begin{kframe}
\begin{alltt}
\hlstd{Dx_pop} \hlkwb{<-} \hlkwd{death_table}\hlstd{(sim_out}\hlopt{$}\hlstd{population,} \hlkwc{ages} \hlstd{= age_grp,} \hlkwc{period} \hlstd{=} \hlnum{0}\hlopt{:}\hlnum{30}\hlstd{)}
\hlstd{Ex_pop} \hlkwb{<-} \hlkwd{exposure_table}\hlstd{(sim_out}\hlopt{$}\hlstd{population,} \hlkwc{ages} \hlstd{= age_grp,} \hlkwc{period} \hlstd{=} \hlnum{0}\hlopt{:}\hlnum{30}\hlstd{)}
\hlstd{mx_pop} \hlkwb{<-} \hlstd{Dx_pop}\hlopt{/}\hlstd{Ex_pop}
\end{alltt}
\end{kframe}
\end{knitrout}

In Figure~\ref{fig:insur}(b), we illustrate the central death rates  in the simulated portfolio at final time. 
Due to the mortality differential between risk class 1 and 2, one would expect to observe more individuals in risk class 2 at higher ages. However, due to exit events, a higher proportion of individuals in risk class 1 exit the portfolio over time, resulting in a greater proportion of individuals in risk class 1 at higher ages than what would be expected in the absence of exit events. Consequently, the mortality rates in the portfolio are more aligned with those of risk class 1 at higher ages. This is a simple example of how composition changes in the portfolio can impact aggregated mortality rates and potentially compensate or reduce an overall mortality reduction (see also \cite{KAAKAI201916}). 

\section{Population with genetically variable traits}
\label{section:ExempleInteraction}
This section provides an example of how to use the \IBMPopSim package to simulate an age-structured population with interactions, based on the model proposed in Example 1 of~\cite{FerTra09}.

In this model, individuals are characterized by their body size at birth \(x_0 \in [0,4]\) and by their physical age \(a \in [0,2]\). 
The body size of an individual $I=(\tau^b,\infty, x_0)$ at time $t$  is a linear function of its age $a(I,t) = t-\tau^b$:
\begin{equation*}
    x(t)= x_0 + ga(I,t),
\end{equation*}
where $g$ is a constant growth rate assumed to be identical for all individuals.

\textbf{Birth events} The birth intensity of each individual $I=(\tau^b, \infty, x_0)$ depends on a parameter $\alpha > 0$ and on its initial size, as given by the equation
\begin{equation}
\label{eq::interaction_birth_intensity}
\lambda^b(t,I) = \alpha (4 - x_0) \leq \bar \lambda^b = 4\alpha.
\end{equation}
Thus, smaller individuals have a higher birth intensity. When a birth occurs, the new individual inherit the same birth size  $x_0$ as its parent with high probability \(1-p\), or  a mutation can occur with probability \(p\), resulting in a birth size given by 
\begin{equation}
\label{eq::interaction_birth_kernel}
    x_0' = \min(\max(0, x_0 + G), 4),
\end{equation}
where $G$ is a Gaussian random variable with mean 0 and variance $\sigma^2$.

\textbf{Death events} Due to competition between individuals, the death intensity of an individual depends on the size of other individuals in the population. Bigger individuals have a better chance of survival. If an individual $I= (\tau^b, \infty, x_0)$ of size $x(t)= x_0 +ga(I,t)$ encounters an individual $J= (\tau^{b}_J, \infty, x_0')$ of size  $x'(t) = x_0'+ ga(J,t)$, then it can die with the intensity
\begin{equation*}
    W(t, I,J) = U(x(t),x'(t)),
\end{equation*}
where the interaction function $U$ is defined by
\begin{equation}
\label{eq::interaction_death_intensity}
    U(x,y) = \beta \left(1- \frac{1}{1+ c\exp(-4(x-y))}\right) \leq \bar W = \beta. 
\end{equation}
The death intensity of an individual  $I$ at time $t$  and in a population $Z$ is the result of interactions with all individuals in the population, including itself, and is given by
\begin{equation*}
    \lambda^d_t(I,Z) = \sum_{J = (\tau^b,\infty, x_0') \in Z}  W (x_0 + g a(I,t), x_0' + g a(J,t)),
\end{equation*}

\subsection{Population}

We use an initial population of 900 living individuals, all of whom have the same size and ages uniformly distributed between 0 and 2 years.

\begin{knitrout}
\definecolor{shadecolor}{rgb}{0.969, 0.969, 0.969}\color{fgcolor}\begin{kframe}
\begin{alltt}
\hlstd{N} \hlkwb{<-} \hlnum{900}
\hlstd{x0} \hlkwb{<-} \hlnum{1.06}
\hlstd{agemin} \hlkwb{<-} \hlnum{0.}
\hlstd{agemax} \hlkwb{<-} \hlnum{2.}
\end{alltt}
\end{kframe}
\end{knitrout}

\begin{knitrout}
\definecolor{shadecolor}{rgb}{0.969, 0.969, 0.969}\color{fgcolor}\begin{kframe}
\begin{alltt}
\hlstd{pop_df} \hlkwb{<-} \hlkwd{data.frame}\hlstd{(}
  \hlstr{"birth"} \hlstd{=} \hlopt{-}\hlkwd{runif}\hlstd{(N, agemin, agemax),} \hlcom{# Uniform age in [0,2]}
  \hlstr{"death"} \hlstd{=} \hlkwd{as.double}\hlstd{(}\hlnum{NA}\hlstd{),} \hlcom{# All individuals are alive}
  \hlstr{"birth_size"} \hlstd{= x0)} \hlcom{# All individuals have the same initial birth size x0}
\hlstd{pop_init} \hlkwb{<-} \hlkwd{population}\hlstd{(pop_df)}
\end{alltt}
\end{kframe}
\end{knitrout}

\subsection{Events}

\paragraph{Birth events}
The parameters involved in a birth event are the probability of mutation \(p\), the variance of the Gaussian random variable and the coefficient \(\alpha\) of the intensity.
\begin{knitrout}
\definecolor{shadecolor}{rgb}{0.969, 0.969, 0.969}\color{fgcolor}\begin{kframe}
\begin{alltt}
\hlstd{birth_params} \hlkwb{<-} \hlkwd{list}\hlstd{(}\hlstr{"p"} \hlstd{=} \hlnum{0.03}\hlstd{,} \hlstr{"sigma"} \hlstd{=} \hlkwd{sqrt}\hlstd{(}\hlnum{0.01}\hlstd{),} \hlstr{"alpha"} \hlstd{=} \hlnum{1}\hlstd{)}
\end{alltt}
\end{kframe}
\end{knitrout}
\noindent The  birth  intensity~\eqref{eq::interaction_birth_intensity} is of class \texttt{individual}. Hence, the event is created by calling the \texttt{mk\_event\_individual} function. The size of the new individual is given in the kernel following~\eqref{eq::interaction_birth_kernel}.
\begin{knitrout}
\definecolor{shadecolor}{rgb}{0.969, 0.969, 0.969}\color{fgcolor}\begin{kframe}
\begin{alltt}
\hlstd{birth_event} \hlkwb{<-} \hlkwd{mk_event_individual}\hlstd{(}
  \hlkwc{type} \hlstd{=} \hlstr{"birth"}\hlstd{,}
  \hlkwc{intensity_code} \hlstd{=} \hlstr{"result = alpha*(4 - I.birth_size);"}\hlstd{,}
  \hlkwc{kernel_code} \hlstd{=} \hlstr{"if (CUnif() < p)
                   newI.birth_size = min(max(0.,CNorm(I.birth_size,sigma)),4.);
                 else
                   newI.birth_size = I.birth_size;"}\hlstd{)}
\end{alltt}
\end{kframe}
\end{knitrout}

\paragraph{Death events}
The death intensity~\eqref{eq::interaction_death_intensity} is of class \texttt{interaction}. Hence, the event is created by calling the \texttt{mk\_event\_interaction} function. The parameters used for this event are the growth rate $g$, the amplitude of the interaction function $\beta$, and the strength of competition $c$.

\begin{knitrout}
\definecolor{shadecolor}{rgb}{0.969, 0.969, 0.969}\color{fgcolor}\begin{kframe}
\begin{alltt}
\hlstd{death_params} \hlkwb{<-} \hlkwd{list}\hlstd{(}\hlstr{"g"} \hlstd{=} \hlnum{1}\hlstd{,} \hlstr{"beta"} \hlstd{=} \hlnum{2.}\hlopt{/}\hlnum{300.}\hlstd{,} \hlstr{"c"} \hlstd{=} \hlnum{1.2}\hlstd{)}
\hlstd{death_event} \hlkwb{<-} \hlkwd{mk_event_interaction}\hlstd{(}
  \hlkwc{type} \hlstd{=} \hlstr{"death"}\hlstd{,}
  \hlkwc{interaction_code} \hlstd{=} \hlstr{"double x_I = I.birth_size + g * age(I,t);
                      double x_J = J.birth_size + g * age(J,t);
                      result = beta*(1.-1./(1.+c*exp(-4.*(x_I-x_J))));"}\hlstd{)}
\end{alltt}
\end{kframe}
\end{knitrout}

\subsection{Model creation and simulation}\label{interaction-simulation}
The model is created using the  \texttt{mk\_model} function. 
\begin{knitrout}
\definecolor{shadecolor}{rgb}{0.969, 0.969, 0.969}\color{fgcolor}\begin{kframe}
\begin{alltt}
\hlstd{model} \hlkwb{<-} \hlkwd{mk_model}\hlstd{(}
    \hlkwc{characteristics} \hlstd{=} \hlkwd{get_characteristics}\hlstd{(pop_init),}
    \hlkwc{events} \hlstd{=} \hlkwd{list}\hlstd{(birth_event, death_event),}
    \hlkwc{parameters} \hlstd{=} \hlkwd{c}\hlstd{(params_birth, params_death))}
\end{alltt}
\end{kframe}
\end{knitrout}

The simulation of one scenario can then be launched with the call of the \texttt{popsim} function, after computing the events bounds $\bar \lambda^b=4 \alpha$  and $\bar W= \beta$. 
\begin{knitrout}
\definecolor{shadecolor}{rgb}{0.969, 0.969, 0.969}\color{fgcolor}\begin{kframe}
\begin{alltt}
\hlstd{sim_out} \hlkwb{<-} \hlkwd{popsim}\hlstd{(}\hlkwc{model} \hlstd{= model,}
    \hlkwc{initial_population} \hlstd{= pop_init,}
    \hlkwc{events_bounds} \hlstd{=} \hlkwd{c}\hlstd{(}\hlstr{"birth"} \hlstd{=} \hlnum{4} \hlopt{*} \hlstd{birth_params}\hlopt{$}\hlstd{alpha,}
                      \hlstr{"death"} \hlstd{= death_params}\hlopt{$}\hlstd{beta),}
    \hlkwc{parameters} \hlstd{=} \hlkwd{c}\hlstd{(params_birth, params_death),}
    \hlkwc{age_max} \hlstd{=} \hlnum{2}\hlstd{,}
    \hlkwc{time} \hlstd{=} \hlnum{500}\hlstd{)}
\end{alltt}
\end{kframe}
\end{knitrout}

\begin{figure}[!ht]
    \begin{subfigure}[b]{0.48\textwidth}
    \includegraphics[width=\textwidth]{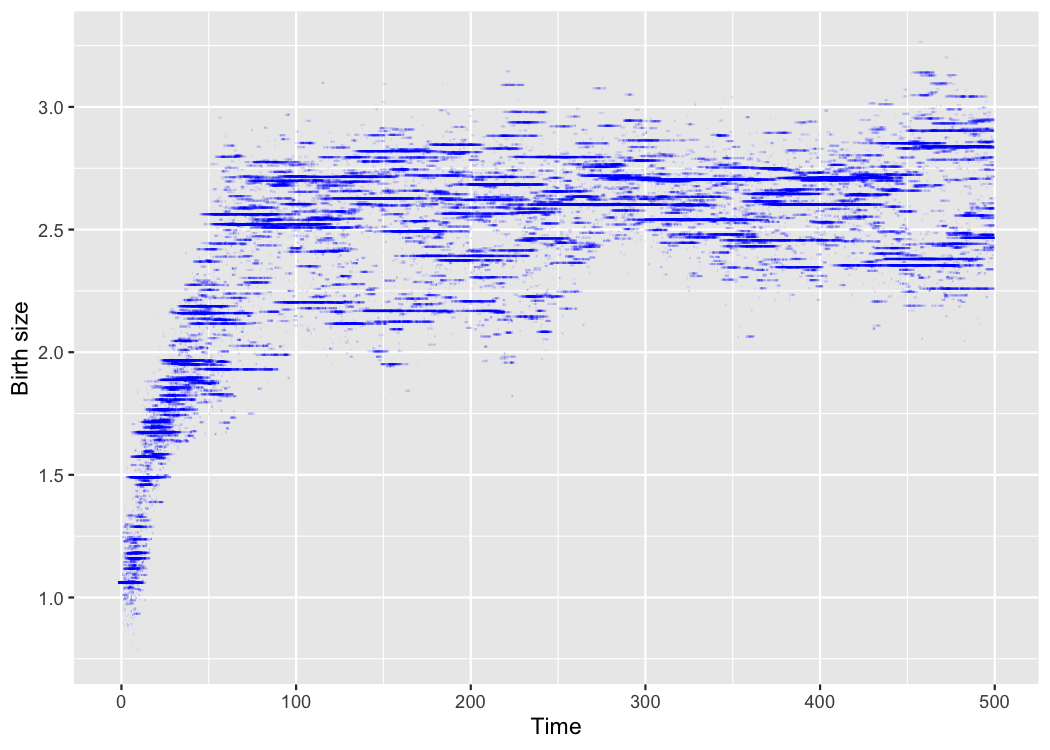}
    \caption{Birth size during life time.}
    \end{subfigure}
    \hfill
    \begin{subfigure}[b]{0.48\textwidth}
        \includegraphics[width=\textwidth]{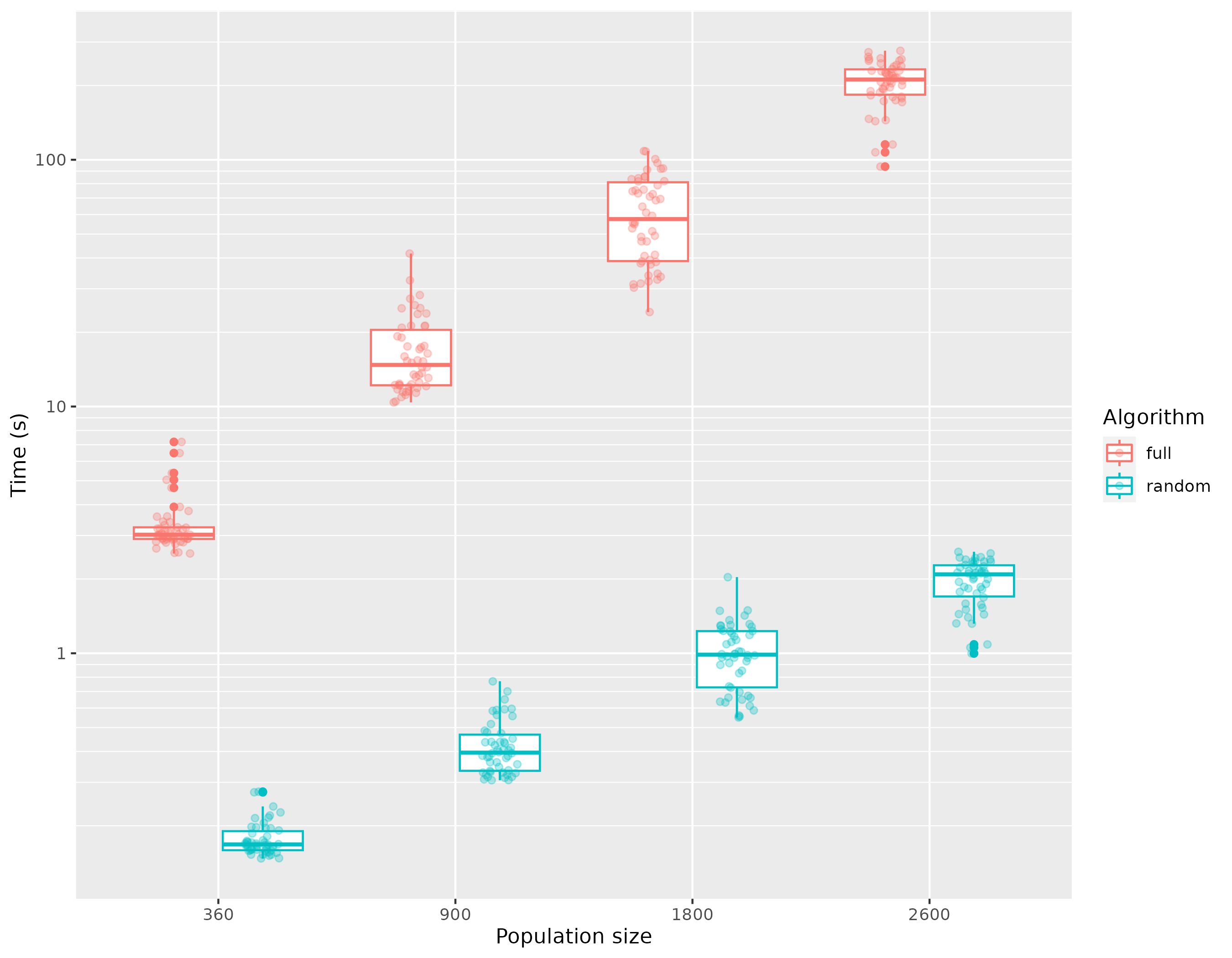}
        \caption{Full vs random algorithm.}
    \end{subfigure}
    \caption{Reproducing the example presented in~\cite{FerTra09} and increasing the population size to observe the difference in computing time between the randomized and full algorithm.}
    \label{fig:interaction}
\end{figure}

Based on the results of a simulation, we can reproduce the numerical results of~\cite{FerTra09}. 
In Figure~\ref{fig:interaction}(a), we draw a line for each individual in the population to represent their birth size during their lifetime. 

The randomized version of Algorithm~\ref{algo:Popinteraction2} allows for much faster computation times than Algorithm~\ref{algo:PopNointeraction2}. 
This is illustrated in Figure~\ref{fig:interaction} (b), where we progressively decrease the value of the mortality rate parameter $\beta$ and increase the birth rate parameter $\alpha$. 
Starting with the values provided in~\cite{FerTra09}, $\alpha=1$ and $\beta=2/300$, resulting in a stationary population size of approximately $N=360$ individuals for a sample of 50 simulations, we can easily increase the stationary population size to approximately $N=2600$ individuals with $\alpha=2$ and $\beta=1/300$
\footnote{The choices $(\alpha, \beta) \in \{(1,2/300),(1, 1/300), (1.5, 1/300), (2, 1/300)\}$ lead to the stationary population sizes $N \in \{360, 900, 1800, 2600\}$. For each set of parameters, we generated a new initial population, which was used for a benchmark of 50 simulations with both randomized and full algorithm. The simulations run on a Intel\textsuperscript{{\textregistered}} Core{\texttrademark}  i7-8550U CPU \@ 1.80GHz × 8 processor, with 15.3 GiB of RAM, under Debian GNU/Linux 11.}. 
In the log-scaled figure, we can observe the trend of computation time as a function of the population size $N$, which is linear for the randomized algorithm and quadratic for the full one (Algorithm \ref{algo:PopNointeraction2}). We can also see that the randomized version of the algorithm is between 17 to 100 times faster than the full one in this example, taking only 2 seconds in average for the randomized version versus 211 seconds for Algorithm \ref{algo:PopNointeraction2} for the biggest population size ($N=2600$) and $T=500$.

\appendix
\section*{Appendix}

\section{Recall on Poisson random measures}
\label{section::preliminaries}
We recall below some useful properties of Poisson random measures, mainly following Chapter~6 of~\cite{Cin11}. We also refer to~\cite{Kal17} for a more comprehensive presentation of random counting  measures.
\begin{definition}[Poisson Random measures]
\label{DefPoissonRandomMeasure}
Let $\mu$ be a $\sigma$-finite diffuse measure on a Borel subspace  $(E,\mathcal{E})$ of $(\R^d, \mathcal{B}(\R^d))$.. A random counting measure $Q= \sum_{k\geq 1} \delta_{X_k}$ is  a Poisson (counting) random measure of {\em mean measure} $\mu$ if
\begin{enumerate}
\item $\forall A \in \mathcal{E}$, $Q(A)$ is a Poisson random variable with $\E[Q(A)]= \mu(A)$.
\item For all  disjoints subsets $ A_1, \dots , A_n \in \mathcal{E}$, $Q(A_1), \dots, Q(A_n)$ are independent Poisson random variables.
\end{enumerate}
\end{definition}
Let us briefly recall here some simple but useful operations  on Poisson measures. In the following, $Q$ is a Poisson measure of mean measure $\mu$, unless stated otherwise. 

\begin{prop}[Restricted Poisson measure]
\label{PropRestrictionPoissonMeasure}
If $B \in \mathcal{E}$, then, the restriction of $Q$ to $B$ defined by
\begin{equation*}
    Q^B = \boldsymbol{1}_B Q = \sum_{k \ge 1} \mathbf{1}_{B}(X_k) \delta_{X_k}
\end{equation*}
is also  a Poisson random measure, of mean measure $\mu^B = \mu(\cdot \cap B)$. %In particular if $B$ is bounded the mean measure $\mu^B$ is finite.
\end{prop}
\begin{prop}[Projection of Poisson measure]
\label{PropProjPoissonMeasure}
If $E = F_1 \times F_2$ is a product space, then the projection
\begin{equation}
Q_1(\d x) = \int_{F_2} Q(\d x , \d y)
\end{equation}
is a Poisson random measure of mean measure $\mu_1 (\d x ) = \int_{F_2} \mu(\d x, \d y)$.
 \end{prop}

\paragraph{Link with Poisson processes}
Let $Q= \sum_{k\geq 1} \delta_{T_k}$ a Poisson random measure on $E=\R^+$ with mean measure $\mu(\d t) = \Lambda (t) \d t$ absolutely continuous with respect to the Lebesgue measure ($\mu(A) = \int_A \Lambda(t) \d t$).
%\begin{equation*}
%    \forall A \in \mathcal{B}(\R^+), \quad \mu(A) = \int_A \Lambda(t) \d t.
%\end{equation*}
The counting process $(N_t)_{t \ge 0}$ defined by
\begin{equation}
\label{eq::inhomogeneous_PP}
    N_t = Q([0,t]) = \sum_{k\geq 1} \boldsymbol{1}_{\{T_k \leq t\}}, \quad \forall \; t\geq 0,
\end{equation}
is an inhomogeneous Poisson process with intensity function (or rate) $t \mapsto \Lambda(t)$.
In particular, when $\Lambda(t) \equiv c$ is a constant, $N$ is a homogeneous Poisson process with rate $c$. Assuming that the atoms  are ordered $T_1< T_2< \dots $, we recall that the sequence $(T_{k+1}-T_k)_{k\geq 1}$ is a sequence of \emph{i.i.d.} exponential variables of parameter $c$.
\paragraph{Marked Poisson measures on $E = \R^+ \times F$}

We are interested in the particular case when $E$ is the product space  $\R^+ \times F$, with $(F,\mathcal{F})$ a Borel subspace of $\R^d$.
Then, a random  counting measure is defined by a random set $S =\{ (T_k, \Theta_k ), k \geq 1\}$. The random variables $T_k\geq 0$ can be considered as time variables, and constitute the jump times of the random measure, while the variables $\Theta_k \in F$ represent space variables.

We recall in this special case the Theorem~VI.3.2 in~\cite{Cin11}.

\begin{prop}[Marked Poisson measure] \label{PropMarkedPoisson}
    Let $m$ be a $\sigma$--finite diffuse measure on $\R^+$, and $K$ a transition probability kernel from $(\R^+,\mathcal{B}(\R^+))$ into $(F, \mathcal{F})$. Assume that the collection $(T_k)_{k \ge 1}$ forms a Poisson process $(N_t) =(\sum_{k\geq 1} \mathsf{1}_{\{T_k \leq t\}})$  with mean $m(\d t) =\Lambda(t) \d t$, and that given $(T_k)_{k \ge 1}$, the variables $\Theta_k$ are conditionally independent and have the respective distributions $K(T_k, \cdot)$.
\begin{enumerate}
\item     Then, $\{ (T_k, \Theta_k) ;\; k \ge 1\}$ forms a Poisson random measure $Q = \sum_{k\ge 1} \delta_{(T_k, \Theta_k)}$ on $(\R^+ \times F, \mathcal{B}(\R^+) \otimes \mathcal{F})$, called a \emph{Marked point process} ,  with mean $\mu$ defined by
            \begin{equation*}
                \mu(\d t, \d y) = \Lambda(t) \d t K(t, \d y).
            \end{equation*}
\item Reciprocally  let  $Q$ be  a Poisson random measure of mean measure $ \mu(\d t, \d y) $, admitting the following disintegration with respect to the first coordinate: $\mu(\d t , \d y) =\tilde  \Lambda(t) \d t \nu(t, \d y)$, with $\nu(t, F)<\infty$.  Let  $K(t, \d y) = \dfrac{\nu(t,\d y) }{\nu(t, F) }$ and $\Lambda(t) = \nu(t, F)\tilde  \Lambda(t)$. Then, $Q = \sum_{k\ge 1} \delta_{(T_k, \Theta_k)}$ is a marked Poisson measure with $(T_k,\Theta_k)_{k\in \N^*}$ defined as above. In particular,  the projection $N= (N_t)_{t\geq0}$ of the  Poisson measure on the first coordinate,
\begin{equation}
    N_t = Q([0,t] \times F) = \sum_{k\geq 1} \boldsymbol{1}_{[0,t] \times F} (T_k, \Theta_k)  = \sum_{k\geq 1} \boldsymbol{1}_{\{T_k \leq t\}}, \quad \forall \; t \geq 0,
\end{equation}
is an inhomogeneous Poisson process of rate $\Lambda(t)= \nu(t, F)\tilde  \Lambda(t)$.
\end{enumerate}

\end{prop}

\begin{remark}
When the transition probability kernel $K$ does not depend on the time: $K(t, A) = \nu(A)$ for some probability measure $\nu$, then the marks
$(\Theta_k)_{k \ge 1}$ form an \emph{i.i.d.} sequence with distribution $\nu$, independent of $(T_k)_{k \ge 1}$. \end{remark}

The preceding proposition thus yields a straight forward iterative simulation procedure for a Marked Poisson process on $[0,T]\times F$ with mean measure $\mu(\d t, \d y) = c \d t K(t, \d y)$ ($c>0$): \\

\begin{algorithm}[H]
    \DontPrintSemicolon
    \SetKwInOut{Input}{Input}\SetKwInOut{Output}{Output}
    \Input{Constant $c$, simulatable kernel $K$ and final time $T$.}
    \Output{Times $(T_1,\dots,T_n)$  and Marks $(Y_1, \dots Y_n)$ of the Marked Poisson measure in $[0,T]\times F$.}
    Initialization: draw $T_1 \sim \mathcal{E}(c)$ and draw $Y_1 \sim K(T_1, \d y)$ \;
    \While{$T_k < T$}{
        increment iterative variable $k \longleftarrow k+1$ \\
        compute next jump time $T_k \longleftarrow T_{k-1} + \mathcal{E}(c)$\\
        draw a conditional mark $Y_k \sim K(T_k, \d y)$
    }
\caption{Simulation of Marked Poisson measure of mean $\mu(\d t, \d y) = c \d t K(t, \d y)$.}
\label{algo:MarkedPoisson}
\end{algorithm}

% A plot of a realization of a such marked Poisson measure given by Algorithm~\ref{algo:MarkedPoisson} is given in Figure~\ref{plot:poisson}.
\begin{figure}[H]
    \centering
    \input{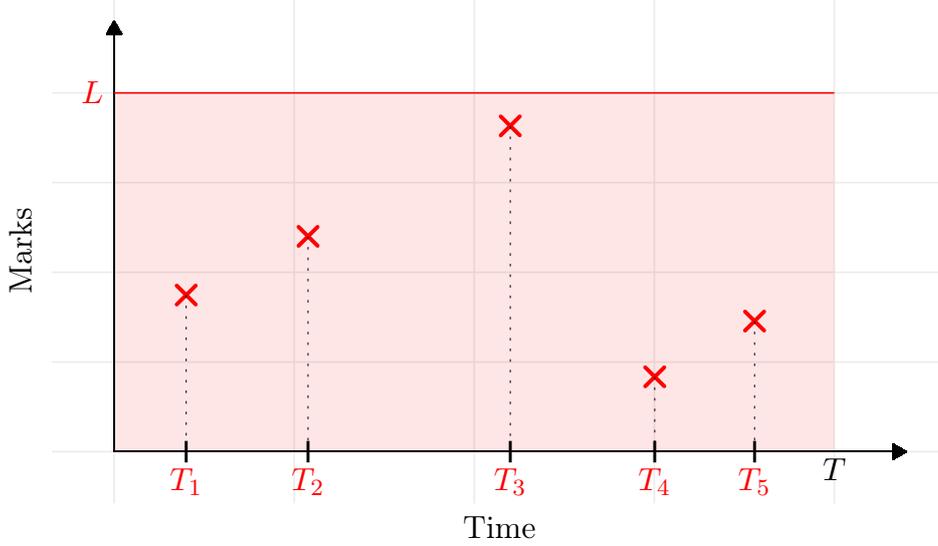}
    \caption{Example of Marked Poisson measure on $[0,T]$ with $m(\d t) = L \d t$ (jump times occur at Poisson arrival times of rate $L$) and with $\nu(\d y) = \frac{1}{L} \mathbf{1}_{[0, L]}(y) \d y$ (marks are drawn uniformly on $[0,L]$).
The mean measure is then $\mu(\d t, \d y) = \d t \boldsymbol{1}_{[0,L]}(y) \d y$.}
\label{plot:poisson}
\end{figure}

%\begin{remark}
%\label{rmk::non_simulatable_P_measure}
%\textcolor{red}{argument à revoir}
%If the measure $\nu$ is only $\sigma$-finite, then $Q([t,t+\epsilon]\times F)=\infty$, for  any $\epsilon >0$ and $t\geq 0$. In this case, the jump times of the Poisson random measure cannot be enumerated increasingly (there is an infinite number of jump times in any time interval of arbitrarily small length), and the Poisson measure cannot be simulated exactly on $[0,T]\times F$.
%\end{remark}

\section{Pathwise representation of IBMs}
\label{pathwiserepresentation}

\paragraph{Notation reminder}  The population's evolution is described by the measure valued process $(Z_t)_{t\geq 0}$. Several types of events $e$  can occur to individuals denoted by $I$.  In an event of type $e$ occur to the individual $I$ at time $t$, then the population state $Z_{t^-}$ is modified by $\phi^e(t,I)$. If $e\in \mathcal{E} \cup \mathcal{E}_W$, then events of type $e$ occur with an intensity $\sum_{k=1}^{N_t} \lambda_t^e(I,Z_t)$, with $\lambda_t^e(I,Z_t)$ defined by \eqref{IndividualIntensity}. If $e \in \mathcal{P}$, then events of type $e$ occur  in the population at a Poisson intensity of $(\mu^e_t)$.

\subsection{Proof of Theorem \ref{ThEqZ}}
\label{ProofThPathwise}
\begin{proof}[Proof of Theorem \ref{ThEqZ}] For ease of notation, we prove the case when $\mathcal{P} =\emptyset$ (there are no events with Poisson intensity).

\noindent \textbf{Step 1} The existence of a  solution to \eqref{SDE_pop} is  obtained by induction.
Let $Z^1$ be the unique solution the thinning equation:
\begin{equation*}
Z_t^1 = Z_0 + \int_0^t \int_{ \J \times \mathbb R^+ }\phi^e (s , I_k)  \mathbf{1}_{\{k \leq N_{0}\} }\mathbf{1}_{\{\theta \leq \lambda_s^e(I_k, Z_{0})\}} Q (\d s ,\d k , \d e, \d \theta ), \quad \forall  0 \leq t \leq T.
\end{equation*}
Let $T_1$ be the first jump time of $Z^1$. Since $Z_{s^-}^1 = Z_{0}$ and $N_{s^-}=N_{0}$ on $[0, T_1]$, $Z^1$ is solution of \eqref{SDE_pop} on $[0,T_1]$.

Let us now assume that \eqref{SDE_pop} admits a solution $Z^n$ on $[0,T_n]$, with $T_n$ the $n$--th event time in the population. Let $Z^{n+1}$ be the unique solution of the thinning equation:
\begin{equation*}
Z^{n+1}_t  =  Z_{t\wedge T_n}^n + \int_{t\wedge T_n}^{t} \int_{ \J \times \mathbb R^+ }\phi^e (s , I_k)  \mathbf{1}_{\{\theta \leq \lambda_s^e(I_k, Z_{T_n}^n)\}} \mathbf{1}_{\{k \leq N_{T_n}^n \} }Q (\d s ,\d k , \d e, \d \theta ).
\end{equation*}
First, observe that  $Z^{n+1}$ coincides with $Z^n$ on $[0,T_n]$. Let $T_{n+1}$ be the $(n+1)$--th jump of   $Z^{n+1}$. Furthermore,  $Z_{s^-}^{n+1} = Z_{T_n}^n$ and $N_{s^-}^{n+1}=N_{T_n}^{n}$ on $[T_n, T_{n+1}]$ (nothing happens between two successive event times),  $Z^{n+1}$ verifies for all $t\leq T_{n+1}$:
\begin{align*}
Z^{n+1}_t  =  Z_{t\wedge T_n}^n +\int_{t\wedge T_n}^{t} \int_{ \J \times \mathbb R^+ }\phi^e (s , I_k)  \mathbf{1}_{\{\theta \leq \lambda_s^e(I_k, Z_{s^-}^{n+1} )\}} \mathbf{1}_{\{k \leq N_{s^-}^{n+1} \} }Q (\d s ,\d k , \d e, \d \theta ).
\end{align*}
Since, $Z^n$ is a solution of \eqref{SDE_pop} on $[0,T_n]$ coinciding with $Z^{n+1}$, this achieves to prove that $Z^{n+1}$ is solution of \eqref{SDE_pop} on $[0,T_{n+1}]$. \\

Finally, let $Z =\lim_{n\to \infty } Z^n$. For all $n\geq 1$, $T_n$ is the $n$--th event time of $Z$,  and $Z$ is solution of \eqref{SDE_pop} on all time intervals $[0,T_n\wedge T]$ by construction. \\
By Lemma \ref{lemma:nonExplosionSDE}, $T_n \underset{n\to \infty}{\longrightarrow} \infty$. Thus, by letting $n\to \infty$ we can conclude that $Z$ is a solution of  \eqref{SDE_pop} on $[0,T]$. \\

\noindent \textbf{Step 2} Let $\tilde Z$ be a solution of  \eqref{SDE_pop}. Using the same arguments than in Step 1, it is straight forward to show that $\tilde Z$ coincides with $Z^n$ on $[0,T_n]$, for all $n \geq 1$. Thus, $\tilde{Z} = Z$, with achieves to prove uniqueness.
\end{proof}

\subsection{Proof of Lemma \ref{lemma:nonExplosionSDE}}
\label{section:proofLemma}
 The proof is obtained using pathwise comparison result, generalizing those obtained in \cite{KaaElK20}.

\begin{proof}[Proof of Lemma \ref{lemma:nonExplosionSDE}]
 Let $Z$ be a solution of \eqref{SDE_pop}. For all $e \in \mathcal{P} \cup \mathcal{E} \cup  \mathcal{E}_W$, let $N^e$ be the process counting the occurrence of events of type $e$ in the population. $N^e$ is a counting process of $\{\mathcal{F}_t\}$-intensity  $(\Lambda_t^e(Z_{t^-}))$, solution of
\begin{align}
 \label{eq:Ne} & N_t^e = \int_0^t \int_{\N\times \R^+} \boldsymbol{1}_{\{k\leq N_{s^-}\}}  \boldsymbol{1}_{\{\theta \leq \lambda_s^e(I_k, Z_{s^-})\}} Q(\d s, \d k, \{e\}, \d \theta), & \quad  \textit{ if } e \in \cE \cup \cE_W, \\
\nonumber& N_t^e = \int_0^t \int_{\R^+} \boldsymbol{1}_{\{\theta \leq \mu^e_s \}} Q^{\mathcal{P}}(\d s, \{e\}, \d \theta), & \quad  \textit{ if } e \in \cal P. \\
\end{align}
By definition,  the jump times of the multivariate counting process $(N^e)_{e \in \mathcal{P} \cup \cE \cup \cE_W}$ are  the population event times $(T_n)_{n\geq 0}$. The idea of the  proof is to show that $(N^e)_{e \in \mathcal{P} \cup \cE \cup \cE_W}$ does not explode in finite time, by pathwise domination with a simpler multivariate counting process. The first steps are to control the population size $N_t = N_0 + N^b_t + N^{en}_t$. \\

\textbf{Step 1}  Let $(\bar N^b, \bar N^{en})$ be the  2-dimensional counting process defined as follows: for $e \in \{b,en\}$, $\bar N^e_0 = 0$ and
\begin{align}
\label{eq:dominatingprocess}
& \bar N_t^e = \int_0^t \int_{\N\times \R^+} \boldsymbol{1}_{\{k\leq N_0 + \bar N_{s^-} \}}  \boldsymbol{1}_{\{\theta \leq f^e(N_0 + \bar N_{s^-})\}} Q(\d s, \d k, \{e\}, \d \theta),  \quad  \textit{ if } e \in \cE \cup \cE_W, \\
& \nonumber  \bar N_t^e = \int_0^t \int_{\R^+} \boldsymbol{1}_{\{\theta \leq \bar \mu^e\}} Q^{\mathcal{P}}(\d s, \{e\}, \d \theta)  \quad  \textit{ if } e \in \cal P,
\end{align}
with $\bar N := \bar N^b + \bar N^{en}$ and $f^e$ the function introduced in Assumption \ref{Assumption:nonExplosion}. \\
- If $b,en \in \cal P$, then $\bar N$ is a inhomogeneous Poisson process. \\
- If $b,en \in \cE \cup \cE_W$, then it is straightforward to show that conditionally to $N_0$, $\bar N$ is  a pure birth Markov process with birth intensity function $g(n) = n\big(f^b(N_0+n) + f^{en}(N_0+n)\big)$. In particular, by Assumption \ref{Assumption:nonExplosion}, $g$ verifies the standard Feller condition for pure birth Markov processes (see e.g. \cite{BanMel15}):
\[ \sum_{n=1}^\infty \frac{1}{g(n)}. \]
- Finally, if $b \in \cE$ and $en \in \cal P$ (or equivalently if $b \in \cal P$ and $en \in \cE$), then one can show easily that $\bar N$ is a pure birth Markov process with immigration, of birth intensity function $g(n)= \bar  \mu^{en} + n f^b(N_0 + n)$ (resp. $g(n)= \bar  \mu^b+ n f^{en}(N_0 + n)$), also verifying the Feller condition.

Therefore, there exists  a non-exploding solution of \eqref{eq:dominatingprocess}, by Proposition 3.3  in \cite{KaaElK20}.\\

\textbf{Step 2} The second step consists in showing that $(N^b, N^{en})$ is strongly dominated by $(\bar N^b, \bar N^{en})$, i.e that all jumps of $(N^b, N^{en})$ are jumps of $(\bar N^b, \bar N^{en})$.  Without loss of generality, we can assume that $f^e:\mathbb{N} \to (0,+\infty)$ is increasing since $f^e(n)$ can be replaced by $\sup_{\{m\leq n \} } f^e(m)$. \\
Let $e\in \{b, en\}$.
If $e \in \mathcal{P}$, then for all $s\in [0,T]$
\[ \{\theta \leq \mu_s^e\} \subset \{ \theta \leq \bar \mu^e\},\]
which yields that all jumps of $N^e$ are jumps of $\bar N^e$.\\%
If  $e \in  \cE \cup \cE_W$, the  proof by induction is analogous to the proof of Proposition 2.1 in \cite{KaaElK20}. Let $T_1^e$ be first jump time  of $N^e$, associated with the marks $(K_1^e,\Theta_1^e)$ of $Q$ (or $Q^{\mathcal{P}}$). Then, by Definition of \eqref{eq:Ne},$K_1^e \leq N_0$ and  $\Theta_1^e \leq \lambda_{T_1^e}^e (I_{K_1^e}, Z_0)$. \\
By Assumption \ref{Assumption:nonExplosion},  we have also
\[ \Theta_1^e \leq \lambda_{T_1^e}^e (I_{K_1^e}, Z_0) \leq f^e(N_0) \leq f^e(N_0 +\bar N_{T_1^{e,-}}), \quad K_1^e \leq N_0 +  \bar N_{T_1^{e,-}}.\]
Thus, $T_1^e$ is also a jump time of $\bar N^e$. By iterating this argument, we obtain that
all jump times of $N^e$ are jump times of $\bar N^e$. \\
Thus, $(N^b, N^{en})$ does not explode in finite time.\\

\textbf{Step 3} It remains to show that  for $e \notin \{b, en\}$,   $N^e$ does not explode.\\ Let $e \neq b, en$. If $e\in \mathcal P$, the proof is the same than in Step 2.  Otherwise, let:
\[ h^e_t(n) = \sup_{I \in \cI,  m \leq n } \lambda^e_t \biggl(I, \sum_{k=1}^{m} \delta_{I_k}\biggr), \quad \forall \; t \in [0,T] \;  n \in \N^*.\]
By Assumptions \ref{AssumptionIntensity1} and \ref{AssumptionIntensity2}, $h^e_t(n) <\infty$, and we can introduce the non exploding counting process $\bar{N}^e$, defined by the thinning equation :
\[ \bar N_t^e = \int_0^t \int_{\N\times \R^+} \boldsymbol{1}_{\{k\leq N_0 + \bar N_{s^-} \}}  \boldsymbol{1}_{\{\theta \leq h^e_s (N_0 + \bar N_{s^-})\}} Q(\d s, \d k, \{e\}, \d \theta),\]
with $\bar N_s = \bar N^b_s + \bar N^{en}_s$. \\
Finally, by Step 2, for $s\in [0,T]$ the population size $N_s = N_0 + N^b_s+ N^{en}_s$ is bounded a.s. by $N_0 + \bar N_s$, since all jumps of $(N^b,N^{en})$ are jumps of $(\bar N^b, \bar N^{en})$. Thus, for all $s\in [0,T]$,
\[ \{k \leq N_{s^-} \}\subset  \{k\leq N_0 + \bar N_{s^-} \}, \text{ and }  \{\theta \leq \lambda_s^e(I_k, Z_{s^-})\} \subset \{\theta \leq h_s^e(N_0 + \bar N_{s^-})\}.\]
This proves that all jumps of $N^e$ are jumps $\bar N^e$, and thus $N^e$ does not explode in finite time.

\end{proof}

\subsection{Alternative pathwise representation}

\begin{theo}
\label{ThEqZrandomized}
Let $\J_{\mathcal{E}} = \mathbb N \times \cE$ and $\J_W  = \mathbb N \times \cE_W$.\\
Let  $Q^{\cE} $ be  a random Poisson measure on $\mathbb R^+ \times \J_\cE \times \mathbb{R}^+$,  of intensity  $ \d t \delta_{\J_{\cE}}(\d k, \d e)  \mathbf{1}_{[0,\bar \lambda^e]} (\theta) \d \theta $, and  $Q^{W} $  a random Poisson measure on $\mathbb R^+ \times \J_W \times \mathbb{N} \times  \mathbb{R}^+$,  of intensity  $ \d t \delta_{\J_{\cE}}(\d k,\d e)) \delta_{\mathbb{N}} (\d j) \mathbf{1}_{[0,\bar W^e]} (\theta)\d \theta $.  Finally, let  $Q^{\mathcal P}$ be a random Poisson measure on $\mathbb R^+ \times \mathcal{P}  \times \mathbb{R}^+$,  of intensity  $ \d t \delta_{\cal P}(\d e)  \mathbf{1}_{[0,\bar \mu^e]} (\theta)\d \theta $.  \\
There exists a unique  measure-valued process $Z$, strong solution on the following SDE driven by Poisson measure:
\begin{align}
\label{SDE_pop_randomized}
\nonumber Z_t  = Z_0 &  + \int_0^t \int_{\J_\cE  \times \mathbb R^+ }\phi^e (s , I_k)  \mathbf{1}_{\{k \leq N_{s^-}\} }\mathbf{1}_{\{\theta \leq \lambda_s^e(I_k, Z_{s^-})\}} Q^\cE (\d s ,\d k , \d e, \d \theta ) \\
&  + \int_0^t \int_{\J_W  \times \N \times  \mathbb R^+ }\phi^e (s , I_k)  \mathbf{1}_{\{k \leq N_{s^-}\} } \mathbf{1}_{\{j \leq N_{s^-}\} }\mathbf{1}_{\{\theta \leq W^e(s, I_k , I_j) \}} Q^W (\d s ,\d k , \d e,  \d j ,\d \theta ),\\
& +  \nonumber    \int_0^t \int_{\mathcal{P} \times \mathbb R^+}  \phi^e(s, I_{s^-}) \mathbf{1}_{\{\theta \leq \mu_s^e \}} Q^{\mathcal{P}} (\d s ,\d e , \d \theta),
\end{align}
with $I_{s^-}$ an individual taken uniformly in $Z_{s^-}$. \\
Furthermore, the solution of \eqref{SDE_pop_randomized} has the same law than the solution of Equation \eqref{SDE_pop}.
\end{theo}

The proof Theorem \ref{ThEqZrandomized} follows the same steps than the proof of Theorem \ref{ThEqZ}.

\section{Proof of Theorem \ref{theorem:algoNoInteraction}}
\label{proof:algonointeraction}
\begin{proof}[Proof of Theorem \ref{theorem:algoNoInteraction}]
For ease of notation, we prove the case when $\mathcal{P} =\emptyset$ (there are no events with Poisson intensity).  \\
Let $Z$ be the population process obtained by Algorithm  \ref{algo:PopNointeraction2}, and $(T_n)_{n\geq 0}$ the sequence of its jump times ($T_0=0$). \\

\noindent \textbf{Step 1} Let $T_1$ be the first event time in the population, with its associated marks defining the type $E_1$ of the event and the individual $I_1$ to which this event occurs. By construction, $(T_1, E_1, I_1)$ is characterized by the first jump of: 
\begin{equation}
Q^0(\d t, \d k , \d e) = \int_{\mathbb R^+} \mathbf{1}_{\{\theta \leq \lambda_{t}^e(I_k,Z_0)\}}\bar Q^0 (\d t ,\d k , \d e, \d \theta ), 
\end{equation}
with $\bar Q^0$ the Poisson measure introduced in the first step of the algorithm described in  Section \ref{sec::simulation_algo}.

Since $T_1$ is the first event time, the population composition stays constant, $Z_t=Z_0$, on $\{t<T_1\}$. In addition, recalling that the first event has the action $\phi^{E_1}(T_1, I_1)$ (see Table \ref{TableEvAction}) on the population $Z$, we obtain that:
\begin{align*}
Z_{t\wedge T_1} & =  Z_0 + \mathbf{1}_{\{t\geq T_1\}} \phi^{E_1} (T_1 , I_1)  \\
 & = Z_0 + \int_0^{t\wedge T_1}  \int_{\J_0} \phi^e (s , I_k)  Q^0 (\d s ,\d k , \d e ) \\
& = Z_0 + \int_0^{t\wedge T_1} \int_{\J_0}  \int_{\mathbb R^+} \phi^e (s , I_k)  \mathbf{1}_{\{\theta \leq \lambda_s^e(I_k,Z_0)\}}\bar Q^0 (\d s ,\d k , \d e, \d \theta ).
\end{align*}
Since $Z_{s^-} = Z_0$ on $\{s \leq T_1\}$, the last equation can be rewritten as
\begin{equation}
\label{eq:DynZT_1}
 Z_{t\wedge T_1}   = Z_0 + \int_0^{t\wedge T_1} \int_{\J_0}  \int_{\mathbb R^+} \phi^e (s , I_k)  \mathbf{1}_{\{\theta \leq \lambda_s^e(I_k,Z_{s^-})\}}\bar Q^0 (\d s ,\d k , \d e, \d \theta ).
\end{equation}

\noindent \textbf{Step 2} The population size at the $n$--th event time $T_n$ is $N_{T_n}$. The $(n+1)$--th event type and the individual to which this event occur are thus chosen in the set
\begin{equation*}
\J_n := \{ 1,\dots, N_{T_n}\} \times (\mathcal{E} \cup \mathcal{E}_W).
\end{equation*}
%Since the population composition is constant on $[T_n, T_{n+1}[$, $Z_t = Z_{T_n}$,  the  $(n+1)$th event is also obtained recursively.
Conditionally to $\mathcal{F}_{T_n}$, let us  first introduce the marked Poisson measure $\bar Q^n$  on $[T_n, \infty) \times \mathcal J_n \times \mathbb R^+$, of intensity:
\begin{align}
\label{eq:barmun}
\bar \mu^n(\d t, \d k, \d e , \d \theta ) & := \mathbf{1}_{\{t > T_n \}}\bar \Lambda (N_{T_n})\d t  \frac{\bar \lambda^e }{\bar \Lambda(N_{T_n})} \delta_{\mathcal J_n}(\d k, \d e) \frac{1}{\bar \lambda^e} \mathbf{1}_{[0,\bar \lambda^e]} (\theta)\d \theta,\\
\nonumber&  = \mathbf{1}_{\{t > T_n \}}\d t  \delta_{\mathcal J_n}(\d k, \d e) \mathbf{1}_{[0,\bar \lambda^e]}(\theta)\d \theta .
\end{align}
By definition, $\bar Q^n$ has no jumps before $T_n$. \\
As for the first event, the triplet $(T_{n+1}, E_{n+1}, I_{n+1})$ is determined by the first jump of the measure $Q^n (\d s ,\d k , \d e) := \int_{\mathbb R^+} \mathbf{1}_{\{\theta \leq \lambda_s^e(I_k, Z_{T_n})\}}\bar Q^n (\d s ,\d k , \d e, \d \theta)$, obtained by thinning of $\bar Q^n$.   Finally, since the population composition is constant on $[T_n, T_{n+1}[$, $Z_t = Z_{T_n}$, the population on $[0,T_{n+1}]$ is defined by:
\begin{align}
\label{EqZ_Tn}
\nonumber Z_{t\wedge T_{n+1}}  &  = Z_{t \wedge T_n }  + \mathbf{1}_{\{t\geq T_{n+1}\}}\phi^{E_{n+1}}(T_{n+1}, I_{n+1}), \\
& = Z_{t \wedge T_n } + \int_{t \wedge T_n}^{t \wedge T_{n+1}} \int_{\J_n
\times \mathbb R^+} \phi^e (s , I_k)  \mathbf{1}_{\{\theta\leq \lambda_s^e(I_k, Z_{s^-})\}}\bar Q^n (\d s ,\d k , \d e, \d \theta ).
\end{align}
Applying $n$ times \eqref{EqZ_Tn} yields that:
\begin{align}
\label{EqZtnRec}
Z_{t\wedge T_{n+1}} = Z_0  + \sum_{l=0}^n \int_{t \wedge T_l}^{t \wedge T_{l+1}}\int_{ \J_l \times  \mathbb R^+} \phi^e (s , I_k)  \mathbf{1}_{\{\theta\leq \lambda_s^e(I_k, \tilde Z_{s^-})\}}\bar Q^l (\d s ,\d k , \d e, \d \theta ).
\end{align}

\noindent \textbf{Step 3} Finally, let $\tilde{Z}$ be the solution of \eqref{SDE_pop}, with $(\tilde T_n)_{n\geq 0}$ the sequence of its event times. Then, we can write similarly for all $n\geq 0$:
\begin{align*}
 \tilde Z_{t\wedge \tilde T_{n+1}} & = Z_0  + \sum_{l=0}^n \int_{t \wedge \tilde  T_l}^{t \wedge \tilde T_{l+1}} \int_{\J \times \mathbb R^+} \phi^e (s , I_k)  \mathbf{1}_{\{\theta\leq \lambda_s^e(I_k,  \tilde Z_{s^-})\}}\mathbf{1}_{ \{k \leq \tilde N_{s^-} \}}   Q(\d s, \d k , \d e , \d \theta ), \\
& = Z_0  + \sum_{l=0}^n \int_{t \wedge \tilde  T_l}^{t \wedge \tilde T_{l+1}} \int_{\J \times \mathbb R^+} \phi^e (s , I_k)  \mathbf{1}_{\{\theta\leq \lambda_s^e(I_k,  \tilde Z_{s^-})\}}\mathbf{1}_{ \{k \leq \tilde N_{\tilde T_l} \}}   Q(\d s, \d k , \d e , \d \theta ),
\end{align*}
since  $\tilde  N_{s^-} = \tilde N_{T_l}$ on $\tilde [T_l, \tilde T_{l+1}]$. \\
For each $l \geq 0$, let
\begin{equation*}
\tilde Q^l(\d t, \d k , \d e , \d \theta )  = \mathbf{1}_{\{t >  \tilde T_l\}} \mathbf{1}_{ \{1, \dots , \tilde N_{\tilde T_l} \}}(k)   Q(\d t, \d k , \d e , \d \theta ).
\end{equation*}
By proposition \ref{PropRestrictionPoissonMeasure}, $\tilde Q^l$ is, conditionally to  $\mathcal{F}_{T_l}$, a  Poisson measure of intensity
\begin{equation*}
\mathbf{1}_{\{t >  \tilde T_l\}}   \d t \mathbf{1}_{ \{1, \dots , \tilde N_{\tilde T_l} \}}(k) \delta_{\J}(\d k, \d e) \d \theta.
\end{equation*}
Noticing that $\mathbf{1}_{ \{1, \dots , \tilde N_{\tilde T_l} \}}(k) \delta_{\J}(\d k, \d e)  =\delta_{\J_l}( \d k , \d e)$, this shows that $\tilde{Q}^l$ has the conditional intensity $\bar \mu^l$ defined in \eqref{eq:barmun} and has thus the same distribution than $\bar Q^l$. Thus, $Z$ in an exact simulation of \eqref{SDE_pop}. 
\end{proof}

\bibliographystyle{alpha}
\bibliography{refs}

\newcommand{\etalchar}[1]{$^{#1}$}
\begin{thebibliography}{BBEK{\etalchar{+}}12}

\bibitem[BBEK{\etalchar{+}}12]{barrieu2012understanding}
Pauline Barrieu, Harry Bensusan, Nicole El~Karoui, Caroline Hillairet,
  St{\'e}phane Loisel, Claudia Ravanelli, and Yahia Salhi.
\newblock Understanding, modelling and managing longevity risk: key issues and
  main challenges.
\newblock {\em Scandinavian actuarial journal}, 2012(3):203--231, 2012.

\bibitem[BCF{\etalchar{+}}16]{billiard2016effect}
Sylvain Billiard, Pierre Collet, R{\'e}gis Ferri{\`e}re, Sylvie
  M{\'e}l{\'e}ard, and Viet~Chi Tran.
\newblock The effect of competition and horizontal trait inheritance on
  invasion, fixation, and polymorphism.
\newblock {\em Journal of theoretical biology}, 411:48--58, 2016.

\bibitem[Ben10]{Ben10}
Harry Bensusan.
\newblock {\em {Interest rate and longevity risk: dynamic model and
  applications to derivative products and life insurance}}.
\newblock Theses, {Ecole Polytechnique X}, 2010.

\bibitem[BM15]{BanMel15}
Vincent Bansaye and Sylvie M{\'{e}}l{\'{e}}ard.
\newblock {\em Stochastic Models for Structured Populations}.
\newblock Springer International Publishing, 2015.

\bibitem[Bou16]{Bou16}
Alexandre Boumezoued.
\newblock {\em {Micro-macro analysis of heterogenous age-structured populations
  dynamics.Application to self-exciting processes and demography.}}
\newblock Theses, {Universit{\'e} Pierre et Marie Curie}, 2016.

\bibitem[Br{\'e}81]{bremaud1981point}
Pierre Br{\'e}maud.
\newblock {\em Point processes and queues: martingale dynamics}, volume~66.
\newblock Springer, 1981.

\bibitem[CFM06]{champagnat2006unifying}
Nicolas Champagnat, R{\'e}gis Ferri{\`e}re, and Sylvie M{\'e}l{\'e}ard.
\newblock Unifying evolutionary dynamics: from individual stochastic processes
  to macroscopic models.
\newblock {\em Theoretical population biology}, 69(3):297--321, 2006.

\bibitem[CHLM16]{costa2016stochastic}
Manon Costa, C{\'e}line Hauzy, Nicolas Loeuille, and Sylvie M{\'e}l{\'e}ard.
\newblock Stochastic eco-evolutionary model of a prey-predator community.
\newblock {\em Journal of mathematical biology}, 72:573--622, 2016.

\bibitem[CIH{\etalchar{+}}20]{calvez2020horizontal}
Vincent Calvez, Susely~Figueroa Iglesias, H{\'e}l{\`e}ne Hivert, Sylvie
  M{\'e}l{\'e}ard, Anna Melnykova, and Samuel Nordmann.
\newblock Horizontal gene transfer: numerical comparison between stochastic and
  deterministic approaches.
\newblock {\em ESAIM: Proceedings and Surveys}, 67:135--160, 2020.

\bibitem[{\c{C}}in11]{Cin11}
Erhan {\c{C}}inlar.
\newblock {\em {Probability and Stochastics}}.
\newblock Springer New York, 2011.

\bibitem[CMM13]{collet2013rigorous}
Pierre Collet, Sylvie M{\'e}l{\'e}ard, and Johan~AJ Metz.
\newblock A rigorous model study of the adaptive dynamics of mendelian
  diploids.
\newblock {\em Journal of Mathematical Biology}, 67:569--607, 2013.

\bibitem[Dev86]{Dev86}
Luc Devroye.
\newblock {\em Nonuniform random variate generation}.
\newblock Springer-Verlag, New York, 1986.

\bibitem[EF11]{JSSv040i08}
Dirk Eddelbuettel and Romain Francois.
\newblock Rcpp: Seamless r and c++ integration.
\newblock {\em Journal of Statistical Software}, 40(8):1--18, 2011.

\bibitem[EHK21]{karoui2021simulating}
Nicole {El Karoui}, Kaouther Hadji, and Sarah Kaakai.
\newblock Simulating long-term impacts of mortality shocks: learning from the
  cholera pandemic.
\newblock {\em arXiv preprint arXiv:2111.08338}, 2021.

\bibitem[FM04]{FouMel04}
Nicolas Fournier and Sylvie M{{\'e}}l{{\'e}}ard.
\newblock A microscopic probabilistic description of a locally regulated
  population and macroscopic approximations.
\newblock {\em Ann. Appl. Probab.}, 14(4):1880--1919, 2004.

\bibitem[FT09]{FerTra09}
R{{\'e}}gis Ferri{{\`e}}re and Viet~Chi Tran.
\newblock Stochastic and deterministic models for age-structured populations
  with genetically variable traits.
\newblock volume~27 of {\em ESAIM Proc.}, pages 289--310. EDP Sci., Les Ulis,
  2009.

\bibitem[HBT{\etalchar{+}}23]{Rdemography}
Rob Hyndman, Heather~Booth Booth, Leonie~Tickle Tickle, John Maindonald,
  Simon~Wood Wood, and {R}~Core Team.
\newblock {\em {demography}: Forecasting Mortality, Fertility, Migration and
  Population Data}, 2023.
\newblock {R}~package version~2.0.

\bibitem[Kal17]{Kal17}
Olav Kallenberg.
\newblock {\em Random measures, theory and applications}, volume~77 of {\em
  Probability Theory and Stochastic Modelling}.
\newblock Springer, Cham, 2017.

\bibitem[KE23]{KaaElK20}
Sarah Kaakai and Nicole {El Karoui}.
\newblock Birth death swap population in random environment and aggregation
  with two timescales.
\newblock {\em Stochastic Processes and their Applications}, 162:218--248,
  2023.

\bibitem[KLAE19]{KAAKAI201916}
Sarah Kaakaï, Héloïse {Labit Hardy}, Séverine Arnold, and Nicole {El
  Karoui}.
\newblock How can a cause-of-death reduction be compensated for by the
  population heterogeneity? a dynamic approach.
\newblock {\em Insurance: Mathematics and Economics}, 89:16--37, 2019.

\bibitem[LS79]{LewShe79}
Peter Lewis and Gerald Shedler.
\newblock Simulation of nonhomogeneous poisson processes by thinning.
\newblock {\em Naval research logistics quarterly}, 26(3):403--413, 1979.

\bibitem[LSA{\etalchar{+}}19]{lavallee2019stochastic}
Fran{\c{c}}ois Lavall{\'e}e, Charline Smadi, Isabelle Alvarez, Bj{\"o}rn
  Reineking, Fran{\c{c}}ois-Marie Martin, Fanny Dommanget, and Sophie Martin.
\newblock A stochastic individual-based model for the growth of a stand of
  japanese knotweed including mowing as a management technique.
\newblock {\em Ecological Modelling}, 413:108828, 2019.

\bibitem[MRR19]{meleard2019birth}
Sylvie M{\'e}l{\'e}ard, Michael Rera, and Tristan Roget.
\newblock A birth--death model of ageing: from individual-based dynamics to
  evolutive differential inclusions.
\newblock {\em Journal of mathematical biology}, 79:901--939, 2019.

\bibitem[RJMR22]{roget2022positive}
Tristan Roget, Pierre Jolivet, Sylvie M{\'e}l{\'e}ard, and Michael Rera.
\newblock Positive selection of senescence through increased evolvability:
  ageing is not a by-product of evolution.
\newblock {\em bioRxiv}, pages 2022--03, 2022.

\bibitem[Tra08]{tran_2008}
Viet~Chi Tran.
\newblock Large population limit and time behaviour of a stochastic particle
  model describing an age-structured population.
\newblock {\em ESAIM: Probability and Statistics}, 12:345--386, 2008.

\bibitem[VKM18]{stmomo}
Andr{\'e}s~M. Villegas, Vladimir~K. Kaishev, and Pietro Millossovich.
\newblock {StMoMo: An R Package for Stochastic Mortality Modelling}.
\newblock {\em Journal of Statistical Software}, 84:1--38, 2018.

\bibitem[VMKH18]{Rstmomo}
Andres Villegas, Pietro Millossovich, and Vladimir Kaishev~Hyndman.
\newblock {\em {StMoMo}: Stochastic Mortality Modelling}, 2018.
\newblock {R}~package version~0.4.1.

\bibitem[ZGHU09]{zinn2009mic}
Sabine Zinn, Jutta Gampe, Jan Himmelspach, and Adelinde~M Uhrmacher.
\newblock Mic-core: A tool for microsimulation.
\newblock In {\em Proceedings of the 2009 Winter Simulation Conference (WSC)},
  pages 992--1002. IEEE, 2009.

\bibitem[Zin14]{Zin14}
Sabine Zinn.
\newblock {The MicSim package of R: an entry-level toolkit for continuous-time
  microsimulation}.
\newblock {\em International Journal of Microsimulation}, 7(3):3--32, 2014.

\end{thebibliography}

\end{document}